\documentclass[a4paper,11pt]{article}

\usepackage[margin=1.2in]{geometry}
\usepackage{xargs}
\usepackage{rotating}
\usepackage{graphicx}
\usepackage{latexsym}
\usepackage{longtable}
\usepackage{verbatim}
\usepackage{float}
\usepackage{fancyhdr}
\usepackage{comment}
\usepackage{lscape}
\usepackage{hyperref}
\usepackage[normalem]{ulem}
\useunder{\uline}{\ul}{}
\usepackage{multirow}

\usepackage[colorinlistoftodos,prependcaption,textsize=normalsize]{todonotes}

\newcommandx{\moharram}[1] {\todo[linecolor=yellow, backgroundcolor=yellow!75, bordercolor=yellow] {#1}}

\title{
(Technical Report)\\
A Systematic Literature Review on Model-driven Engineering for Cyber-Physical Systems
\footnote{
Mustafa Abshir Mohamed would like to thank Turkish government for Türkiye Scholarships (YTB) program. Moharram Challenger and Geylani Kardas would like to thank the European Cooperation in Science \& Technology (COST) Action networking mechanisms and their support of COST Action IC1404: Multi-Paradigm Modelling for Cyber-Physical Systems (MPM4CPS). COST is supported by the EU Framework Programme Horizon 2020.
}}

\author{
\\Mustafa Abshir Mohamed,\\
\textit{mustafaxoodiye@gmail.com,}\\
Ege University, Izmir, Turkey\\
\\
Geylani~Kardas,\\
\textit{geylani.kardas@ege.edu.tr,}\\
Ege University, Izmir, Turkey\\
\\
Moharram~Challenger,\\
\textit{moharram.challenger@uantwerpen.be,}\\
University of Antwerp and Flanders Make, Belgium\\
\\
}

\date{Dec. 02, 2020}

\begin{document}


\maketitle

\begin{abstract}
This technical report presents a Systematic Literature Review (SLR) study that focuses on identifying and classifying the recent research practices pertaining to CPS development through MDE approaches. The study evaluates 140 research papers published during 2010–2018. Accordingly, a comprehensive analysis of various MDE approaches used in the development life-cycle of CPS is presented. Furthermore, the study identifies the research gaps and areas that need more investigation. The contribution helps researchers and practitioners to get an overall understanding of the research trends and existing challenges for further research/development.\\

\end{abstract}

\pagebreak
\tableofcontents

\noindent\hrulefill

\vspace{10mm}

\section{Introduction} \label{introduction}

For the last 2 decades, one of the most profitable and productive businesses is software production industry. Despite the profitability of software production, the economic cost that arose as a result of the inadequacies in the design and architecture cannot be understated. One of the major catalysts for ineffective software and high costs of production is the complexity inherent in the software itself.

To overcome the complexity aspect of the software, engineers and researchers continuously seek to raise the abstraction level of software development by providing solutions like Model-driven Engineering (MDE). To develop a large-scale complex systems such as Embedded systems, Internet of Things (IoT), and Cyber-physical Systems (CPS) that are composed of various software and hardware components, the system need to be designed and implemented using higher-level of abstraction.

\subsection{Cyber-physical systems}

\begin{figure}[H]
\centering
	\includegraphics[scale=0.6]{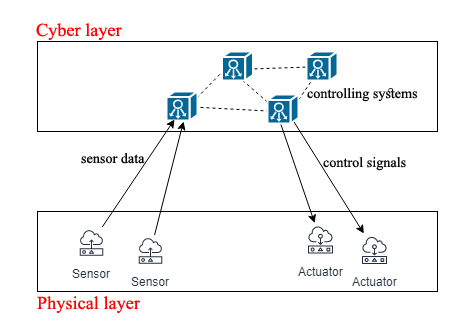}
	\caption{General CPS architecture}
	\label{fig:CPS_architecture}
\end{figure}

The first emerge for the term "Cyber-physical system" (CPS) was in 2006 at the National Science Foundation \cite{lee2015past}. CPS is a system that its computational and communication components control and monitor physical phenomena \cite{lee2008cyber}. In CPS, sensors monitor and measure certain physical phenomena, like pressure, temperature, light, touch, etc. The measured data are transferred to the controllers/software through communication elements (i.e. wired/wireless network). The controllers/software make decisions/actions based on the received data from the sensors and send them through communication elements to actuators which in return make changes to the physical phenomena \cite{sanislav2017dependability}, the feedback loop. The overall architecture of a CPS is depicted in (Figure \ref{fig:CPS_architecture}).

Applications of CPS include, but are not limited to, monitoring complex real-world phenomena, smart manufacturing (i.e. industry 4.0), smart building, critical infrastructures, like chemical and power plants, smart grids, natural gas distribution systems, transportation systems etc. \cite{lee2008cyber}. 

Significant challenges come across developers of CPS due to its heterogeneous nature, such as the need for knowledge and skills from multiple academic disciplines, the integration of the artifacts of those various disciplines, and the difficulty of the maintenance activities of such heterogeneous artifacts. As a practical example, developing a CPS requires a group of developers from different disciplines like software engineering, electric engineering, electronic engineering, and so on. This induces communication challenges among these engineers due to the different tools and abstractions used in each discipline. Another challenge can be the time consumed to comprehend and integrate the codes written by these different developers. The maintenance of such a code with various levels of abstractions is executive or very difficult. In order to eliminate these challenges and reduce the complexity of CPS development, one of the key approaches is Model-driven Engineering (MDE), which is frequently used in many business domains for software development \cite{france2007model}.

\subsection{Model-driven engineering}

In general, models have two features. Reduction feature where the models focus on the main properties of a system and neglect the details to keep the representation of the system relatively easier, and mapping feature whereby models are generalized from a prototype of the original system. Models could be used for different purposes such as sketches, blueprints, or programs. There is an increasing need for the use of such models in software development for the following reasons \cite{brambilla2017model};

\begin{itemize}
    \item the increasing demand for accelerating the development process.
    \item the increasing complexity of software artifacts relative to the software functionality demanding abstraction level to be raised to facilitate the maintenance process or future upgrading.
    \item the need for a medium language between the non-developers (e.g., customers, managers, business stakeholders, etc.) and the software developers.
\end{itemize}

The use of models as the basic building blocks for the development of software artifacts is called Model-driven Engineering (MDE). MDE paradigm raises the abstraction level of software/system development from low-level artifacts to a higher-level of models. MDE bridges the gap between problem identification and software/system implementation phases. This can be done by thoroughly/partially generating either software implementations (C++, Java, and C\#) or deployment artifacts (XML-based configuration) from models that describe the system at multiple levels of abstraction, and from a variety of perspectives \cite{france2007model}.

\subsection{Methodologies for secondary studies}

Generally, to conduct a secondary study, there are several methodologies that can be followed. Survey research is one of them. As it is proposed by \cite{pinsonneault1993survey}, there is a significant difference between survey research and survey. A survey is a quantitative method that aims to collect data about features, behaviors or opinions of a specific group of people as a representative for a target population. On the other hand, survey research is a methodology that is specified to conduct surveys for advanced scientific knowledge or research purposes.

Systematic Mapping (SM) is a method of collecting previously written research papers, articles, conference papers, book chapters etc. in a certain area of study based on a set of questions made by the researchers. Collected documents can be later reviewed, analyzed and structured in different categories to provide a wider view of the research area. Thereby, it could be much easier to determine those areas that need more research studies to be done. In this regard, it serves as a valuable basis for future researchers \cite{petersen2008systematic}.

Systematic Literature Review (SLR), also known as a systematic review, is a form of a secondary review that aims to identify, evaluate and interpret all the available researches that are relevant to particular research questions, topic area, or phenomenon of interest. It collects and critically scrutinizes data from the studies included in the review (aka. primary studies) \cite{kitchenham2007guidelines}.

The procedures used in conducting SM and SLR are nearly identical. However, they are different in terms of their goals, processes, and results. Firstly, SM focuses on categorizing the conducted studies based on a thematic analysis. Whereas, SLR focuses on empirical evidence in its categorization. Secondly, SM studies show the research gaps in a specific area of study. In addition to these features of SM, SLR makes an in-depth comparison between tools, techniques, and approaches proposed by the primary studies. In terms of the process, in the inclusion and exclusion phase of the SM, thematic analysis (e.g., reading the title, abstract, introduction, and conclusion sections of the publication) is used. Meanwhile in SLR, the studies shall be read in-depth (e.g. by additionally reading the methodology and other required sections in the publication).

In conclusion, SLR can be an intensive and more-in-depth version of SM. In this study, the SLR is adopted as the research methodology. This study is achieved by following the process proposed in \cite{de2007scientific} and \cite{siddaway2014systematic} and using guidelines defined in \cite{kitchenham2007guidelines}.

\subsection{Problem statement }

A unified development methodology for CPS has not been standardized yet. The abundance of different hardware platforms available makes the development of such systems very complex. There is a need for a unified methodology that permits efficient raise of the abstraction level to overcome issues of heterogeneity induced by the multidisciplinary nature of the system. Towards this goal, many researchers believe that MDE is an alternative solution to overcome challenges such as development complexity, heterogeneity, adaptability, and reuse and they propose various applications of MDE for CPS development. However, no secondary study highlighting 1- previous researches 2- current research efforts 3- open challenges related to applying MDE approaches for CPS has been done yet. This overview would be helpful to both researchers and practitioners for discovering the pros and cons for applying MDE for CPS and for identifying interesting research directions. Without such a secondary study, it is cumbersome to determine what was proposed in the literature, what has been successfully completed and what rather has failed or missing.

\subsection{Contributions}

The aim of this study is conducting an SLR of the studies used MDE techniques such as Domain-specific modeling, Metamodeling, Validation \& Verification, Model Transformation, and Artifact Generation for CPS. Evaluation of research questions and analysis of the proposed approaches and toolchain in the primary studies is performed as the result of the systematic review in the scope of this study. Furthermore, in this work, trends, bibliometrics and demographics are provided to help collecting important information such as; the active authors/researchers in this domain, the publication per year for each country and other information.

The results of this study may help the researchers to easily reach the desired class of studies and related publications considering the tools, technologies, approaches, and best practices used. This study also enables researchers avoid unnecessary duplications of trial and error. Finally, it identifies research gaps and areas need more investigations and determine best practices, tools, techniques, and languages which can be used.

The remainder of the paper is organized as follows: Section 2 reports the related work. Section 3 describes the research methodology and protocol definition to carry out this SLR, then, the procedure of conducting the SLR is discussed in Section 4. Section 5 elaborates the results. Finally the discussion of the results and the conclusions are presented in sections 6 and 7 respectively.

\newpage
\section{Related Work} \label{RelatedWork}

Since the scope of this study is to present an SLR on the state-of-the-art of MDE for CPS, the related secondary studies (surveys, SMs, SRLs, Tertiary Studies) are addressed in this section as the related work. Although, there is no secondary study exactly on this topic, the following studies are relevant to the topic. Table \ref{tab:my-table} presents a summary of these related work.

\begin{table}[H]
\centering
\caption{Summarized Related work}
\label{tab:my-table}
\begin{tabular}{|p{0.1\textwidth}|p{0.15\textwidth}|p{0.25\textwidth}|p{0.12\textwidth}|p{0.09\textwidth}|p{0.15\textwidth}|}
\hline
\textbf{Study} & \textbf{Methodology} & \textbf{Domain} & \textbf{Paradigm} & \textbf{\# of Papers} & \textbf{Studies b/w years} \\ \hline
\cite{barivsicsystematic} & SLR & CPS & MPM & 265 & 2006-2017 \\ \hline
\cite{rashid2015toward} & SLR & Embedded Systems & MBSE & 61 & 2008-2014 \\ \hline
\cite{queiroz2014development} & SLR & Safety-critical embedded systems & PLE+MDE & 19 & - \\ \hline
\cite{casalaro2015model} & SM & Mobile robot systems & MDE & 69 & After 2000 \\ \hline
\cite{liebel2018model} & Survey & Embedded systems & MBE & - & - \\ \hline
\cite{agner2013brazilian} & Survey & Embedded software & UML+MDA & - & - \\ \hline
Our work & SLR & CPS & MDE & 140 & 2010-2018 \\ \hline
\end{tabular}
\end{table}

In \cite{barivsicsystematic}, an SLR on multi-paradigm modelling for cyber-physical systems is presented where authors focuses on studies promoting multi-modeling, multi-view and multi-formalism approaches for the development of CPS. The study reported the most used approaches and tools in the primary studies for multi-paradigm modeling as well as indicating the type of formalism presented, and which language/tool is used for implementing it. Furthermore, they report the actors and stakeholders involved in the modeling process and their background knowledge. 

The authors of \cite{rashid2015toward} conducted SLR of the development of embedded systems using model-based system engineering (MBSE) approach. The study reviewed 61 research papers published in one of the four renowned scientific databases (IEEE, SPRINGER, ELSEVIER, and ACM) during the years 2008-2014. Subsequently, primary studies are grouped into six categories according to their relevance to the corresponding model-based system engineering activity namely general category, modeling category, model transformation category, model verification category, simulation category, and property specification category. As the result, the study presents 28 tools which support modeling, model transformation, validation, and verification activities. The study examined the utilization of UML and SysML/MARTE profiles, and it also analyzed the application of both model-to-model and model-to-text transformations.

Another SLR is presented in \cite{queiroz2014development} in which the authors investigate studies combining Product Line Engineering (PLE) and MDE for the development of safety-critical embedded systems. This study further examined whether there are empirical studies applied the aforementioned techniques in the development process of safety-critical embedded systems. The study expose that in recent years, use of MDE combined with PLE techniques to build safety-critical embedded systems is gradually growing. The study also states that the proposed approaches in the primary studies are not compared with any other related studies, besides, these approaches do not explicitly differentiate between the software and hardware variabilities.

An SM study is presented in \cite{casalaro2015model}. This study investigates the implementations of MDE in the field of mobile robot systems (MRS). In this study, 69 research papers were selected, and as a result, the authors found out that many domain-specific modeling languages (DSMLs) are supported with tools which are mostly built ad-hoc. Also, they reported that the solutions based on UML and using Eclipse-based tools were less preferred in this field.

A survey is presented in \cite{liebel2018model} in which the quantitative data from 113 subjects were collected to provide the current state-of-practice (SoP) and challenges faced by the domain of embedded systems due to weaknesses in model-based engineering (MBE). The survey has two research questions, the first question is related to capturing the state of MBE practice in the embedded systems domain, how much activities concern MBE compared to non-MBE, and understanding the pros and cons of adopting and deploying MBE. The second question is about estimating whether there are important variations in the SoP between different groups in the embedded systems domain. As a result, the study provides information about the used methods and tools, purposes of models, effects of using it, and weaknesses of MBE. Furthermore, answers to the survey show that most of the participants believe that the positive outcomes of MBE distinctly exceeded the negative outcomes. Nonetheless, survey participants mentioned weaknesses such as the interoperability challenges amongst existing tools, and MBE needs high efforts to train the developers.

Another survey is presented in \cite{agner2013brazilian}. The study introduces statistical findings about the use of UML modeling and model-driven approaches for the design of embedded software in Brazil. The goal of this study is to identify gaps in the knowledge of how exactly UML and MDE or Model-driven Architecture (MDA) are used in industry, and to provide an understanding of how social and organizational factors impact the use of UML and MDE/MDA.

Unlike the work presented in \cite{rashid2015toward} and \cite{queiroz2014development}, the current study focuses on conducting SLR on studies concerning the development of CPS using the MDE paradigm. The work herein and the SLR given in \cite{barivsicsystematic} both consider the development of CPS. However, the current study differentiates from \cite{barivsicsystematic} in terms of the results since the current work finds MDE approaches and tools used for the development of CPS, the most addressed MDE phase, developed tools and languages in this regard, and also presenting reported CPS challenges by the primary studies. These were not taken into account in \cite{barivsicsystematic}.

\newpage
\section{Methodology} \label{methodology}

In this section, the applied methodology for the conducted SLR is discussed. In the following sections, initially, the process to be followed during this work is described, followed by defining research questions, search and studies selection strategy, then specifying inclusion and exclusion criteria and quality assessment and self-assessment criteria, and finally determining the procedure to follow for data extraction.

\subsection{Process}

\begin{figure}
\centering
	\includegraphics[scale=0.16]{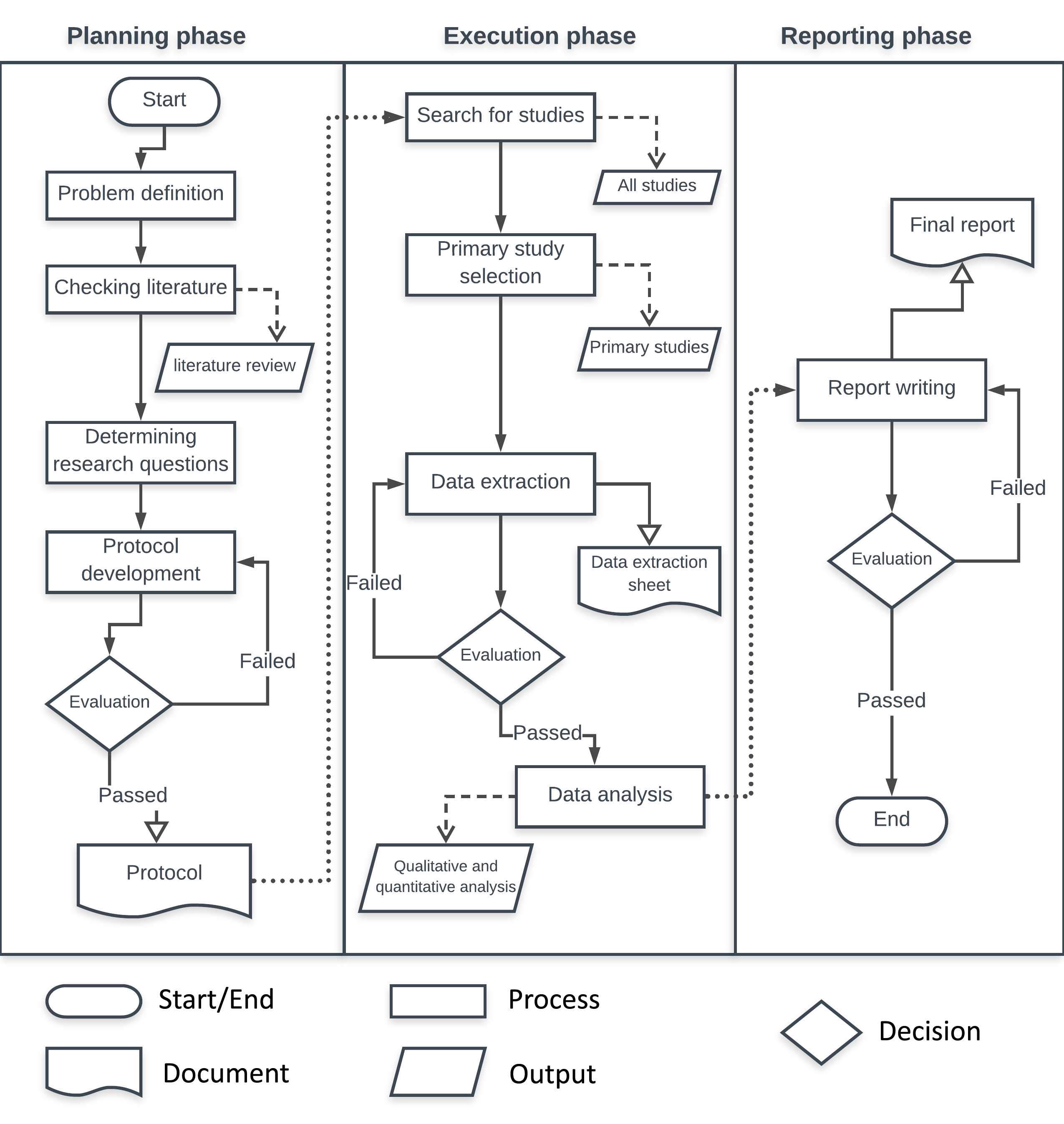}
	\caption{Overview of the systematic literature review process}
	\label{fig:SLR_process}
\end{figure}

The procedure of the systematic review was developed by following the guidelines defined in \cite{kitchenham2007guidelines}. Figure \ref{fig:SLR_process} shows an overview of the followed process. SLR composes three main phases; review planning, review execution, and review reporting \cite{de2007scientific}.

\begin{enumerate}
    \item Planning phase
    \begin{enumerate}
        \item Protocol development: Research scope and review protocol are developed; the protocol is subject to improvement in later stages through an iterative manner.
        \item Determining research questions: Research questions are defined following the PICOC guidelines \cite{kitchenham2007guidelines}. 
    \end{enumerate}
    \item Execution phase
    \begin{enumerate}
        \item Collecting studies: Search keywords are formalized, and then, studies are collected.
        \item Determining proper studies: Publications are included or excluded based on the inclusion and exclusion criteria defined in the protocol.
        \item Extracting data from studies: Information is extracted from the studies according to the research questions.
        \item Data analysis: Data extracted from primary studies are analyzed to answer the research questions.
    \end{enumerate}
    \item Reporting phase: This phase involves the systematic discussion and reports the outcomes of the analysis.
\end{enumerate}

\subsection{Team}

The team in this study consists of the following 3 people:
\begin{itemize}
    \item \textbf{Primary researcher:} 
    Mustafa Abshir Mohamed, a graduate student in Information Technologies at Ege University, Turkey, with knowledge of CPS and MDE. He conducted the SLR and carried out most of the tasks of the SLR study. 
    \item \textbf{Secondary researcher:} 
    Dr. Geylani Kardaş, an associate professor at Ege University, Turkey. He is a senior researcher with many years of expertise in MDE. He supported the primary and the secondary researchers during protocol definition, analysis of the findings, and report writing. He also helped resolving the conflicts between the findings of the primary and the secondary researchers. He is also the supervisor of the primary researcher.
    \item \textbf{Secondary researcher:} 
    Dr. Moharram Challenger, a research professor at University of Antwerp, Belgium. He is active in the fields of IoT, CPS, and MDE. He also has knowledge in conducting SLR and SM studies. He regularly supported and reviewed the work performed by the primary researcher. He is also the co-supervisor of the primary researcher.
\end{itemize}

\subsection{Research questions}

In this study, the state-of-the-art MDE techniques in CPS are taken into consideration. To address them, research questions were identified by following PICOC criteria \cite{kitchenham2007guidelines} (see Table \ref{tab:PICOC_definition}).

\begin{table}[H]
\centering
\caption{PICOC criteria definition.}
\label{tab:PICOC_definition}
\begin{tabular}{|p{0.15\textwidth}|p{0.85\textwidth}|}
\hline
\textbf{Population} & CPS \\ \hline
\textbf{Intervention} & MDE techniques for CPS \\ \hline
\textbf{Comparison} & not applicable \\ \hline
\textbf{Outcome} & report on the current state-of-the-art   approaches, languages, tools, and challenges of MDE for CPS. \\ \hline
\textbf{Context} & peer-reviewed publications. \\ \hline
\end{tabular}
\end{table}

Following the defined PICOC mentioned above, the research questions of this study are determined as below:

\begin{itemize}
    \item \textbf{RQ1:} Are any of MDE approaches or techniques used in/for the development of the studied CPS?
    \textbf{Objective:} With answering this question, the existing MDE approaches for CPS, modeling purpose, and the MDE phase addressed are reported.
    \begin{itemize}
        \item \textbf{RQ1.1:} What is the modeling approach presented/used in the study?
        \item \textbf{RQ1.1.1:} What is the purpose for which the models were used?
        \item \textbf{RQ1.2:} Which phase(s) of the system development is/are addressed in the study (using MDE)?
    \end{itemize}
    \item \textbf{RQ2:} Is/Are there any tool(s) used to apply MDE in/for CPS in the study?
    \textbf{Objective:} With answering this question, the used languages and tools, also, developed tool(s) by the study, its availability and the used language to develop the tool are reported.
    \begin{itemize}
        \item \textbf{RQ2.1:} Which language/tool(s) is/are presented/used in each phase of the system development? 
        \item \textbf{RQ2.2:} If any tool(s) is/are developed in the study, is/are this/these tool(s) reported?
        \item \textbf{RQ2.3:} Is/are the developed tool(s) available and/or accessible?
        \item \textbf{RQ2.4:} What is/are the framework(s) or programming language(s) for the development of this/these tools?
    \end{itemize}
    \item \textbf{RQ3:} What is/are the CPS component(s) addressed in the study? 
    \textbf{Objective:} This question is aimed to report on the CPS component such as sensor, cyber component, physical component, actuator etc. that is modeled. 
    \item \textbf{RQ4:} Does the study present any application domain?	
    \textbf{Objective:} It is aimed to report the CPS domain like critical infrastructure, Smart Buildings, Industry 4.0 etc. which the primary studies are targeting. 
    \begin{itemize}
        \item \textbf{RQ4.1:} What is the application domain?
        \item \textbf{RQ4.2:} What is the use case?
    \end{itemize}
    \item \textbf{RQ5:} Is there any evaluation presented in the study?	
    \textbf{Objective:} Reporting on the evaluation method followed by these primary studies such as case study, use case, example and empirical study.
    \begin{itemize}
        \item \textbf{RQ5.1:} What is the evaluation approach?
        \item \textbf{RQ5.2:} If the evaluation is based on a case study, what is the case study?
    \end{itemize}
    
    \item \textbf{RQ6:} Does the study address any challenge(s)?	
    \textbf{Objective:} Reporting on the CPS challenges which primary studies are addressing, also, challenges addressed during tool development/usage by the primary studies. 
    \begin{itemize}
        \item \textbf{RQ6.1:} Which CPS challenge(s) does the paper address?
        \item \textbf{RQ6.2:} Does the study report challenges addressed during developing the approach/tool?
    \end{itemize}

\end{itemize}
 	
\subsection{Search and selection strategy}

This stage could be considered as one of the most important and critical stages when conducting a secondary study (i.e. in this case, SLR). Therefore, it should be carefully defined since the search of primary studies should ensure the comprehensive coverage of the topic under consideration. For a search strategy to be optimal, it needs to simultaneously include utmost relevant primary studies (i.e. recall) and exclude irrelevant ones (i.e. precision).  One can deduce that an optimal search strategy must have 100\% recall and/or 100\% precision. Nevertheless, it is unpromising that a search strategy gives 100\% in both/either recall and/or precision. Accordingly, one should come up with a gratifying trade-off search strategy (i.e. good enough), that results in not many relevant studies missed, and a manageable quantity of irrelevant studies included \cite{skoglund2009reference}.

\begin{figure} [H]
\centering
	\includegraphics[scale=1.1]{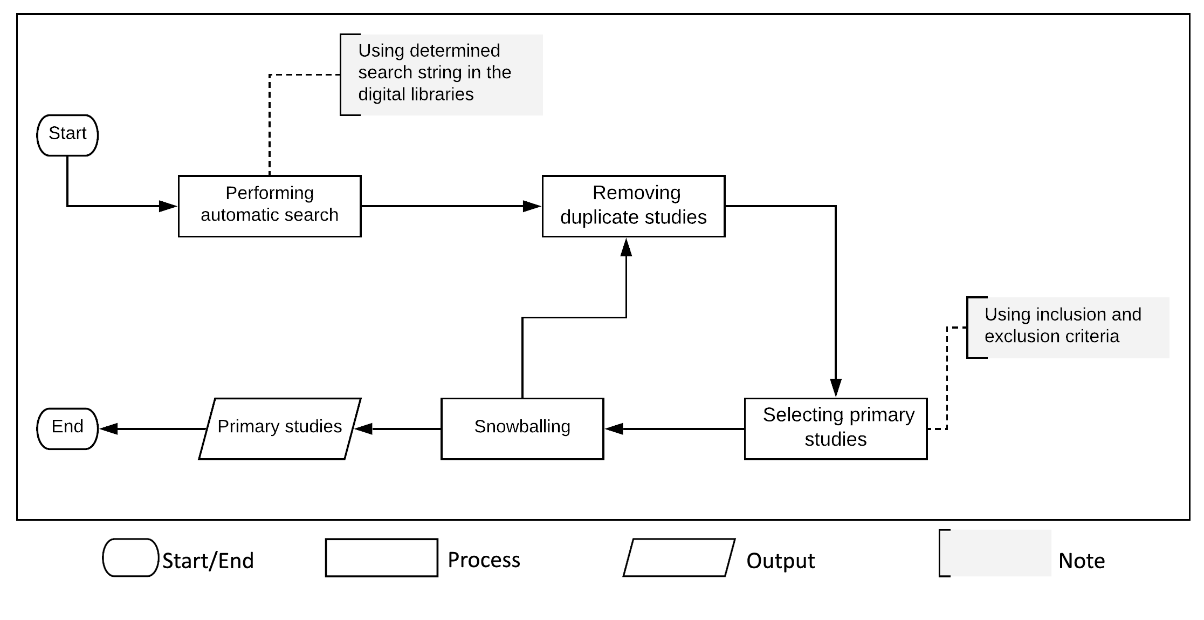}
	\caption{Search and selection strategy}
	\label{fig:Search_and_selection_strategy}
\end{figure}

The search strategy, developed in this thesis, composes four stages. Firstly, an automatic search over the most relevant scientific digital libraries was performed. Secondly, all duplicate papers were removed. Thirdly, following predetermined criteria of inclusion, only papers related to the topic were considered. Eventually, further studies were searched by forward snowballing. The composition of the search and selection strategy followed in this work is shown in Figure \ref{fig:Search_and_selection_strategy}.

\textbf{Stage 1: Performing automatic search}

To get as many related primary studies as possible, an automatic search on the digital libraries shown in Table \ref{tab:Digital_libraries} was performed.

\begin{table}[H]
\centering
\caption{Digital Libraries}
\label{tab:Digital_libraries}
\begin{tabular}{|l|l|l|}
\hline
\multicolumn{1}{|c|}{\textbf{Digital Library}} & \multicolumn{1}{c|}{\textbf{URL}} & \multicolumn{1}{c|}{\textbf AccessDate} \\ \hline
ACM & https://dl.acm.org/ & Oct/2018 \\ \hline
Dblp & https://dblp.uni-trier.de/ & Oct/2018 \\ \hline
IEEE Xplore & https://ieeexplore.ieee.org/ & Oct/2018 \\ \hline
ScienceDirect & https://www.sciencedirect.com/ & Oct/2018 \\ \hline
Scopus & https://www.scopus.com/ & Dec/2018 \\ \hline
Web of Science & https://www.webofknowledge.com/ & Oct/2018 \\ \hline
\end{tabular}
\end{table}

PICOC criteria \cite{kitchenham2007guidelines} were used to define the keywords shown in Table \ref{tab:Keywords_definition}, which leads to form “good enough” search strings. These search strings are used when performing automatic search in the aforementioned digital libraries. 

\begin{table}[H]
\centering
\caption{Keywords definition based on PICOC criteria}
\label{tab:Keywords_definition}
\begin{tabular}{|p{0.15\textwidth}|p{0.85\textwidth}|}
\hline
\textbf{Population} & "cyber-physical system*" OR  "cyber physical system*" OR "smart system*" OR  "cyberphysical systems" OR "cps" \\ \hline
\textbf{Intervention} & MDE OR MDD OR MDA OR "model-driven *" OR "model driven *" OR "code generation" OR  "generative approach*" OR "model-based approach*" OR  "domain specific model*" OR metamodel* OR "meta-model*"  OR "meta model*" OR "modeling approach*" \\ \hline
\textbf{Comparison} & Not applicable \\ \hline
\textbf{Outcome} & Report on the current state-of-the-art approaches, languages, tools and challenges of MDE for CPS. \\ \hline
\textbf{Context} & Peer-reviewed publications. \\ \hline
\end{tabular}
\end{table}

The overall search string is as follows:

("model-driven development" OR "model-driven engineering" OR "model-driven architecture" OR "code generation" OR "generative approach" OR "model-based approach" OR "model-driven approach" OR "domain specific model*" OR metamodel OR "meta-model" OR "meta model" OR "modeling approach") 
AND
("cyber-physical system*" OR "cyber physical system*" OR "smart system*" OR "cyberphysical systems" OR "cps")

Due to the different syntax of each digital library, a specific search string for each of these libraries was created. This is to ensure including as much relevant primary studies as possible. 

\textbf{Stage 2: Removing duplicate studies}

Initially, pool of primary studies was kept in Mendeley  reference manager \footnote{https://www.mendeley.com}. Also, this repository was used to facilitate the process of determining duplicate studies. Two papers are considered as duplicate if: 

\begin{itemize}
    \item their title, author(s), publication date and venue are the same. In case of different versions of the same paper, the most recent is kept.
    \item the same paper is published in different venues, one of them is selected (the most recent).
    \item the same study has both journal and conference publications, the journal publication is considered as it contains the extended study and provides more information.
\end{itemize}

\textbf{Stage 3: Selecting primary studies}

In this stage, primary studies are selected following predefined inclusion and exclusion criteria (see Section \ref{I&E_criteria}). Only those studies matching the criteria are included in the final pool of the research. The criteria were applied considering the reading of TITLE, ABSTRACT, KEYWORDS, INTRODUCTION sections, however, if it is not enough for reaching a decision, other parts like METHODOLOGY and CONCLUSION are considered. 

\textbf{Stage 4: Forward Snowballing}

To ensure no left potential primary studies, papers which might not have been reached by automatic searching or published in the predefined digital libraries were also searched. According to \cite{wohlin2014guidelines}, forward snowballing process is realized by identifying other publications citing any of the primary studies. Google Scholar is used to find those papers. Consequently, newly found and selected papers are added to the final pool.

\subsection{Inclusion \& Exclusion Criteria (Selection Criteria)} \label{I&E_criteria} 

Once all potentially relevant papers are gathered, their relevance must be assessed. Selection criteria are intended for the purpose of identifying those papers (primary studies) directly related to the research questions as suggested in \cite{kitchenham2007guidelines}. The inclusion and exclusion criteria must be based on the research questions. These criteria are applied when selecting the primary studies and when performing forward snowballing. To reduce the potentiality of a bias to occur, these criteria should be documented in the protocol definition stage. The selection criteria might be revised during the search process. Inclusion and exclusion criteria are applied to a paper/study by reading sections like title, abstract, introduction, and conclusion.

\begin{itemize}
    \item a paper is included in the primary studies pool only if it meets all the inclusion criteria and none of the exclusion criteria. 
\end{itemize}

\subsubsection{Inclusion criteria}

\begin{itemize}
    \item \textbf{IC1:} Study must propose at least one of the MDE approaches or techniques for CPS.
    \item \textbf{IC2:} Study must target CPS or at least one of its application domains.
    \item \textbf{IC3:} Study must be peer-reviewed journal papers, workshop papers or conference papers.
    \item \textbf{IC4:} Models presented by the study must not be used only for documentation and design purposes.
    \item \textbf{IC5:} Paper publication period must be between 2010 and 2018.
    \item \textbf{IC6:} Study must be available in full-text and published in a renowned digital library.
\end{itemize}
	
\subsubsection{Exclusion criteria}

\begin{itemize}
    \item \textbf{EC1:} Study is a secondary study (survey, systematic mapping, systematic review, etc.).
    \item \textbf{EC2:} Study is irrelevant to CPS or any of its application domains and the field of software engineering.
    \item \textbf{EC3:} The study is a summarized version of a complete work already in the SLR pool.
    \item \textbf{EC4:} Study is a kind of educational, editorial, tutorial, or other material (i.e., not a scientific paper).
    \item \textbf{EC5:} Study was written in other languages than English.
\end{itemize}

\subsection{Study quality assessment and self-assessment}

Quality assessment (QA) and self-assessment (SA) questions similar to \cite{barivsicsystematic} are formed. The QA questions (see Table \ref{tab:Quality_self-assessment}), are defined to assess the quality of the studies.  QA-1 measures the degree of clarity to which the primary studies define the problem they are addressing. QA-2 answers whether the primary studies reported the contributions of their study clearly. QA-3 reports whether the study presents any future works or not. The results of the quality assessment questions are presented in Section \ref{results}.

SA question SA-1, see Table \ref{tab:Quality_self-assessment}, is used to validate the understanding of the primary reviewer regarding the reviewed paper. If the primary reviewer gives a self-assessment score below 50\% (i.e. not very confident about the paper) for a study, the secondary reviewer revises the extracted data from the study.

\begin{table}[H]
\centering
\caption{Quality assessment and self-assessment}
\label{tab:Quality_self-assessment}
\begin{tabular}{|p{0.25\textwidth}|p{0.07\textwidth}|p{0.65\textwidth}|}
\hline
\textbf{Quality Assessment} &  &  \\ \hline
 & QA-1 & What is the level of fineness in which the problem of the study defined? \\ \hline
 & QA-2 & To which extent the promises of the study are explicit? \\ \hline
 & QA-3 & Does the study present any future work(s)? \\ \hline
\textbf{Self-assessment} &  &  \\ \hline
 & SA-1 & What is the reader's trust level regarding the paper? \\ \hline
\end{tabular}
\end{table}

\subsection{Data extraction}

Initially, the final pool of the primary studies is stored in Mendeley. Next, Google sheet is used for the data extraction stage. In the sheet, research questions are represented in columns, whereas, primary studies are presented in rows. The process of data extraction in this study goes through 3 phases. Data extraction form is shown in Table \ref{tab:Data_extraction_form}. 

\begin{itemize}
    \item \textbf{Phase 1:} The primary reviewer starts extracting data from the primary studies (answering research questions). Extracted data for each study is represented in a row where each row has a key that refers to the study in Mendeley. Data extraction of each paper is followed by answering quality and self-assessment questions.
    \item \textbf{Phase 2:} The secondary reviewer starts reviewing primary studies with self-assessment score below 50\%. After evaluating the study, if the secondary reviewer agrees with the answers given by the primary reviewer, the study is marked as agreed on, else, it goes through phase 3.
    \item \textbf{Phase 3:} In this phase, primary and secondary reviewers discuss the paper disagreed upon in an effort to reach a common ground.
\end{itemize}

\begin{table}[H]
\centering
\caption{Data extraction form}
\label{tab:Data_extraction_form}
\begin{tabular}{|p{0.03\textwidth}|p{0.17\textwidth}|p{0.6\textwidth}|p{0.08\textwidth}|}
\hline
\multicolumn{1}{|c|}{\textbf{\#}} & \multicolumn{1}{c|}{\textbf{Study data}} & \multicolumn{1}{c|}{\textbf{Description}} & \multicolumn{1}{c|}{\textbf{RQ}} \\ \hline
1 & Study ID & unique identifier for the study & - \\ \hline
2 & Bibliometric \& demographics & Authors' name, Title of the study, Year of publication, Authors affiliated country, number of citations & - \\ \hline
3 & Source & IEEE Xplore, ACM, Scopus, Science Direct etc. & - \\ \hline
4 & Article type & Conference, Journal, Workshop etc. & - \\ \hline
5 & Modeling approach & used modeling approach(s) by the study & RQ 1.1 \\ \hline
6 & Modeling purpose & The purpose for which the study used models & RQ 1.2 \\ \hline
7 & MDE phase & The MDE phase (i.e. system design, simulation, transformation, V\&V) the study addressed & RQ 1.3 \\ \hline
8 & Tools/Languages & used or developed tools/languages by the study & RQ 2 \\ \hline
9 & CPS component & The CPS component (i.e. physical component, cyber component, sensor, actuator) the study addressed & RQ 3 \\ \hline
10 & CPS application domain & The CPS application domain the study targeted & RQ 4 \\ \hline
11 & Type of evaluation & The type of evaluation (i.e. case study, use case, empirical study) the study presented & RQ 5 \\ \hline
12 & CPS challenges & The type of CPS challenge(s) the study addressed & RQ 6 \\ \hline
13 & Quality assessment & problem statement, contribution, future work & QA 1, QA 2, QA 3 \\ \hline
14 & Self-assessment & Reviewer’s level of understanding of the reviewed paper & SA 1 \\ \hline
\end{tabular}
\end{table}

\newpage
\section{Performing Systematic Review} \label{performing-SLR}

In this section, the process followed to conduct this systematic review is explained. Related process for performing the execution phase was applied by following the protocol defined in Section \ref{methodology}.

\subsection{Performing automatic search}

The digital libraries, indicated in Section \ref{methodology}, were automatically searched, as it is also advised in \cite{kitchenham2007guidelines}. In order to perform an automatic search, search strings are developed. These search strings must fit the syntax of the targeted search engine. They should be “good-enough” to include as many relevant studies as possible, and concurrently, exclude irrelevant ones. Keyword and overall search string are defined using PICOC criteria as discussed in Section \ref{methodology}. Table \ref{tab:Search_strings} shows searched digital libraries and the corresponding search string(s) used. After concluding the automatic search, 646 studies were obtained.

Many challenges were encountered while using the digital libraries, one main challenge of using these digital libraries is the lack of guidelines explaining how to use the advanced search feature of these digital libraries. Also, the number of allowed terms of the search string is limited in digital libraries like ScienceDirect and DBLP, which lead to splitting the search string into multiple search strings. Also, wild cards are not supported in ScienceDirect. Another challenge is that the digital libraries like ACM and IEEE do not provide the capability to restrict the search to more than one specific area at once, e.g. title, abstract, and keywords combined.

\begin{table}[]
\centering
\caption{Search strings}
\label{tab:Search_strings}
\begin{tabular}{|l|l|p{10cm}|}
\hline
\multicolumn{1}{|c|}{\textbf{Digital Library}} & \multicolumn{1}{c|}{\textbf{Results}} & \multicolumn{1}{c|}{\textbf{Search query}} \\ \hline
IEEE & 164 & ("model-driven development" OR "model-driven engineering" OR "model-driven architecture" OR "code generation" OR "generative approach" OR "model-based approach" OR "model-driven approach" OR "domain specific model*" OR metamodel OR "meta-model" OR "meta model" OR "modeling approach") AND ("cyber-physical system*" OR "cyber physical system*" OR "smart system*" OR "cyberphysical systems" OR "cps") \\ \hline
ACM & 55 & recordAbstract:(("model-driven development" OR "model-driven engineering" OR "model-driven architecture" OR "code generation" OR "generative approach" OR "model-based approach" OR "model-driven approach" OR "domain specific model*" OR metamodel OR "meta-model" OR "meta model" OR "modeling approach") AND ("cyber-physical system*" OR "cyber physical system*" OR "smart system*" OR "cyberphysical systems" OR "cps"))  \\ \hline
Web of Science & 16 & TI=(("model-driven development" OR "model-driven engineering" OR "model-driven architecture" OR "code generation" OR "generative approach" OR "model-based approach" OR "model-driven approach" OR "domain specific model*" OR metamodel OR "meta-model" OR "meta model" OR "modeling approach") AND ("cyber-physical system*" OR "cyber physical system*" OR "smart system*" OR "cyberphysical systems" OR "cps"))) AND LANGUAGE: (English)  \\ \hline
Scopus & 363 & See the online repository at: \href{https://docs.google.com/spreadsheets/d/1YAX2mRoZE7Zchv1FCw2o9fCQMltFxPg2Jw2-4Vh1DiA/edit#gid=730724762}{Here} \\ 
\hline
\multirow{3}{*}{ScienceDirect} & 23 & ("code generation" OR "generative approach" OR "domain specific modelling" OR "modelling approach") AND ("cyber-physical systems" OR "cyber physical systems" OR "smart systems" OR cps OR "cyberphysical systems") \\  
& 12 & ("model-driven development" OR "model-driven engineering" OR "model-driven architecture" OR "model-based approach" OR "model-driven approach") AND ("cyber-physical systems" OR "cyber physical systems" OR "smart systems" OR "cyberphysical systems") \\  
& 9 & (metamodel OR "meta-model" OR "meta model") AND ("cyber-physical systems" OR "cyber physical systems" OR "smart systems" OR "cyberphysical systems") \\ \hline
\multirow{3}{*}{DBLP} & 4 & (metamodel $|$ "meta-model" $|$ "meta model") ("cyber-physical systems" $|$ "cyber physical systems" $|$ "smart systems" $|$ "cyberphysical systems") \\  
& 0 & ("model-driven development" $|$ "model-driven engineering" $|$ "model-driven architecture" $|$ "model-based approach" $|$ "model-driven approach") ("cyber-physical systems" $|$ "cyber physical systems" $|$ "smart systems" $|$ "cyberphysical systems") \\ 
& 0 & ("code generation" $|$ "generative approach" $|$ "domain specific modelling" $|$ "modelling approach") ("cyber-physical systems" $|$ "cyber physical systems" $|$ "smart systems" $|$ cps $|$ "cyberphysical systems") \\ 
\hline
\end{tabular}
\end{table}

\subsection{Removing duplicates}

All 646 studies were stored in Mendeley which detected duplicate studies from which one was manually removed. The process of duplication-checking goes until further stages (i.e., forward snowballing). The eliminated duplicate papers were 113 studies, so, 533 studies remained. 

\subsection{Selecting primary studies}

\begin{figure} []
\centering
	\includegraphics[width=1.1\textwidth]{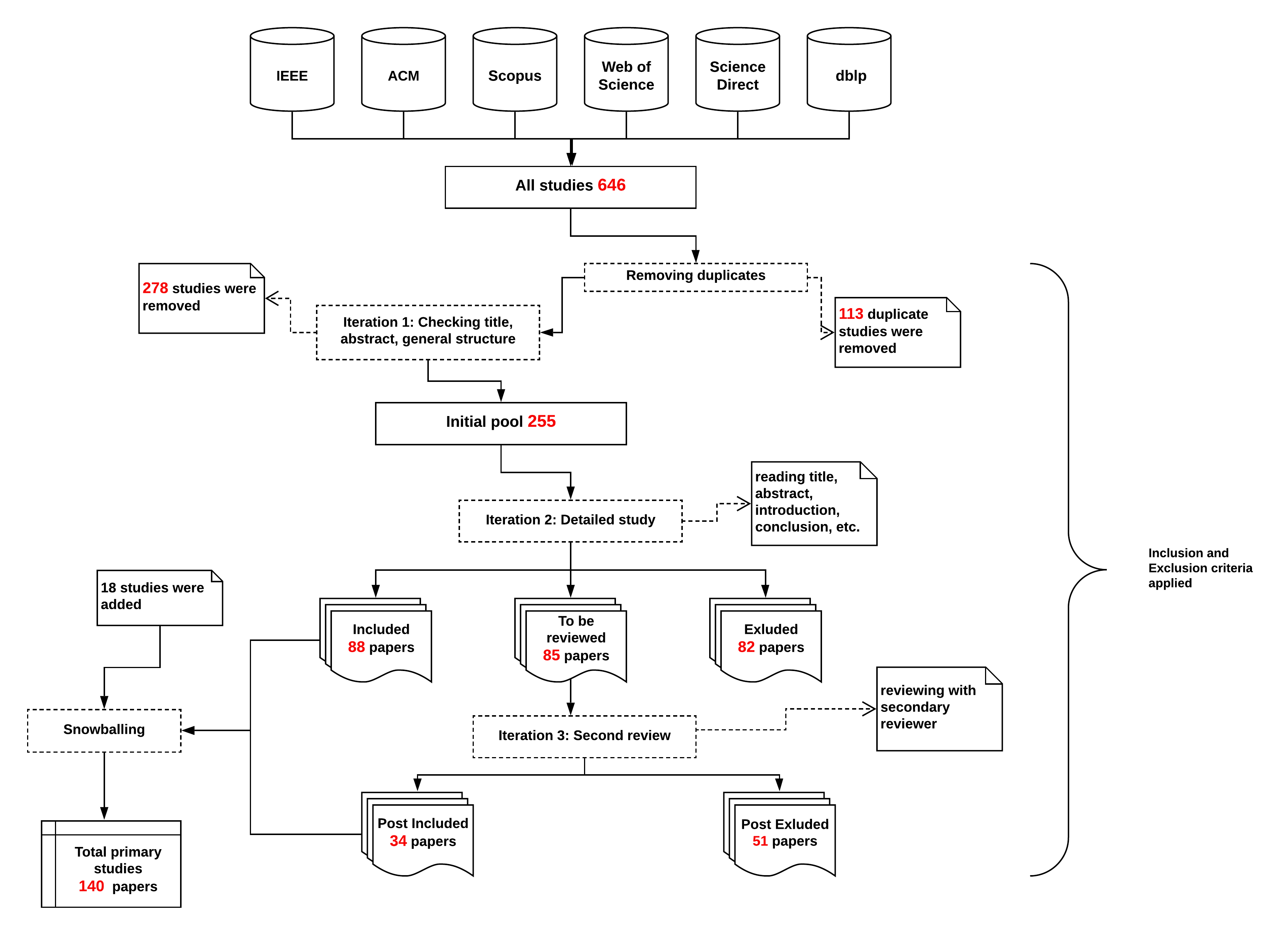}
	\caption{Search and selection process}
	\label{fig:Search_and_selection_process}
\end{figure}

Selection of studies was based on the inclusion and exclusion criteria defined in Section 3. The process of selecting primary studies is shown in Figure \ref{fig:Search_and_selection_process}. This stage is applied on 533 studies. The inclusion or exclusion of studies are performed in several iterations: 

\begin{itemize}
    \item \textbf{Iteration 1:} The primary reviewer went through each study reading its title, abstract, and checking the general content (figure, models, tables, etc.). Studies which meet the inclusion and exclusion criteria passed to the next iteration (278 studies were removed in this iteration).
    \item \textbf{Iteration 2:} All the studies which passed iteration 1 were read in more detail by further reading the related paper’s introduction and conclusion sections and if necessary other sections (e.g. methodology and case study). This iteration resulted in including 88 papers and excluding 82 papers. 85 papers left undecided “to be reviewed”.
    \item \textbf{Iteration 3:} The 85 undecided papers in iteration 2 were again reviewed with a secondary reviewer. In this stage, both reviewers agreed on either including or excluding the paper. As a result, 34 papers were later included, whereas 51 papers were later excluded. 
\end{itemize}

To sum up, 88 papers were included from iteration 2 and 34 papers were included from iteration 3, forming a pool of 122 primary studies.

\subsection{Forward Snowballing}

After defining the primary studies, forward snowballing was performed during data extraction phase. The process of conducting forward snowballing is stated in Section 3. As a result, 18 papers were included to the pool of the primary studies, making a total of 140 papers considered in this study.

\subsection{Realizing Data extraction}

In this stage, Google spreadsheets were used for extracting data from the primary studies. The final version of this spreadsheet is available on IEEE Data Port\footnote{\url{https://dx.doi.org/10.21227/zbkz-6461
}} \cite{abshir-dataset}. Each study is given a unique key in order to match it with the original paper in Mendeley. The full text of each paper was read to answer the research questions as well as the quality and self-assessment questions. A detailed qualitative and quantitative analysis was derived from the outcome of this section. 

Unlike in the selection stage, papers were read in a meticulous manner according to the protocol defined in this study. 

\subsection{Data analysis}
Analyzing the extracted data from the primary studies was the last stage in the execution phase. As stated in section \ref{methodology}, data analysis is encompass both quantitative and qualitative analysis. The results of the analysis are presented in Section \ref{results}.

\newpage
\section{Results} \label{results}

In this section, the results and the findings of the SLR are presented. The section starts with bibliometrics and demographics analysis, followed by quality assessment analysis, and finally, analysis of research questions.

\subsection{Bibliometrics \& Demographics}

\subsubsection{Publication trend per year}

\begin{figure} [H]
\centering
	\includegraphics[scale=1.2]{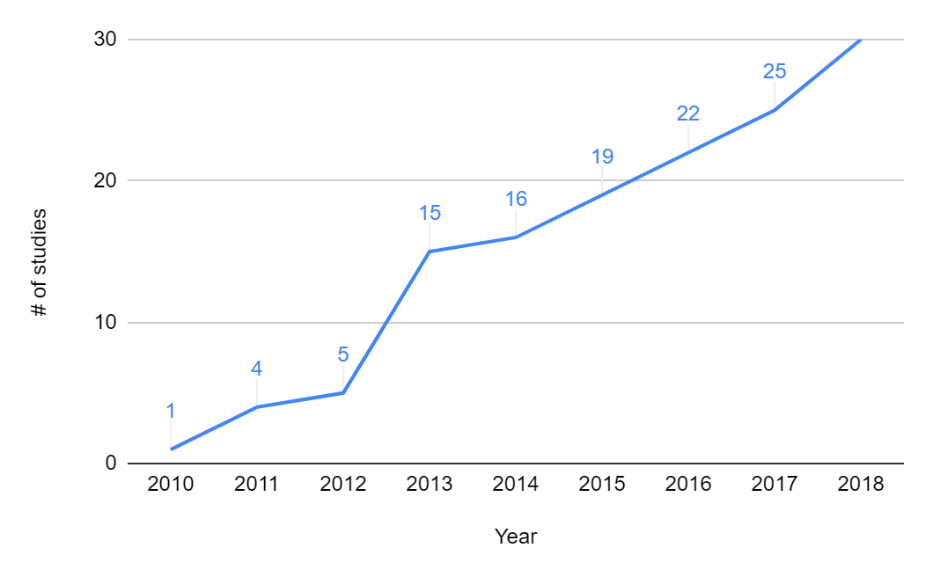}
	\caption{Publication trend per year}
	\label{fig:Publication_trend_per_year}
\end{figure}

Basically, Figure \ref{fig:Publication_trend_per_year} depicts the increase in the number of research papers on this topic. Between the years 2010-2018, researchers' interest in the domain of applying MDE for CPS had grown continuously for the period under observation.

\subsubsection{Citation analysis and top-cited studies}

\begin{figure} [H]
\centering
	\includegraphics[scale=0.8, angle =90]{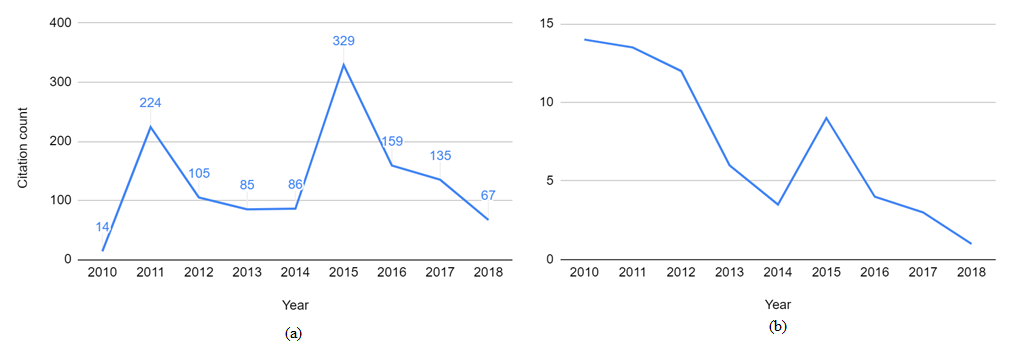}
	\caption{Number of citation per year}
	\label{fig:Number_of_citation_per_year}
\end{figure}

In this section, results related to the citation distribution over the year of publication is presented. The number of citations was obtained using Google Scholar. Figure \ref{fig:Number_of_citation_per_year}(a) shows distribution of citations over publication years, where Figure \ref{fig:Number_of_citation_per_year}(b) shows the median number of all papers' citations published in a given year.  Only 15\% of the primary studies are never cited. The 3 most cited papers are listed in (Table \ref{tab:Most_cited_papers}).

\begin{table}[H]
\centering
\caption{Most cited papers}
\label{tab:Most_cited_papers}
\begin{tabular}{|p{0.08\textwidth}|p{0.6\textwidth}|p{0.08\textwidth}|p{0.15\textwidth}|}
\hline
\multicolumn{1}{|c|}{\textbf{Study}} & \multicolumn{1}{c|}{\textbf{Title}} & \multicolumn{1}{c|}{\textbf{Year}} & \multicolumn{1}{c|}{\textbf{Citations\#}} \\ \hline
\cite{Chen2011} & Petri Net Modeling of Cyber-Physical Attacks on Smart Grid & 2011 & 196 \\ \hline
\cite{Xin2015} & Cyber-Physical Modeling and Cyber-Contingency Assessment of Hierarchical Control Systems & 2015 & 84 \\ \hline
\cite{Seiger2015} & Modelling complex and flexible processes for smart cyber-physical environments & 2015 & 67 \\ \hline
\end{tabular}
\end{table}

\subsubsection{Active researchers in the domain}

\begin{figure} []
\centering
	\includegraphics[scale=1.2]{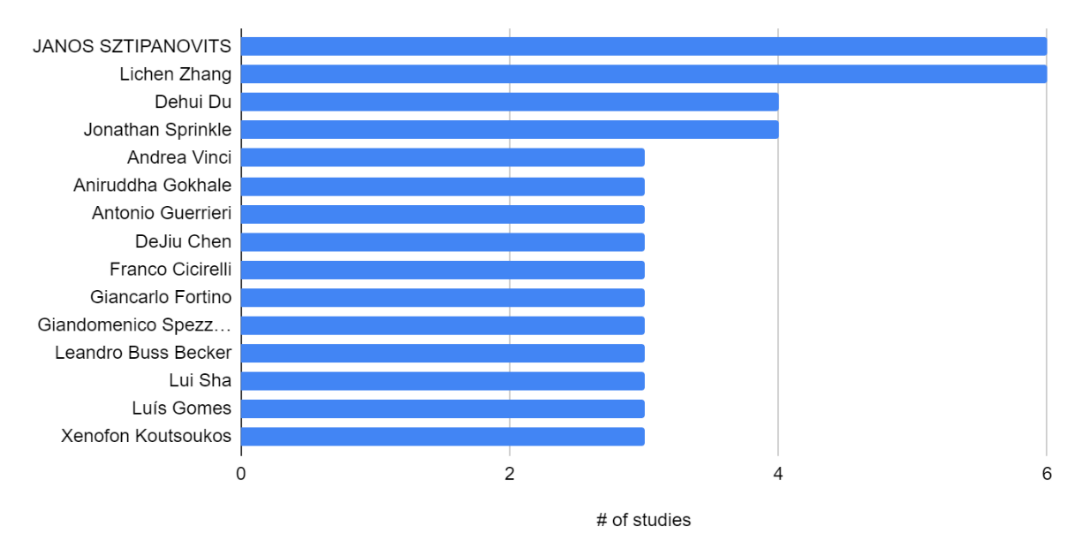}
	\caption{Authors with at least three papers}
	\label{fig:Authors_with_at_least_three_papers}
\end{figure}

To get an overview of the most active researchers in this domain, the number of papers published by each author are counted. To keep the brevity of the ranking results, Figure \ref{fig:Authors_with_at_least_three_papers} shows only researchers who published at least three papers in the pool of primary studies. The authors "Lichen Zhang" and "Janos Sztipanovits" have the greatest number of publications, each with 6 papers. Followed by "Dehui Du" and "Jonathan Sprinkle" with 4 papers each. The complete list of authors can be found in the Appendixes (Table \ref{App:authors}).


\subsubsection{Countries contributing to the field (based on author affiliations)}

Similar to \cite{garousi2013systematic} that presented bibliometric studies in software engineering, most active countries are listed based on authors’ affiliation, that is authors who have published papers in the field of applying MDE for CPS. If a researcher moved between two or more countries, we assigned each of his/her papers to the exact affiliation information on top of each paper. If a paper was written by researchers from more than one country, the counters for each of those countries were incremented by one.

\begin{figure} []
\centering
	\includegraphics[scale=1.2]{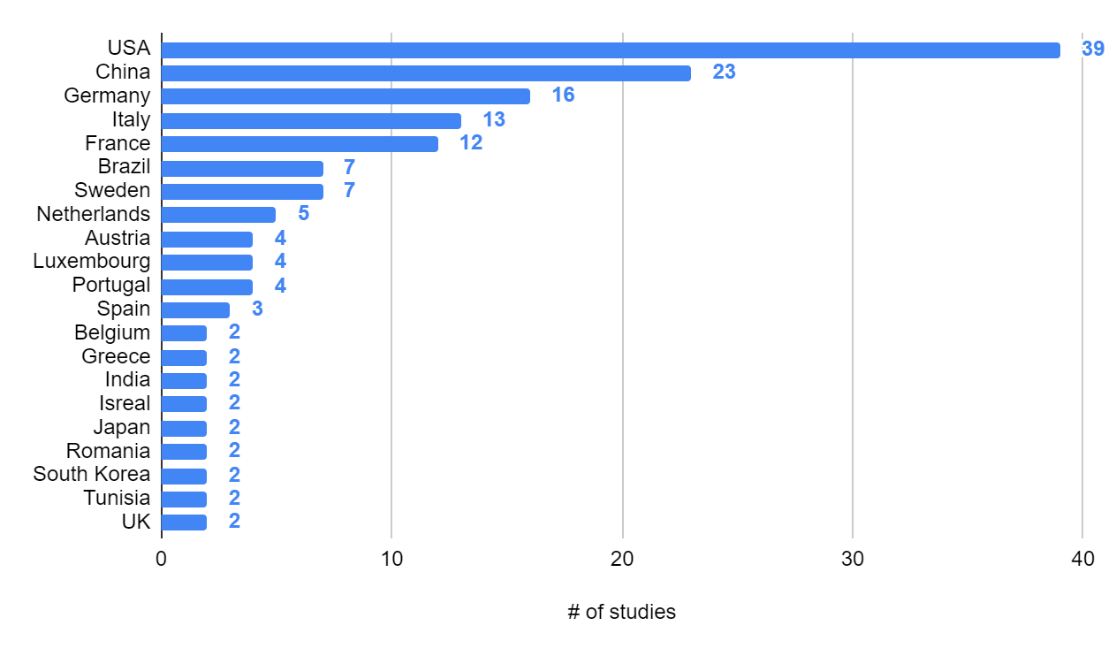}
	\caption{Countries contributing to the field (based on author affiliations)}
	\label{fig:Countries_contributing_to_the_field}
\end{figure}

 For better comprehension of the chart in Figure \ref{fig:Countries_contributing_to_the_field} that shows the ranking of countries with at least two publications -The complete list of the countries can be found in the Appendixes (Table \ref{app:Countries}). The top 5 countries are; USA (with 39, 25.16\%), China (with 23, 14.84\%), Germany (with 16, 10.32\%), Italy (with 13, 8.39\%), and France (with 12, 7.74\%). According to the analysis, 112 (80\%) of the papers were written by the author(s) affiliated to one country, while 28 papers (20\%) were jointly written by authors from more than one country. In terms of internationally authored papers and the collaborating nations, the collaboration between China and the USA is the highest \cite{Guo2017,Jiang2018,Xin2015}, followed by Sweden and Italy \cite{Angelo2018,Sapienza2014}, and Tunisia and France \cite{B2018,Graja}.
 
\subsubsection{Publication venues}

90 of the studies (64.75\%) were conference papers, while 33 (23.74\%) and 14 (10.07\%) studies were journals and workshop papers respectively. Table \ref{tab:Venues} shows the ranking of the top venues with at least two studies. The complete list of publication fora can be found in the appendixes (Table \ref{App:venues}). There are 16 venues in Table \ref{tab:Venues}: 10 conferences/symposia, 5 journals, and 1 workshop. Interestingly, one can see that journals are at the bottom of the list with 2 publications each. That is, researchers in this field seem most likely preferring conferences than journals.

\begin{table}[H]
\centering
\caption{Venues with at least two papers}
\label{tab:Venues}
\begin{tabular}{|p{0.1\textwidth}|p{0.6\textwidth}|p{0.08\textwidth}|}
\hline
\multicolumn{1}{|c|}{\textbf{Venue type}} & \multicolumn{1}{c|}{\textbf{publication venue}} & \multicolumn{1}{c|}{\textbf{\# of studies}} \\ \hline
Conference & International Conference on Emerging Technologies and Factory Automation (ETFA) & 6 \\ \hline
Workshop & Workshop on Domain-specific modeling & 6 \\ \hline
Conference & ACM/IEEE International Conference on Cyber-Physical Systems (ACM/IEE ICCPS) & 4 \\ \hline
Conference & Industrial Cyber-Physical Systems (ICPS) & 3 \\ \hline
Conference & International Conference on Engineering of Complex Computer Systems (ICECCS) & 3 \\ \hline
Conference & International Conference on Industrial Informatics (INDIN) & 3 \\ \hline
Conference & Annual Computer Software and Applications Conference (IEEE COMPSAC) & 2 \\ \hline
Conference & Brazilian Symposium on Computing Systems Engineering (SBESC) & 2 \\ \hline
Conference & International Conference on Networking, Sensing and Control (ICNSC) & 2 \\ \hline
Conference & International Systems Conference (SysCon) & 2 \\ \hline
Conference & ACM Symposium on Applied Computing (ACM SAC) & 2 \\ \hline
Journal & Advanced Engineering Informatics & 2 \\ \hline
Journal & IEEE Transactions on Smart Grid & 2 \\ \hline
Journal & IFAC Symposium on Information Control Problems in Manufacturing INCOM & 2 \\ \hline
Journal & International Journal of Critical Infrastructure Protection & 2 \\ \hline
Workshop & IFAC Workshop on Intelligent Manufacturing Systems & 2 \\ \hline
\end{tabular}
\end{table}

\subsection{Quality assessment}

When reading a research paper, the reader must be able to easily identify sections like: 1- the problem the paper address, 2- proposed contribution by the study, and  3- possible future work, as the clarity of these sections increase the quality and the overall understanding of the study. Therefore, it is meaningful to present statistics regarding how the primary studies stated these three sections. The results are given in Table \ref{tab:Quality_assessment}.

In terms of the clarity of stating the problem which the study addresses, 74 studies (52.86\%) clearly and precisely described the problem, while the problem description of the other 66 studies (47.14\%) is partially obscure. Regarding stating the contribution, no contribution is stated by 26 studies (18.57\%), 55 studies (39.29\%) clearly stated their contribution, while the other 59 studies (42.14\%) vaguely stated their contribution. Concerning the future work, 112 studies (80\%)  reported the possible future work, while the remaining 28 studies (20\%) do not report any future work.

\begin{table}[H]
\centering
\caption{Quality assessment results analysis}
\label{tab:Quality_assessment}
\begin{tabular}{|l|l|l|}
\hline
\multicolumn{3}{|c|}{\textbf{Quality assessment}} \\ \hline
Quality & \% & \# of studies \\ \hline
\multicolumn{3}{|c|}{\textbf{QA-1: Problem statement}} \\ \hline
Not stated & 0 & 0\% \\ \hline
Partially stated & 47.14\% & 66 \\ \hline
Fully stated & 52.86\% & 74 \\ \hline
\multicolumn{3}{|c|}{\textbf{QA-2: Contribution}} \\ \hline
Not presented & 18.57\% & 26 \\ \hline
Presented and Clear & 39.29\% & 55 \\ \hline
Presented but not clear & 42.14\% & 59 \\ \hline
\multicolumn{3}{|c|}{\textbf{QA-3: Future work}} \\ \hline
No & 20.00\% & 28 \\ \hline
Yes & 80.00\% & 112 \\ \hline
\end{tabular}
\end{table}

\subsection{Research questions Analysis }

It is worth mentioning that some of the studies fit more than one group, that is, some papers reported more than one modeling approach, the purpose of modeling (modeling activity) and targeted MDE technique/phase. Therefore, in this work, each study is assured not to be limited to only one group, and instead assign it to every possible group reported.

\subsubsection{Modeling approaches employed for applying MDE in CPS} \label{RQ1}

In this section, the results and findings for \textbf{RQ1: Are any of MDE approaches or techniques used in/for the development of the studied cyber-physical system?} and its sub-questions are presented.	

\textbf{RQ1.1: What is the modeling approach presented/used in the study?}

\begin{figure} [H]
\centering
	\includegraphics[scale=0.4]{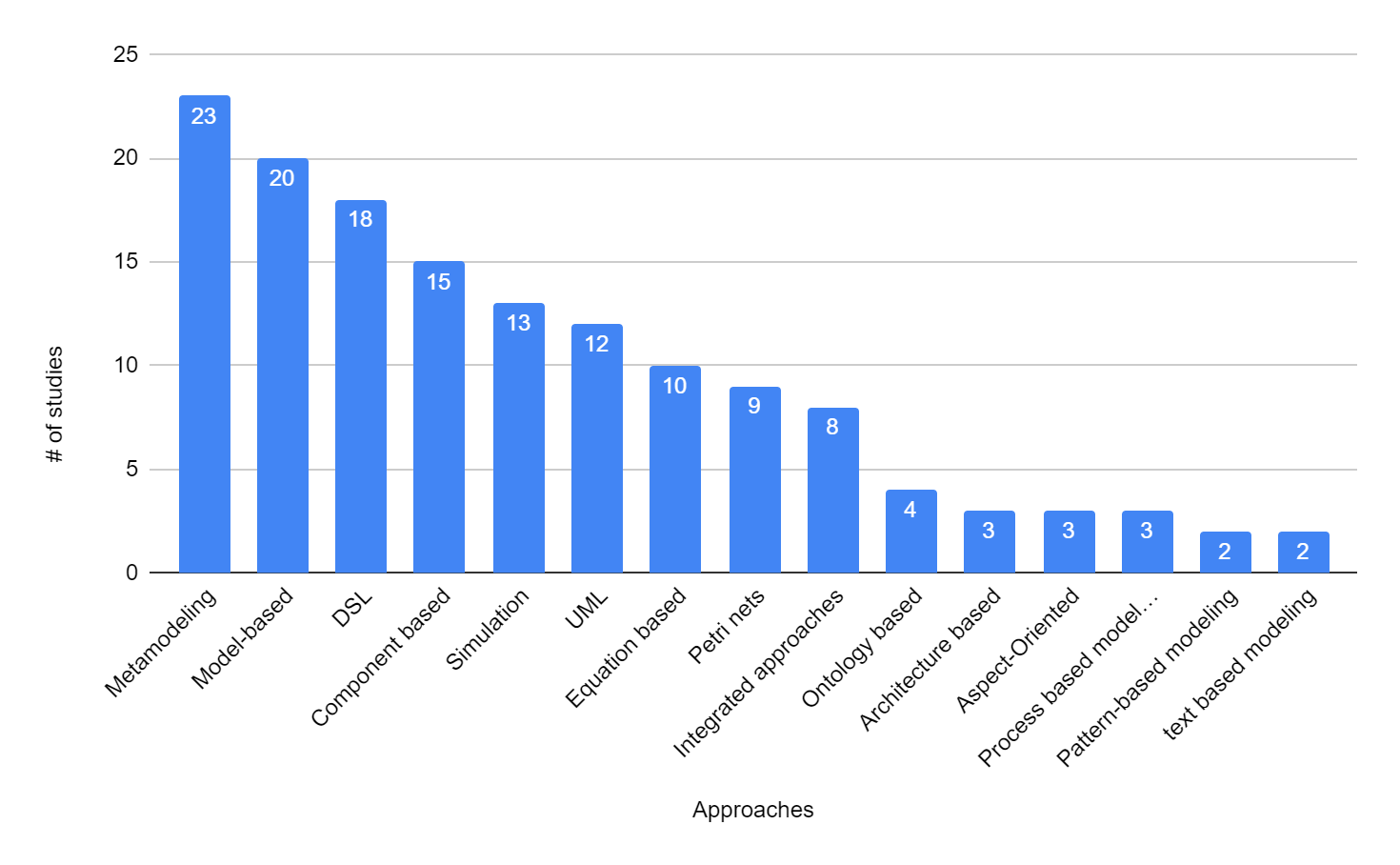}
	\caption{Reported modeling approaches}
	\label{fig:Reported_modeling_approaches}
\end{figure}

As shown in (Figure \ref{fig:Reported_modeling_approaches}), the most used approach is metamodeling. 15.86\% of the primary studies (23 papers) reported metamodeling as the approach used in their studies. This is followed by model-based approach with 20 papers (13.79\%), DSL with 18 papers (12.41\%) and component-based approach with 15 papers (10.34\%). Other used approaches include; State Machine based modeling, Model Driven Development, Signal-based Modeling, Models@run time, Agent-oriented modeling, Dynamic Constraint Feedback (DCF), Properties Modeling, Stochastic Occurrence Hybrid Automata (SOHA)-based modeling, Model-Integrated-Computing (MIC), Microservice-based development and Theory-based (e.g. modeling theory based on fuzzy logic). 

Integrated approaches category comprises studies which promote either the integration of multiple approaches or multi-domain modeling approach. Studies employing integrated approaches are \cite{Walch2017a,Goncalves2017}. On the other hand, studies which used multi-modeling approaches are  \cite{Zhang2013a,Zhang2013c,Zhang2014d,Kuesap2008,Pagliari2018,Broenink2016}.

\begin{figure} [H]
\centering
	\includegraphics[scale=1.2]{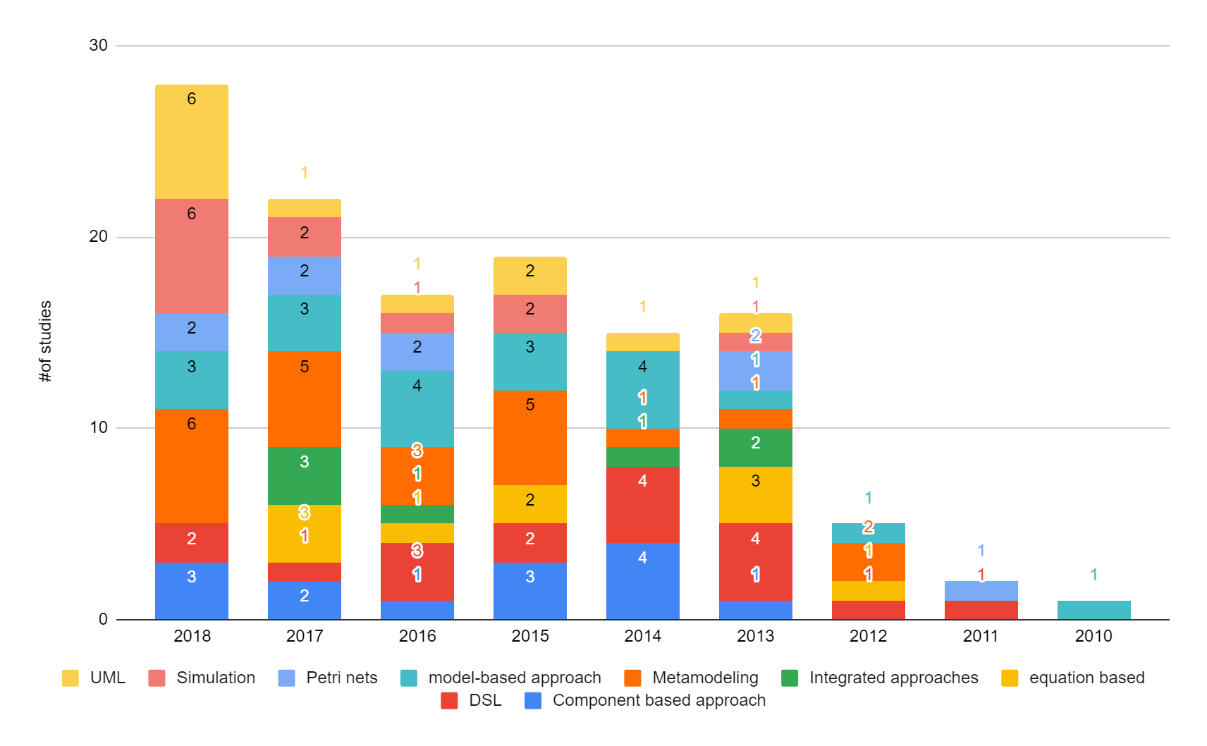}
	\caption{Distribution of the reported modeling approaches over the years}
	\label{fig:Distribution_of_the_reported_modeling_approaches_over_the_years}
\end{figure}

Figure \ref{fig:Distribution_of_the_reported_modeling_approaches_over_the_years} shows the distribution of modeling approaches over the years. For better comprehension of the chart, the most used approaches reported by more than 5 studies are given only. 

The most consistently used approach within the period of the study (2010-2018) was DSL except for 2010. This approach was at least reported by one paper between years 2011-2018. However, its growth fluctuates. Metamodeling and Model-based approach also showed a consistent presence between 2012-2018, while UML and Component-based approach were present continuously between 2013-2018. Although Metamodeling approach had minor reduction in its usage between the years 2012 and 2016, it always increased. For the years between 2015 and 2018, it is clearly observed that the Metamodeling approach was always amongst the top-most used 3 approaches.

For further understanding of the modeling approaches, the distribution of the most reported approaches over the countries is shown in Figure \ref{fig:Reported_modeling_approaches_vs.Authors_countries}, which leads to find the followings:

\begin{figure} [H]
\centering
	\includegraphics[scale=1.15]{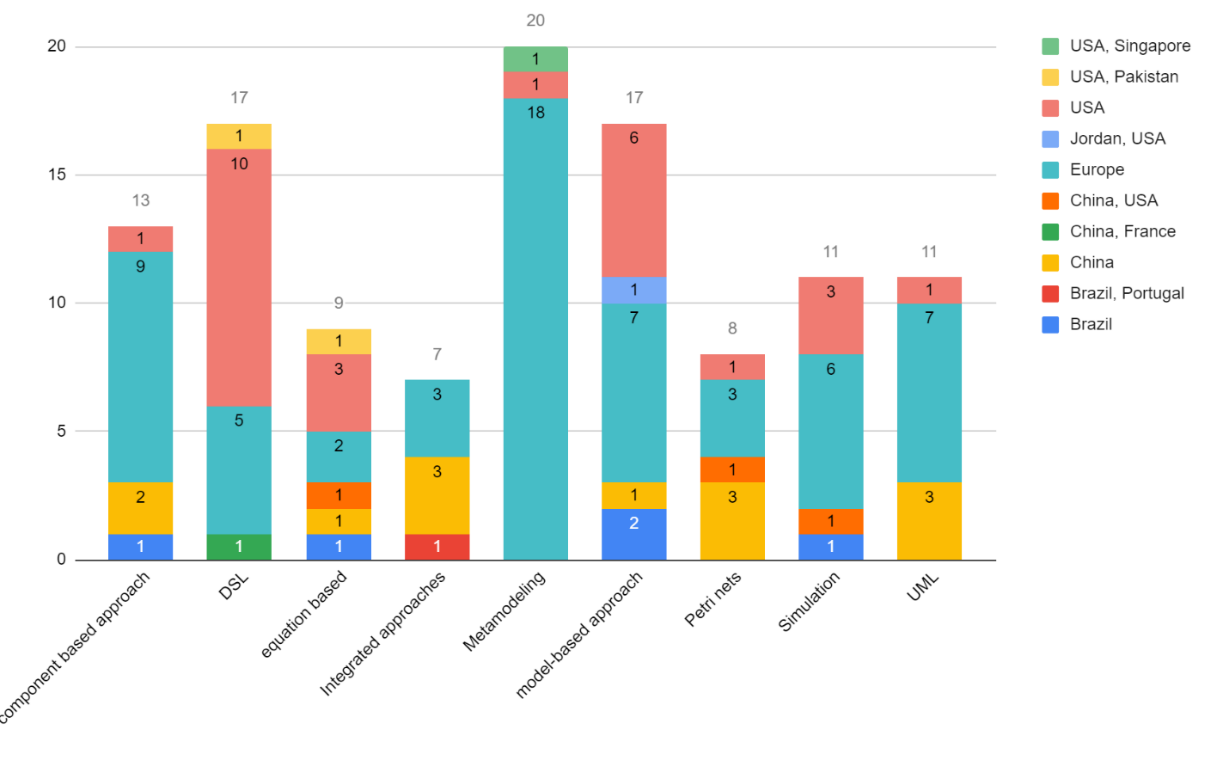}
	\caption{Reported modeling approaches vs. Authors' countries}
	\label{fig:Reported_modeling_approaches_vs.Authors_countries}
\end{figure}

\begin{itemize}
    \item \textbf{Metamodeling:} is the most reported modeling approach with a total of 23 studies. 18 of those studies were written by authors affiliated to Europe, 2 by authors affiliated to Korea, 1 study by researchers affiliated to USA, 1 study was jointly written by authors from USA and Singapore, and 1 study was jointly written by authors from Malaysia, India, Europe(Austria).
    
    \item \textbf{Model-based approach:} reported by 20 studies, 6 studies are affiliated to USA, 1 study was jointly written by the authors from USA and Jordan, 7 studies are affiliated to Europe, and the rest of the studies are distributed amongst; Brazil (2 studies), China (1 study), Israel (1 study), Taiwan (1 study), and 1 study written jointly by New Zealand, Europe(Finland), and China.
    
    \item \textbf{DSL:} Reported by 18 studies in total, 10 papers written by USA affiliated researchers, 1 study jointly written by USA and Pakistan researchers, 5 studies written by authors affiliated to Europe. Others include 1 study by KSA, and 1 study jointly written by China and Europe (France) affiliated researchers.
    
    \item \textbf{Component-based approach:} Among 15 studies following this approach, 9 studies were from Europe and 2 studies were from China. Others: 1 study by Brazil, 1 study by India, and 1 study jointly prepared by the researcher from Morocco and Europe (Latvia, Spain, Czech Republic, Netherlands, Italy, Romania).
    
    \item \textbf{Simulation:} This approach was reported by 13 studies in which 6 studies were written by authors affiliated to Europe, 3 from USA affiliated researchers and 1 study was written jointly by USA and China affiliated authors. Others: 1 study by Brazil, 1 study by Israel, and 1 study jointly written by authors affiliated to Israel and Japan.
    
    \item \textbf{UML:} Reported by 12 studies in total; 7 studies were written by authors affiliated to Europe, 3 studies by China affiliated researchers, 1 study by USA, and 1 study by Korea.
    
    \item \textbf{Equation-based modeling:} 10 studies used this approach. 3 studies were written by USA affiliated authors, 2 studies had joint authorship, one by USA and Pakistan and the other one by USA and China affiliated researchers. 2 studies are from Europe, 1 study from China, 1 study from Brazil, and 1 study from Iran.
    
    \item \textbf{Petri nets:} Used by 9 studies. 3 studies are from Europe affiliated researchers, 3 studies were written by researchers affiliated to China, 1 study was jointly written by China and USA affiliated authors. Others: 1 study by USA, and 1 study jointly written by USA and Europe (Spain, UK).
    
    \item \textbf{Integrated approaches:} This approach was reported by 8 studies. 3 studies are from Europe, 3 studies were written by authors affiliated to China, 1 paper jointly written by Brazil and Europe (Portugal) affiliated authors, and 1 study jointly written by authors affiliated to Japan and Thailand. 

\end{itemize}

In summary, it can be seen that the Metamodeling and Model-based approaches are mostly used in Europe. On the other hand, DSL approach is mostly used in the USA and its usage surpasses all the European countries combined.

Further, it is important to mention that although equation-based approach is reported by 10 studies, it was used jointly with other approaches in 5 out of the 10 studies. \cite{Tariq2012} used equation-based modeling with DSL where they developed DSML for the performance analysis purpose. \cite{Simko2013a} also used DSL with equation-based approach to develop a DSML for simulation. \cite{Silva2015} along with equation-based modeling used a simulation-based approach and used the Ptolemy II modeling tool and Simulink Design Verifier (SLDV) for Model-based Testing and formal verification. \cite{Qian2013} used equation-based modeling with Petri nets based modeling approach. The study used discrete/continuous Petri nets for scheduling the analysis. \cite{Mezhuyev2013} used Metamodeling based approach with equation-based modeling for the development of meta-models using Visual Environment for Cyber-Physical Modelling (VE-CPM). The remaining studies, which used equation-based modeling as their only approach, did not present any tool/language except \cite{Low2017} that presented a tool HA-SPIRAL for code generation. To this end, the equation-based modeling approach is somewhat useful as a supporting approach rather than as an independent approach in this field – applying MDE for CPS.

\textbf{RQ1.1.1: What is the purpose for which the models were used?}

\begin{figure} []
\centering
	\includegraphics[scale=1.6, angle =90]{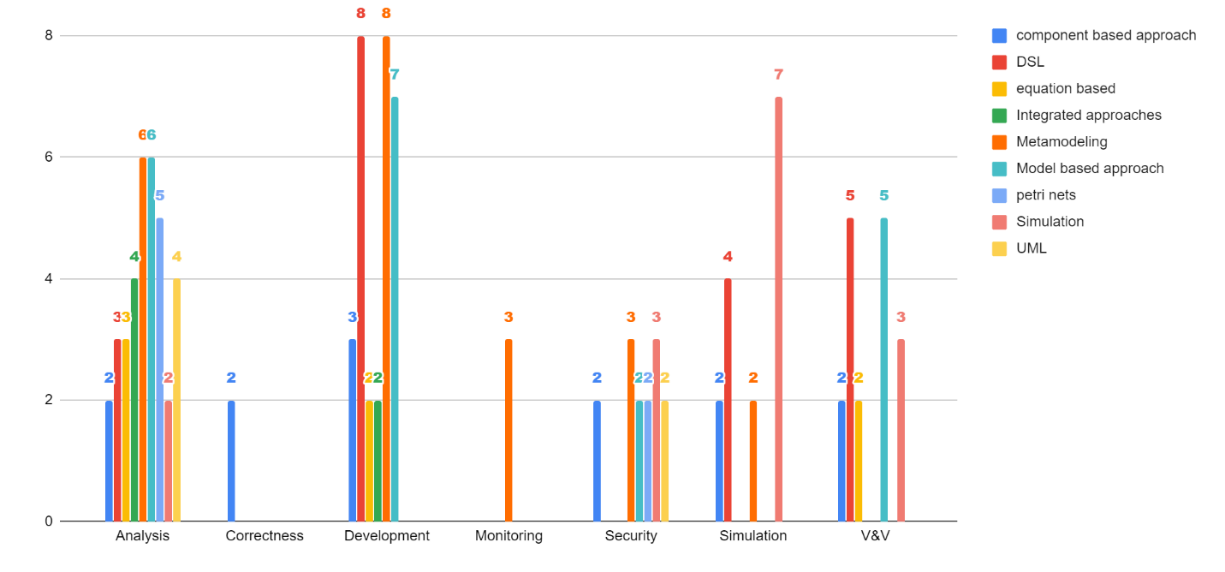}
	\caption{Modeling approaches distribution over modeling activities}
	\label{fig:Modeling_approaches_distribution_over_modeling_activities}
\end{figure}

Out of the 140 studies, 136 of them report their purpose for using the models, while 4 do not state their purpose. From the 136 studies, 111 papers reported only one activity, while 22 reported two activities, 2 reported three activities, and the last paper reported four activities. Figure \ref{fig:Reported_modeling_activities} shows activities reported by at least 2 studies for better comprehension of the chart. Figure \ref{fig:Modeling_approaches_distribution_over_modeling_activities} represents the distribution of modeling approaches over the modeling activities. All modeling activities together with the approaches used and the studies reported them are shown in the Appendixes (Table \ref{App:modelingactivities}). Reported activities are as follows:

\begin{figure} []
\centering
	\includegraphics[scale=0.45]{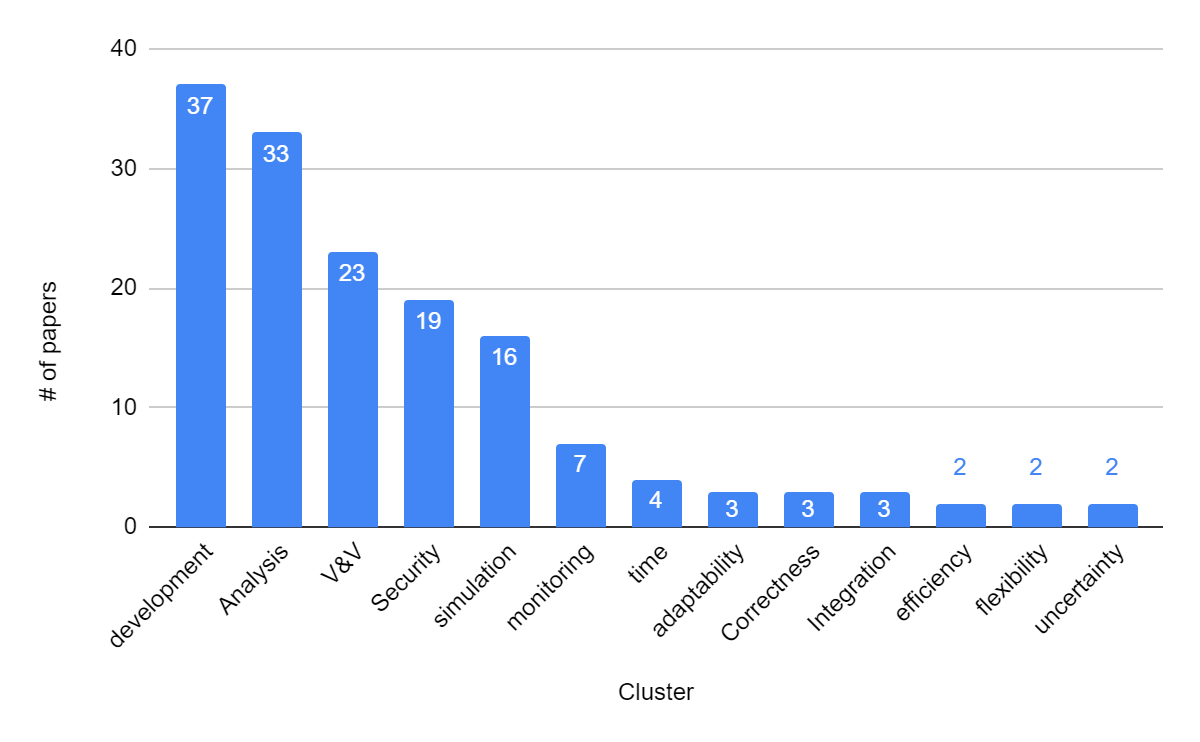}
	\caption{Reported activities/purposes for which modeling approaches used}
	\label{fig:Reported_modeling_activities}
\end{figure}

\begin{itemize}
    \item \textbf{Development:} 37 papers (22.42\%) are grouped under this category. These studies can be put into two categories: firstly, papers that developed DSL, Metamodel, tool, or language, secondly, studies that aim at automating the development process of a system, and perform tasks like transformation, code generation, building libraries, design process, and others. The most used approaches for this activity are Metamodeling and DSL, 8 papers used each of the two approaches. Model-based approach was used by 7 studies, while 3 papers reported Component-based approach. Further, Equation based approach, Integrated approaches, and Architecture-based approach reported by 2 studies each, while the rest of the approaches were reported by 1 study each. 
    
    \item \textbf{Analysis:} Reported by 33 studies (20\%). Here, the aim of the studies is mainly focused on analyzing an existing system (DSL, metamodel, tool) for various activities. The most reported ones include: safety analysis, performance analysis, requirement analysis, security analysis, cost and energy consumption analysis, dependability analysis, and so on. Metamodeling and Model-based approaches are the most reported approaches for this activity with 6 studies for each, followed by Petri nets with 5 studies, Integrated approaches and UML each reported by 4 studies, 3 studies each for DSL and equation based approach, 2 studies reported Simulation and Component based approach for each, while the rest of the approaches were reported by 1 study for each.
    
    \item \textbf{Validation and Verification:} 23 studies (13.94\%), studies in this group conducted V\&V activities regarding DSML validation, metamodel verification, behavior verification, verification of correctness, safety properties verification, model-based testing, formal verification and so on. Approaches used for this activity are distributed as follows: 5 studies reported DSL and model-based approach for each, followed by Simulation based approach with 3 studies. Equation-based, Component-based and Ontology-based approaches reported by 2 papers each. The rest of the approaches were reported by 1 study each.

    \item \textbf{Security:} 19 studies (11.52\%) are concerned about the security of the system from different aspects. Studies reported about safety are also grouped in this set. Activities conducted by this group includes threat modeling, attack modeling, analyzing cyber-attacks, security evaluation and experimentation, safety guarantees of the generated code, and safe reconfiguration. The most used approaches for this activity are Metamodeling and Simulation reported by 3 studies each, followed by Model based approach, Component based approach, UML, Petri nets, Pattern-based modeling reported by 2 studies, while the rest of the approaches were reported by 1 study each.
    
    \item \textbf{Simulation:} The aim of the studies in this group (16 studies (9.70\%)) is the use of simulations for various purpose like using simulations for verification reasons or accompanying it with DSML, while other studies used it for analysis purpose. Mostly, studies reported simulation along with other activities like V\&V, Analysis and Development. Obviously, the most used approach for this activity is Simulation based approach which is reported by 7 studies. It is followed by 4 studies reporting DSML, 2 studies for Metamodeling and 2 studies for Component based approach. 
    
    \item \textbf{Monitoring:} 7 studies (4.24\%) reported about system monitoring or management activities, such as performance monitoring, runtime behavior monitoring, process monitoring, monitoring simulation activities and results. The most reported approach in this group is Metamodeling with 3 studies. Other existing approaches were reported by 1 study each.
    
    \item \textbf{Time:} 4 studies (2.42\%) seek to improve the time aspect of the system to increase productivity.
    
    \item \textbf{Adaptability:} 3 studies (1.82\%) support the implementation of self-adaption aspect of the system.
    
    \item \textbf{Correctness:} 3 studies (1.82\%) support the correctness of the system (DSML, metamodel, tool), often in terms of the correctness of operations or the generated code.
	
    \item \textbf{Integration:} 3 studies (1.82\%) seek to combine different aspects of CPS and support their integration.
\end{itemize}

See the Appendixes (Table \ref{App:modelingactivities}) for other activities that were reported by only one study. For a deeper understanding of how studies addressed modeling approaches and the activities for which they were used, studies can be grouped into three categories:

\begin{itemize}
    \item Studies which presented one modeling approach and used it for one modeling purpose, e.g.  \cite{Neema2018,Zhou2018,Garamvolgyi2018}
    
    \item Studies which presented one modeling approach and used it for more than one modeling purpose. Studies using the same approach for two different modeling purposes are \cite{Cicirelli2017a,Chen2016,Angelo2018} while studies using the same modeling approach for more than two modeling purposes are  \cite{Jiang2018,Chen2017}.
    
    \item -	Studies which presented more than one modeling approach and used it for one modeling purpose/reason. For instance, \cite{Mezhuyev2013} presented metamodeling and equation-based modeling approaches for the development purposes. \cite{Qian2013} used petri nets and equation-based approach for the analysis reasons, \cite{Tariq2012} also used DSL and equation-based approaches for analysis activities, and finally \cite{Motii2017} presented UML and pattern-based modeling approaches for security purposes.
\end{itemize}

\textbf{RQ1.2: Which phase(s) of the system development is/are addressed in the study (using MDE)?}

In this section, in addition to answering RQ1.2, a correlation analysis of RQ1.2 with RQ2 was carried out to find out the used or developed tools/languages in each of the MDE phases.

\begin{figure} [H]
\centering
	\includegraphics[scale=0.9]{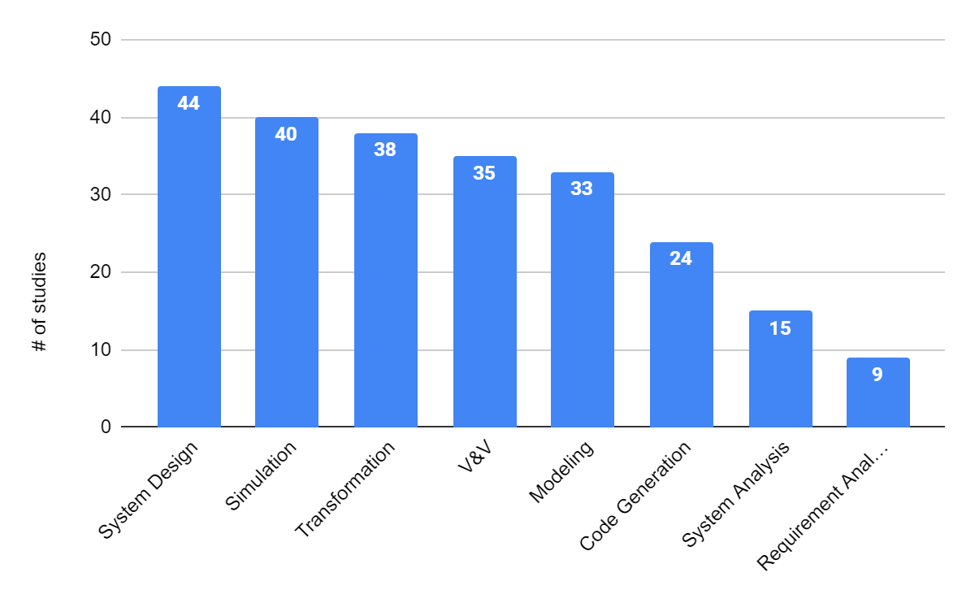}
	\caption{Reported MDE phases/activities}
	\label{fig:Reported_MDE_phases}
\end{figure}

Figure \ref{fig:Reported_MDE_phases} shows the reported MDE phases and their use frequencies. While it is possible to design and/or define MDE phases in numerous ways, we adopted the MDE phases defined in \cite{barivsicsystematic} in our work since it also relates with the modeling of CPS. As can be seen in Figure \ref{fig:Reported_MDE_phases}, the studies differ in the number of the MDE phases they addressed. 72 studies addressed 1 phase, 46 studies reported 2 phases, 17 studies reported about 3 phases, and 5 studies reported 4 phases. Discussion on all these MDE phases is given in the following where they are sorted from most reported to less reported in these studies

\vspace{3mm}
\textbf{System design}
\vspace{3mm}

System design is the most reported MDE phase, it is reported by 44 studies (18.41\%). 15 studies presented DSL, 14 studies developed metamodel, 4 studies developed tools, 4 studies developed extensions, and the other 7 studies either develop a new modeling approach \cite{Goncalves2016,Kuesap2008}, or combine MDE with existing approaches for CPS development \cite{Ollinger2013}. Table Table \ref{tbl:SystemDesign} summarizes developed tools/languages and the corresponding papers. 

\vspace{2mm} 

Studies which developed DSLs are given as follows: \cite{Angelo2018} proposed a DSML called CyPhEF that supports the development and validation of self-adaptive CPS. \cite{Aziz2016} developed a simple graphical DSML for CPS while a DSML for irrigation networks was developed in \cite{Tariq2012}. In \cite{Nagele2017}, authors developed a DSL that helps in quick construction of co-simulations for CPS, the grammar of the DSL was implemented in Xtext, while the code generation implementation was defined in Xtend. A framework called Advanced Vessel Simulation (AVS) was developed in \cite{Lavigne2018} which supports design and evaluation of racing sailboat simulations. The AVS metamodel was developed in EMF, and Sirius was used for developing the graphical editor. A textual DSL named CHARIOT was created with using Xtext in \cite{Pradhan2015}.

\vspace{2mm} 

A DSL for managing different sensor configurations for a self-driving mini vehicle was developed in \cite{Mamun2013}, the domain knowledge, static semantics, and the abstract syntax of a sensor management DSL were defined with the Eclipse Modeling Framework (EMF). \cite{Koutsoukos2012} developed a DSML for the design of networked control systems (NCS) using passivity for separating the NCS control design from uncertainties (i.e. time delays and packet loss). In \cite{Tariq2014}, authors used an ecore-based meta-model to define the abstract syntax of the proposed DSML, and the concrete syntax was implemented as an extension of Simulink standard blocks. 

\vspace{2mm}

\cite{Alrimawi2018} developed two meta-models for representing and sharing incident knowledge of CPS. Meta-models were developed as Eclipse plugins. A metamodel for a systematic analysis of CPS threat modeling was developed in \cite{Martins2015} using MetaGME, while \cite{Maksuti2017} developed a metamodel using ADOxx and UML and they used it for the description of an end-to-end communication use case. A meta-model for the development of a smart cyber-physical environment was presented in \cite{Cicirelli2016}.

\vspace{2mm}

\cite{Bougouffa2018} developed a meta-model for flexibility and dynamic reconfiguration of the automated production systems by using Eclipse Modeling Framework (EMF). \cite{Thramboulidis2018} used UML profile to develop a meta-model for modeling cyber-physical assembly systems. Also, in \cite{Cicirelli2017a}, UML class-diagrams were used to develop a meta-model called Smart Environment Metamodel (SEM) to design smart cyber-physical environments. In \cite{Huang2018}, the authors extended some metamodels of SysML/MARTE for capturing the characteristics of CPS like continuous behavior and stochastic behavior. The approach was implemented in GEMOC. They defined the abstract syntax using EMF, graphical concrete syntax in Sirius, and the textual concrete syntax using Xtext. Meta-models conforming to ISA95 and ISA88 standards were developed in \cite{Irisarri2016} for monitoring the process of an oil production industry.

\vspace{2mm}

In \cite{Walch2017a}, they used ADOxx to develop the modeling tool Cyber-Physical Systems for Industry (CPS4I) for the connection of CPS and conceptualizations of industrial applications in integrated models. A tool, called FTOS, based on openArchitectureWare8 (oAW) was developed in \cite{Buckl2010}, which provides code generation for designing fault-tolerant automation systems. In \cite{Graja}, the authors presented BPMN4CPS which is an extension of BPMN 2.0 for handling CPS features. New extensions for MechatronicUML were developed in \cite{Heinzemann2017slrcps}.

\begin{table}[]
\centering
\scriptsize
\caption{developed DSLs, Meta-models, tools, and extensions}
\label{tbl:SystemDesign}
\begin{tabular}{|l|l|l|}
\hline
\multicolumn{3}{|c|}{\textbf{System Design}}                                                                                                                               \\ \hline
\multicolumn{1}{|c|}{\textbf{Type}} & \multicolumn{1}{c|}{\textbf{Used framework/language}} & \multicolumn{1}{c|}{\textbf{Relevant studies}}                               \\ \hline
DSML                                & Eclipse Modeling Framework (EMF)                      &
\cite{Angelo2018,Mamun2013,Lavigne2018,Tariq2014}                                \\ \hline
                                    & Generic Modeling Environment (GME)                    &
\cite{Tariq2012,Zhang2013b,Koutsoukos2012,Bao2016,Zhang2014,An2011}            \\ \hline
                                    & Xtext                                                 &
\cite{Nagele2017,Pradhan2015}                                                      \\ \hline
                                    & Papyrus tool                                          & 
\cite{Aziz2016}                                                                     \\ \hline
                                    & Xtend                                                 & \cite{Nagele2017}                                                                   \\ \hline
                                    & Graph Rewriting and Transformation (GReAT)            &
\cite{Bao2016,Zhang2014}                                                           \\ \hline
                                    & Simulink                                              & \cite{Tariq2014}                                                                    \\ \hline
                                    & Web Generic Modeling Environment (WebGME)             & 
\cite{Bunting2016}                                                                  \\ \hline
                                    & GEMOC                                                 & \cite{Guan2018}                                                                     \\ \hline
                                    & Sirius                                                &
\cite{Guan2018,Lavigne2018}                                                        \\ \hline
                                    & Kermeta                                               & 
\cite{Guan2018}                                                                     \\ \hline
Meta-model                          & Eclipse Modeling Framework (EMF)                      & 
\cite{Bougouffa2018,Huang2018,Seiger2015}                                         \\ \hline
                                    & UML                                                   & 
\cite{Cicirelli2017a,Thramboulidis2018,Cicirelli2018,Maksuti2017,Cicirelli2016}
                        \\ \hline
                                    & GEMOC                                                 & 
\cite{Huang2018}                                                                    \\ \hline
                                    & Xtext                                                 & 
\cite{Huang2018}                                                                    \\ \hline
                                    & Sirius                                                & 
\cite{Huang2018}                                                                    \\ \hline
                                    & Kevoree Modeling Framework (KMF)                      & 
\cite{Hartmann2015}                                                                 \\ \hline
                                    & GME (MetaGME)                                         & 
\cite{Martins2015}                                                                  \\ \hline
                                    & ADOxx                                                 & 
\cite{Maksuti2017}                                                                  \\ \hline
                                    & SysML                                                 & 
\cite{Barbieri2016}                                                                 \\ \hline
                                    & ISA95 and ISA88 based                                 & 
\cite{Irisarri2016}                                                                 \\ \hline
                                    & -                                                     & 
\cite{Alrimawi2018}                                                                 \\ \hline
                                    & -                                                     & 
\cite{SampathKumar2015}                                                             \\ \hline
Tool                                & MetaEdit+                                             & 
\cite{Chen2015}                                                                     \\ \hline
                                    & ADOxx                                                 & \cite{Walch2017a}                                                                   \\ \hline
                                    & openArchitectureWare (oAW)                            & \cite{Buckl2010}                                                                    \\ \hline
                                    & -                                                     & \cite{Jeon2012}                                                                     \\ \hline
Extention                           & -                                                     &
\cite{Bocciarelli2017,Goncalves2016,Graja,Heinzemann2017slrcps}                  \\ \hline
\end{tabular}
\end{table}

\vspace{3mm}
\textbf{Simulation}
\vspace{3mm}

 40 of the studies (16.74\%) reported simulation. 11 studies addressed exactly the simulation process. They can be summarized as follows: Only 1 study \cite{Tundis2017} developed a simulator. 2 studies developed meta-models \cite{Merschak2018,Ciavotta2018}, and 8 studies 
 \cite{Barve2018,Hahn2015,Housh2018,Matinnejad2018,Mordecai2018,VanAcker2015,Neema2018,Rashid2018} used existing tools for modeling and simulation. Remaining 29 studies incorporating simulation addressed the other phases (i.e. system design, transformation, V\&V, etc.).

\vspace{2mm}

Table \ref{tbl:Simulation} shows the tools and languages used for the simulation activities. Studies presented different reasons for using simulation, e.g. \cite{Angelo2018} presented simulation as a feature of the developed DSL and used it for efficiency and time analysis via MECSYCO co-simulation engine. Also \cite{Tariq2012} presented the simulation as a feature of the developed DSL and used it for performance analysis via MATLAB and EPANET. In \cite{Nagele2017}, authors developed a DSL for constructing HLA-based co-simulations. \cite{Koutsoukos2012} used Simulink for time and network delay analysis. \cite{Bao2016} benefited from Robocode simulator to simulate a reconfigurable conveyor system's behavior and run it in the background (used it as a background simulation) to output time information and the coordinate for the generation of Java animation. Similar to [63], \cite{DiNatale2016} used Simulink for time performance analysis. \cite{Fan2018} is another example for utilizing simulations for analysis purposes in which CPS Safety Analysis and simulation Platform (CP-SAP) was developed. Simulations were also used for security experimentation purposes like in \cite{Yan2009,Hahn2015,Rashid2018}.

\begin{table}[]
\centering
\scriptsize
\caption{Reported simulation tools and languages}
\label{tbl:Simulation}
\begin{tabular}{|l|l|l|}
\hline
\multicolumn{3}{|c|}{\textbf{Simulation}}                                                                                                                                                                                                                                                                                           \\ \hline
\multicolumn{1}{|c|}{Existing tool}                                                    & \multicolumn{1}{c|}{Developed tool} & \multicolumn{1}{c|}{Relevant studies}                                                                                                                                                                \\ \hline
Simulink/Stateflow                                                                     & -                                   & \begin{tabular}[c]{@{}l@{}}
\cite{Koutsoukos2012,Matinnejad2018,DiNatale2016,Pajic2012,Jiang2018,An2011,VanDeMortel-Fronczak2014,VanAcker2015,Yan2009,Palachi2013,Rashid2018}\end{tabular} \\ \hline
MATLAB                                                                                 & -                                   & \cite{Tariq2012}                                                                                                                                                                                            \\ \hline
Modelica                                                                               & -                                   & \cite{Simko2013a}                                                                                                                                                                                           \\ \hline
ModelicaML                                                                             & -                                   & \cite{Zhang2014c}                                                                                                                                                                                           \\ \hline
C2WT                                                                                   &                                     & \cite{Barve2018,Neema2018}                                                                                                                                                                                 \\ \hline
Ptolemy-II                                                                             & -                                   & \cite{Parveen2018,Silva2014,Fan2018}                                                                                                                                                                      \\ \hline
EPANET                                                                                 & -                                   & \cite{Tariq2012,Housh2018}                                                                                                                                                                                 \\ \hline
-                                                                                      & Meta-model                          & \cite{Merschak2018}                                                                                                                                                                                         \\ \hline
MECSYCO                                                                                & -                                   & \cite{Angelo2018}                                                                                                                                                                                           \\ \hline
POOSL                                                                                  & -                                   & \cite{Nagele2017}                                                                                                                                                                                           \\ \hline
Robocode                                                                               & -                                   & \cite{Bao2016}                                                                                                                                                                                              \\ \hline
\begin{tabular}[c]{@{}l@{}}Embedded Systems Modeling\\   Language (ESMoL)\end{tabular} & -                                   & \cite{Neema2014}                                                                                                                                                                                            \\ \hline
IOPT-Flow simulator                                                                    & -                                   & \cite{Pereira2016}                                                                                                                                                                                          \\ \hline
SystemC                                                                                & -                                   & \cite{Weissnegger2016a}                                                                                                                                                                                     \\ \hline
CPGAME                                                                                 & -                                   & \cite{Mordecai2018}                                                                                                                                                                                         \\ \hline
Verilog-AMS                                                                            & -                                   & \cite{Lora2017}                                                                                                                                                                                             \\ \hline
-                                                                                      & Smart Grid Simulator (SGS)          & \cite{Tundis2017}                                                                                                                                                                                           \\ \hline
-                                                                                      & Meta-model                          & \cite{Ciavotta2018}                                                                                                                                                                                         \\ \hline
CPAL                                                                                   & -                                   & \cite{Navet2016}                                                                                                                                                                                            \\ \hline
DEVS-Suite Simulator                                                                   & -                                   & \cite{Alshareef2018}                                                                                                                                                                                        \\ \hline
JSBSim                                                                                 & -                                   & \cite{Hahn2015}                                                                                                                                                                                             \\ \hline
ScicosLab                                                                              & -                                   & \cite{Hahn2015}                                                                                                                                                                                             \\ \hline
CIF                                                                                    & -                                   & \cite{Reijnen2017}                                                                                                                                                                                          \\ \hline
\end{tabular}
\end{table}

\vspace{3mm}
\textbf{Transformation}
\vspace{3mm}

38 studies (15.90\%) presented transformations (listed in Table \ref{tab:Transformation}). 30 studies covered one transformation type (either M2M or M2T), 3 studies considered two transformation types, while 1 study \cite{Gerking2015} showed 3 different transformation types namely M2M, M2T, T2M. the transformation types presented by the other 4 studies was not clarified. Therefore, 39 transformations were presented in total and they are as follow: 28 M2M transformations, 10 M2T transformations, and one T2M transformation. Studies implemented M2M transformations can be categorized into two:

\vspace{2mm}

First category covers the studies using existing tools and languages. \cite{Guo2018} used Y2U tool to transform Statechart models to UPPAAL timed automata model. \cite{Goncalves2016a} presented M2M transformation by transforming Simulink simulation models to AADL architectural models using Assisted Transformation of Models engine. AADL and Modelica were used in \cite{Zhang2013a,Zhang2014d}, where both Modelica and AADL were transformed to each other. In \cite{Drago2013}, authors used Critical Infrastructure Protection - Vulnerability Analysis and Modeling (CIP VAM) UML profile to transform UML models to Bayesian Network (BN) models. In \cite{Ollinger2013}, the authors transformed UML models to Distributed Embedded Real-time Compact Specification (DERCS) models with using GenERTiCA. In \cite{Bakirtzis2018}, the authors transformed a SysML model to a graph by employing GraphML. Other studies include: \cite{Gerking2015,DiNatale2016,Motii2017} used QVT, \cite{Dell2014} implemented M2M transformation using Xtext, \cite{Bao2016,Zhang2014,Whitsitt2014} used Graph Rewriting and Transformation (GReAT) for M2M transformation, while EXTEND is used for the M2M transformation in \cite{Buckl2010}.

\vspace{2mm}

Studies which developed metamodel, tool, or language for the M2M, M2T, and T2M transformations, constitute the second category. In \cite{Son2012a}, they proposed a transformation method that transforms Simulink model to ECML model by designing metamodels for both Simulink and ECML. \cite{Pajic2012} developed the model translation tool UPP2SF that transforms UPPAAL timed automata models to Simulink/Stateflow. On the contrary, \cite{Jiang2018} developed a tool named STU that translates Simulink/Stateflow model into UPPAAL timed automata model. In \cite{Goncalves2017}, the authors developed a tool named ECPS Verifier that was used for the transformation of AADL models to UPPAL timed automata. In \cite{Passarini2014}, they presented a tool named Simulink/AADL Translator Tool (AS2T) that automates the transformation of the simulation models of Simulink to AADL models. \cite{Cheng2015} presented a framework called Modana that helps transforming SysML and MARTE models into Reactive Modules Language (RML) and Modelica models. 

\vspace{2mm}

The studies bellow presented the implementations of M2T transformations using existing tools like Acceleo \cite{Mamun2013,DiNatale2016}, Xtend \cite{Gerking2015,Dell2014}, IOPT tools \cite{Brito2017}, and GenERTiCA \cite{Ollinger2013}, except for one study \cite{Zhou2018} that presented meta-models of HybridUML and Quantified Hybrid Program (QHP) and then they used ATLAS Transformation Language (ATL) for defining the transformation rules. \cite{Gerking2015} was the only study presenting T2M transformation using Xtext.

\begin{table}[]
\caption{Reported model transformations, and used transformation tools and languages}
\label{tab:Transformation}
\begin{tabular}{|l|l|l|l|}
\hline
\multicolumn{4}{|c|}{\textbf{Transformation tools}}                                                                                                                                                                            \\ \hline
\multicolumn{1}{|c|}{Transformation Type} & \multicolumn{1}{c|}{Existing tool} & \multicolumn{1}{c|}{Developed tool}                                                                   & \multicolumn{1}{c|}{Relevant studies} \\ \hline
M2M                                       & -                                  & metamodel                                                                                             & \cite{Son2012a}                              \\ \hline
                                          & Y2U tool                           & -                                                                                                     & \cite{Guo2018}                               \\ \hline
                                          & AST Engine                         & -                                                                                                     & \cite{Goncalves2016a}                        \\ \hline
                                          & QVT                                & -                                                                                                     & \cite{Gerking2015,DiNatale2016,Motii2017}  \\ \hline
                                          & AADL,Modelica                      & -                                                                                                     & \cite{Zhang2013a,Zhang2014d}                \\ \hline
                                          & CIP VAM UML                        & -                                                                                                     & \cite{Drago2013}                             \\ \hline
                                          & EXTEND                             & -                                                                                                     & \cite{Buckl2010}                             \\ \hline
                                          & GreAT                              & -                                                                                                     & \cite{Bao2016,Zhang2014,Whitsitt2014}      \\ \hline
                                          & UML4IoT                            & -                                                                                                     & \cite{Thramboulidis2018}                     \\ \hline
                                          & -                                  & UPP2SF                                                                                                & \cite{Pajic2012}                             \\ \hline
                                          & ME+ tool                           & -                                                                                                     & \cite{Chen2016}                              \\ \hline
                                          & -                                  & STU                                                                                                   & \cite{Jiang2018}                             \\ \hline
                                          & GenERTiCA                          & -                                                                                                     & \cite{Ollinger2013}                          \\ \hline
                                          & -                                  & ECPS Verifier                                                                                         & \cite{Goncalves2017}                         \\ \hline
                                          & Xtext                              & -                                                                                                     & \cite{Dell2014}                              \\ \hline
                                          & GraphML                            & -                                                                                                     & \cite{Bakirtzis2018}                         \\ \hline
                                          & -                                  & AS2T                                                                                                  & \cite{Passarini2014}                         \\ \hline
                                          
                                          & -                                  & Modana                                                                                                  & \cite{Cheng2015}                         \\ \hline
                                          & Other                                  & -                                                                                                & \cite{Nikiforova2017a,Liu2017a,Tariq2014,Jarus2016,Chen2018}  
                                          
                                          \\ \hline
                                         
M2T                                       & Acceleo                            & -                                                                                                     & \cite{Mamun2013,DiNatale2016}               \\ \hline
                                          & Xtend                              & -                                                                                                     & \cite{Gerking2015,Dell2014}                 \\ \hline
                                          & ATL                                & metamodel                                                                                             & \cite{Zhou2018}                              \\ \hline
                                          & IOPT tools                         & -                                                                                                     & \cite{Brito2017}                             \\ \hline
                                          & GenERTiCA                          & -                                                                                                     & \cite{Ollinger2013}                          \\ \hline
                                          & Other                          & -                                                                                                     & \cite{Bougouffa2018,Weissnegger2016a,Goncalves2016}                          \\ \hline
T2M                                       & Xtext                              & -                                                                                                     & \cite{Gerking2015}                           \\ \hline
Other                                     &                                    &                                                                                                       & \cite{Tuo2017,Alshareef2018,B2018}         \\ \hline
\end{tabular}
\end{table}

\vspace{3mm}
\textbf{Validation and Verification (V\&V)}
\vspace{3mm}

35 studies (14.64\%) reported V\&V activity. Only 2 studies developed a tool \cite{Weissnegger2016a,B2018}, and 1 study developed an ontology \cite{Lynch2016}. Table \ref{tab:VV} presents studies which implemented V\&V as part of their work.

\vspace{2mm}

Studies using UPPAAL for verification include the followings: \cite{Guo2018}  used UPPAAL to formally verify the safety properties of medical guideline. A Domain-specific model checking (DSMC) for MECHATRONICUML using UPPAAL model checker was presented in \cite{Gerking2015}. A pacemaker was modeled and verified using UPPAAL in \cite{Pajic2012}. In \cite{Jiang2018}, the authors used the UPPAAL tool for the verification of SIMULINK/STATEFLOW models after being transformed to UPPAAL timed automata. In \cite{Goncalves2017}, UPPAAL was used for the formal verification (i.e. model checking) of AADL models.

\vspace{2mm}

Other tools and languages used for model checking for verification include; Simple Promela Interpreter (SPIN) model checker was used in \cite{Zhang2014} to verify the Promela code. Also, in \cite{He2018}, SPIN was used as a model checker to verify the PrT net models after translating it to a Promela code. In \cite{Pagliari2018}, they used a probabilistic model checker called PRISM. The authors in \cite{Pohlmann2014} verified their protocols via timed model checking MECHATRONICUML.

\vspace{2mm}

Simulink/Stateflow was used in \cite{VanDeMortel-Fronczak2014} to verify supervisory controllers for hierarchical systems. Simulink Design Verifier (SLDV) was used for the verification of the simulation models in \cite{Silva2015}. In \cite{Murugesan}, they used SLDV to verify the behavioral models developed in Simulink in order to meet the requirements modeled. Furthermore, Object Constraint Language (OCL) was used for the verification of the static semantics of a meta-model presented in \cite{Mamun2013}. Also, in \cite{Angelo2018}, OCL was used for defining and validating metamodel constraints. In \cite{Zhou2018}, a verification of KeYmaera-QHP code in KeYmaera, a hybrid verification tool, was presented. In \cite{Neema2014}, the authors used FORMULA for metamodel analysis and verification. Frama-C was used in \cite{Cohen2015} to prove and verify a developed C code library. In \cite{Murugesan}, Assume Guarantee REasoning Environment (AGREE) tool was utilized to verify that the AADL architectural models satisfy the system requirements.

\vspace{2mm}

Studies, presenting validation, are summarized as follows: \cite{Walch2017a} developed a modeling tool (CPS4I) and a modeling method (SeRoIn) then validated them using Open Models Laboratory (OMiLAB). CHECK validation language was used in \cite{Buckl2010} to formulate tests for the detection of semantic design errors in the developed models. A generated code in \cite{Banerjee2014} was tested and analyzed using Frama-C. In \cite{Reijnen2017}, the authors implemented simulation-based validation. 

\vspace{2mm}

A tool, named Simulation and Verification of Hierarchical Embedded Systems (SHARC), was developed in \cite{Weissnegger2016a} for the verification of the behavior of automotive safety-critical systems. In \cite{Lynch2016}, the authors developed an ontology and used it as the validation mechanism.

\begin{table}[]
\caption{Reported V\&V and the used tools}
\label{tab:VV}
\begin{tabular}{|l|l|l|}
\hline
\multicolumn{3}{|c|}{\textbf{V\&V}}                                                                                                                                                                                                                        \\ \hline
\multicolumn{1}{|c|}{Existing Tool/Language}                                               & \multicolumn{1}{c|}{Developed Tool/Language} & \multicolumn{1}{c|}{Relevent studies}                                                                          \\ \hline
UPPAAL                                                                                     & -                                            & \cite{Guo2018,Gerking2015,Guo2017,Pajic2012,Jiang2018,Goncalves2017}\\ \hline
Simulink Design Verifier (SLDV)                                                            & -                                            & \cite{Silva2015,Whalen2014,Murugesan}                                                                               \\ \hline
SPIN (Simple Promela Interpreter)                                                          & -                                            & \cite{Zhang2014,He2018}                                                                                              \\ \hline
Object Constraint Language (OCL)                                                           & -                                            & \cite{Mamun2013,Angelo2018}                                                                                          \\ \hline
\begin{tabular}[c]{@{}l@{}}Clock Constraint Specification Language\\   (CCSL)\end{tabular} & -                                            & \cite{Peters2015}                                                                                                     \\ \hline
Open Models Laboratory (OMiLAB)                                                            & -                                            & \cite{Walch2017a}                                                                                                     \\ \hline
KeYmaera                                                                                   & -                                            & \cite{Zhou2018}                                                                                                       \\ \hline
CHECK                                                                                      & -                                            & \cite{Buckl2010}                                                                                                      \\ \hline
-                                                                                          & Ontology                                     & \cite{Lynch2016}                                                                                                      \\ \hline
\begin{tabular}[c]{@{}l@{}}Linux Driver Verification tool\\   (LDV)\end{tabular}           & -                                            & \cite{Kothari2018}                                                                                                    \\ \hline
FORMULA                                                                                    & -                                            & \cite{Neema2014}                                                                                                      \\ \hline
EAST-ADL                                                                                   & -                                            & \cite{Chen2015}                                                                                                       \\ \hline
-                                                                                          & SHARC                                        & \cite{Weissnegger2016a}                                                                                               \\ \hline
Web Generic Modeling Environment (WebGME)                                                  & -                                            & \cite{Bunting2016}                                                                                                    \\ \hline
Protégé                                                                                    & -                                            & \cite{Sanislav2017}                                                                                                   \\ \hline
Simulink                                                                                   & -                                            & \cite{VanDeMortel-Fronczak2014}                                                                                       \\ \hline
PIPE+                                                                                      & -                                            & \cite{He2018}                                                                                                         \\ \hline
Frama-C                                                                                    & -                                            & \cite{Cohen2015}                                                                                                      \\ \hline
PRISM                                                                                      & -                                            & \cite{Pagliari2018}                                                                                                   \\ \hline
MECHATRONICUML                                                                             & -                                            & \cite{Pohlmann2014}                                                                                                   \\ \hline
AGREE                                                                                      & -                                            & \cite{Murugesan}                                                                                                      \\ \hline
-                                                                                          & BPMN4CPS Tool                                & \cite{B2018}                                                                                                          \\ \hline
\end{tabular}
\end{table}

\vspace{3mm}
\textbf{Modeling}
\vspace{3mm}

33 studies (13.81\%) reported about modeling. This category encompasses studies which used existing languages/tools for modeling, wherein studies which developed a language/tool for modeling were included in the group of system design. Only 4 studies \cite{Kothari2018,Orojloo2017,Cheng2014,Xin2015} did not report any tool, instead, they either proposed an approach for modeling CPSs or used equational models. Table \ref{tab:Modeling} shows various languages/tools used for modeling by the studies.

\vspace{2mm}

In \cite{Silva2015,Silva2014}, authors used Ptolemy II to model Medical CPS, while in \cite{Barbieri2016} the behavioral model of production nominal resource was modeled using Ptolemy II. In \cite{Parveen2018}, authors modeled a Holter Monitor. They used Ptolemy II to model the device's functionality and UPPAAL for modeling system's state space and the transitions between them. A modeling approach called time-constrained aspect-oriented Petri net was presented in \cite{Qian2013}. The approach combines discrete/continuous Petri nets and aspect-orientation for modeling CPS. In the work presented in \cite{Liu2017a}, colored Petri nets were extended to probabilistic colored Petri nets for modeling and analyzing CPS attacks. Petri nets were also used in \cite{Chen2011} for modeling smart grid threats.

\vspace{2mm}

View oriented approach was adopted in \cite{Zhang2013a} for the description of different aspects of an aerospace CPS. Modelica was used for modeling the overall architecture of a lunar rover robot and the lunar rover robot's body structure model while AADL was used for modeling the navigation system of the lunar rover. Similarly, authors in \cite{Zhang2013c} integrated AADL, UML and Modelica to model the requirements of a vehicular ad-hoc network. Further, in \cite{Zhang2014c}, AADL and Modelicaml were integrated to model big data-driven CPS. In \cite{Murugesan}, Simulink/Stateflow was adopted to model a generic patient-controlled analgesia infusion pump system for analyzing logical requirements and behaviors, while AADL was used for developing the architectural model of the system.

\vspace{2mm}

Yakindu statechart tools was adopted in \cite{Guo2018} to model and simulate a stroke statechart model. Likewise, in \cite{Guo2017}, they used Yakindu statecharts for the modeling of a simplified cardiac arrest. The study in \cite{Garamvolgyi2018}, created a UML statechart model for an envisioned CPS scenario using YAKINDU statechart modeling tool again. In \cite{Motii2017}, Papyrus tool was used for creating the UML models. The authors in \cite{Kacem2017} presented a methodology for knowledge representation of CPS using the modeling tool Papyrus. \cite{Sanislav2017} used UML for defining the dependability analysis models. Implementations of modeling CPS using HybridUML was presented in \cite{Zhou2018}.

\vspace{2mm}

Finite state machine (FSM) was adopted in \cite{Brandenbourger2016} to model the behavior of automation components. \cite{Navet2016} used FSM to describe the logic of a servo tester. GME was used in \cite{Li2015} to build Lathe CNC System models and export models' data as an XML file. ASLan++ was used in \cite{Rocchetto2017} for modeling water treatment plant and attack model. \cite{Cheng2014} presented a new formalism named Stochastic Occurrence Hybrid Automata and a modeling approach to model the stochastic behavior in CPSs.

\begin{table}[]
\centering
\caption{Reported tools/languages used for modeling and the studies used them.}
\label{tab:Modeling}
\begin{tabular}{|l|l|}
\hline
\multicolumn{2}{|c|}{\textbf{Modeling}}                                                                                           \\ \hline
\multicolumn{1}{|c|}{Tool/Language} & \multicolumn{1}{c|}{Relevant studies}                                                       \\ \hline
Ptolemy II                          & \cite{Silva2015,Parveen2018,Barbieri2016,Silva2014} \\ \hline
Petri nets                          & \cite{He2018,Chen2011,Qian2013,Liu2017a}                                                        \\ \hline
AADL                                & \cite{Zhang2013a,Zhang2013c,Zhang2014c,Murugesan}                                               \\ \hline
YAKINDU Statechart Tools            & \cite{Guo2018,Guo2017,Garamvolgyi2018}                                                           \\ \hline
Papyrus tool                        & \cite{DiNatale2016,Kacem2017,Motii2017}                                                          \\ \hline
UML                                 & \cite{Zhang2013c,Sanislav2017}                                                                    \\ \hline
Modelica                            & \cite{Zhang2013c,Zhang2014c}                                                                      \\ \hline
Simulink                            & \cite{Whalen2014,Murugesan}                                                                       \\ \hline
finite state machine (FSM)          & \cite{Brandenbourger2016,Navet2016}                                                               \\ \hline
SysML                               & \cite{DiNatale2016,Huang2018}                                                                     \\ \hline
HybridUML                           & \cite{Zhou2018}                                                                                    \\ \hline
MARTE                               & \cite{Huang2018}                                                                                   \\ \hline
UPPAAL                              & \cite{Parveen2018}                                                                                 \\ \hline
GME (Generic Modeling Environment)  & \cite{Li2015}                                                                                      \\ \hline
MetaEdit+                           & \cite{Chen2016}                                                                                    \\ \hline
web ontology language (OWL)         & \cite{Petnga2016}                                                                                  \\ \hline
ADVISE Meta tool                    & \cite{Cheh2017}                                                                                    \\ \hline
\end{tabular}
\end{table}

\vspace{3mm}
\textbf{Code generation}
\vspace{3mm}

24 studies (10.04\%) reported about language/tools used or developed for code generation purposes. Table \ref{tab:Code-generation} lists used and developed languages/tools for code generation. These studies are categorized here into studies which used existing tools for code generation, and studies which developed tools for code generation.

\vspace{2mm}

In \cite{Pajic2012}, they used Simulink Real-Time Workshop Embedded Coder (RTWEC) to generate C code from a pacemaker Stateflow chart. Likewise, C code and VHDL code were generated in \cite{Jiang2018,Murugesan} from the Stateflow models using Simulink coder. Moreover, a tool named GeneAuto was presented in \cite{Cohen2015} that generates C or ADA code from Simulink models. In \cite{Neema2014}, built-in code generator for Embedded Systems Modeling Language (ESMoL) was used to generate functional C code. IOPT-Flow tool framework was used in \cite{Pereira2016} to generate C and Javascript code or VHDL hardware descriptions.

\vspace{2mm}

Programmable Logic Controllers (PLCs) Code Generation was presented in \cite{Gritzner2018} using Scenario Modeling Language (SML). In the same manner, implementation of PLC code generation was presented in \cite{Reijnen2017} using Compositional Interchange Format (CIF). Clock Constraint Specification Language (CCSL) constraints were utilized in \cite{Peters2015} for code generation purposes. The authors in \cite{Nagele2017} used Xtend for the support of code generation for OpenRTI. Kevoree Modeling Framework (KMF) was used in \cite{Hartmann2015} for the generation of Java API in order to create and manipulate the runtime models.

\vspace{2mm}

In \cite{Ataide2018}, they developed a tool named I2C4IOPT for automatic code generation of globally asynchronous and locally synchronous systems (GALS) - supported by Arduino boards. An ISA88 editor was implemented in EMF in \cite{Bougouffa2018} to generate a programmable logic controllers (PLC) control code. A code generator was developed in \cite{Koutsoukos2012} for the generation of Simulink models and network-scripts. In \cite{Banerjee2014}, a model-based code generator for medical CPS was presented. An interpreter was developed in \cite{Zhang2014} to translate finite-state machine (FSM) models and constraints into Promela code.

\begin{table}[]
\caption{Reported code generation languages/tools and studies used them}
\label{tab:Code-generation}
\begin{tabular}{|l|l|l|}
\hline
\multicolumn{3}{|c|}{\textbf{Code generation}}                                                                                                                                                        \\ \hline
\multicolumn{1}{|c|}{Existing Tool/Language}                                                                   & \multicolumn{1}{c|}{Developed Tool/Language} & \multicolumn{1}{c|}{Relevant studies} \\ \hline
Simulink coder                                                                                                 & -                                            & \cite{Pajic2012,Jiang2018,Murugesan}       \\ \hline
-                                                                                                              & I2C4IOPT                                     & \cite{Ataide2018}                            \\ \hline
Acceleo                                                                                                        & -                                            & \cite{Lavigne2018}                           \\ \hline
\begin{tabular}[c]{@{}l@{}}Clock Constraint Specification Language\\   (CCSL)\end{tabular}                     & -                                            & \cite{Peters2015}                            \\ \hline
Xtend                                                                                                          & -                                            & \cite{Nagele2017}                            \\ \hline
-                                                                                                              & code generator                               & \cite{Koutsoukos2012,Banerjee2014,Li2015}  \\ \hline
XPand                                                                                                          & -                                            & \cite{Buckl2010}                             \\ \hline
Scenario Modeling Language (SML)                                                                               & -                                            & \cite{Gritzner2018}                          \\ \hline
-                                                                                                              & ISA88 editor                                 & \cite{Bougouffa2018}                         \\ \hline
\begin{tabular}[c]{@{}l@{}}Embedded Systems Modeling\\   Language (ESMoL)\end{tabular}                         & -                                            & \cite{Neema2014}                             \\ \hline
IOPT-Flow tool framework                                                                                       & -                                            & \cite{Pereira2016}                           \\ \hline
-                                                                                                              & interpreter                                  & \cite{Zhang2014}                             \\ \hline
\begin{tabular}[c]{@{}l@{}}Generation of Embedded Real-Time Code based\\   on Aspects (GenERTiCA)\end{tabular} & -                                            & \cite{Ollinger2013}                          \\ \hline
KMF                                                                                                            & -                                            & \cite{Hartmann2015}                          \\ \hline
Rhapsody code generator                                                                                        & -                                            & \cite{Palachi2013}                           \\ \hline
HA-SPIRAL                                                                                                      & -                                            & \cite{Low2017}                               \\ \hline
\begin{tabular}[c]{@{}l@{}}Compositional Interchange Format\\   (CIF)\end{tabular}                             & -                                            & \cite{Reijnen2017}                           \\ \hline
GeneAuto                                                                                                       & -                                            & \cite{Cohen2015}                             \\ \hline
\end{tabular}
\end{table}

\vspace{3mm}
\textbf{System Analysis}
\vspace{3mm}

15 studies (6.28\%) reported language/tools which used or was developed for system analysis reasons. Table \ref{tab:System-Analysis} lists the used and developed languages/tools for system analysis. Studies can be categorized into ones directly using existing tools for system analysis, and others developing new tools for system analysis purposes.

\vspace{2mm}

In \cite{Zhao2017}, meta-models for operational analysis and system analysis were developed. They also used TTool for safety analysis. A knowledge-based approach using Failure Models, Effects and Criticality Analysis (FMECA) techniques was presented in \cite{Ali2018}. The authors first modeled FMECA using UML class diagram, then the FMECA metamodel was expressed in Protégé, which was then used to build an ontology-based KB. A metamodel was developed to enable the management of application requirements and business constraints for CPS in \cite{Sapienza2014}. CPS meta-model for knowledge formalization was presented in \cite{Lezoche2018}, where they also implemented formal concept analysis in their work. CPS Safety Analysis and Simulation Platform (CP-SAP) was developed in \cite{Fan2018} for Human-machine interaction (HMI) safety analysis of CPS. A framework called Modana was presented in \cite{Cheng2015} that aims to model and analyze the non-functional aspect (i.e. time, energy, etc) of Energy-Aware CPS.

\vspace{2mm}

In \cite{Qian2013}, a modeling approach based on discrete/continuous Petri nets was proposed for schedulability analysis. Also, a modeling approach was presented in \cite{Liu2017a} that supports both qualitative and quantitative analysis of CPS attacks using probabilistic colored Petri nets. CPS dependability analysis was presented in \cite{Hu2013} using Stochastic Petri Net (SPN). An approach for specification and analysis of automotive CPS was presented in \cite{Zhang2013c} where Modelica was used for analyzing engine model and AADL was used for End to End Delay Analysis from brake-pedal to throttle actuator. Security analysis tool CL-AtSe was used in \cite{Rocchetto2017} for analyzing and discovering potential attacks on Industrial Control Systems.

\begin{table}[H]
\centering
\caption{Reported System analysis languages/tools}
\label{tab:System-Analysis}
\begin{tabular}{|l|l|l|}
\hline
\multicolumn{3}{|c|}{\textbf{System Analysis}}                                                                                                                                                                                                                          \\ \hline
  Existing Tool/Language & Developed   Tool/Language & Relevant   studies\\ \hline
EAST-ADL                                                                                 & -                                                                                        & \cite{Chen2017,Chen2016}                                                                 \\ \hline
Petri   nets                                  & -                                                                                        & \cite{Qian2013,Liu2017a,Hu2013}                                                          \\ \hline
AADL                                                                                     & -                                                                                        & \cite{Zhang2013c}                                                                        \\ \hline
Modelica                                                                                 & -                                                                                        & \cite{Zhang2013c}                                                                        \\ \hline
-                                                                                        & Metamodel                                                                                & \cite{Sapienza2014,Morozov2018,Zhao2017,Lezoche2018,Ali2018}                             \\ \hline
-                                                                                        & CP-SAP                                                                                   & \cite{Fan2018}                                                                           \\ \hline
-                                                                                        & Modana                                                                                   & \cite{Cheng2015}                                                                         \\ \hline
CL-AtSe                                                                                  & -                                                                                        & \cite{Rocchetto2017}                                                                     \\ \hline
TTool                                                                                    & -                                                                                        & \cite{Zhao2017}                                                                          \\ \hline
HiP-HOPS tool                                                                            & -                                                                                        & \cite{Chen2016}                                                                          \\ \hline
\end{tabular}
\end{table}

\subsubsection{Developed/used MDE tools for CPS}

In this section, the results and findings of \textbf{"RQ2: Is/Are there any tool(s) used/developed to apply MDE in/for cyber-physical systems in the study?"} and its sub-questions are presented.

Out of the 140 studies, 13 studies did not present or develop any tool/language. The other 127 studies are as follows: 

\begin{itemize}
    \item 68 studies used existing tools/languages for modeling CPS.
    \item 59 studies developed DSLs, Metamodels, tools in addition to using existing tool/languages. 
\end{itemize}

\textbf{RQ2.1: Which tool(s) is/are presented/used in each phase of the system development?}

\begin{figure} []
\centering
	\includegraphics[scale=1.15]{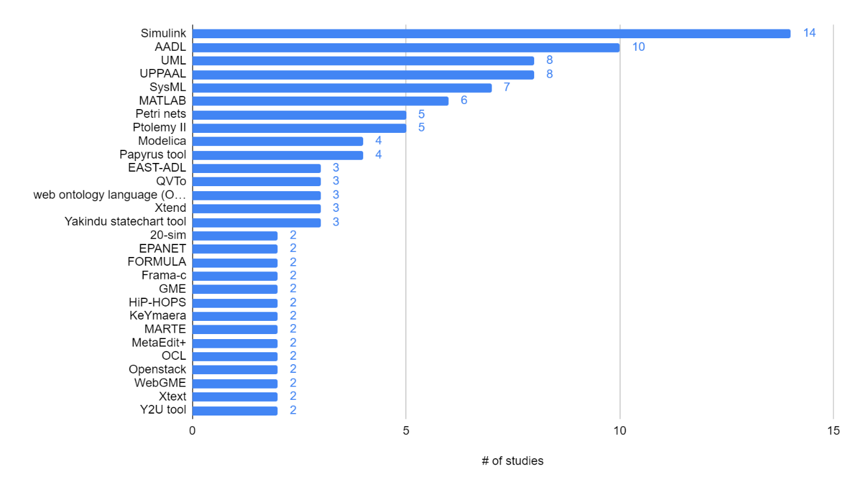}
	\caption{Most used tools/languages which are at least reported by 2 studies.}
	\label{fig:Most_used_tools_languages}
\end{figure}

Figure \ref{fig:Most_used_tools_languages} shows tools and languages used by the primary studies. For better understanding of the tools/languages, and to give the reader a clear idea about the MDE phase/activity the tool/language was used for, a correlation analysis between RQ1.2 and RQ2 is presented in section \ref{RQ1}. Therefore, in this section, the most used tool/languages are briefly discussed.

\vspace{2mm}

The study found that Simulink is the most used tool. Majority of the reviewed studies used Simulink for simulation purposes, listed in (Table \ref{tbl:Simulation}). Simulink was also used for modeling \cite{Whalen2014,Murugesan}. Simulink coder is used for code generation purposes in \cite{Pajic2012,Jiang2018,Murugesan}. Simulink Design Verifier (SLDV) was adopted for the verification of models \cite{Silva2015,Whalen2014,Murugesan}. AADL follows Simulink as the most used tool. It was used for modeling the cyber part of the system \cite{Zhang2013a}, developing architectural models \cite{Murugesan}, or for system analysis \cite{Zhang2013c}.

\vspace{2mm}

UML is used by various studies for building metamodels, listed in Table \ref{tbl:SystemDesign}, it was also used for modeling activities like defining dependability analysis models \cite{Sanislav2017}. The vast majority of the studies used UPPAAL for verification, see Table \ref{tab:VV}. For instance, \cite{Parveen2018} used UPPAAL for modeling system's state space and the transitions between them.

\textbf{RQ2.2: If any tool(s) is/are developed in the study, is/are this/these tool(s) reported?}

\begin{figure} [H]
\centering
	\includegraphics[scale=0.8]{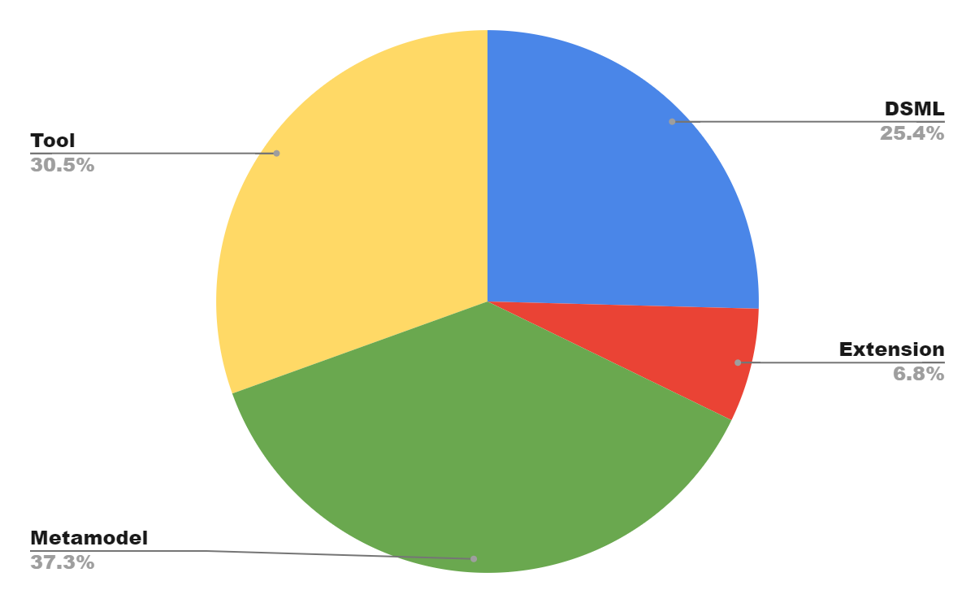}
	\caption{Frequency and type of the developed tool/language.}
	\label{fig:Frequency_and_type_of_the_developed_tool_language}
\end{figure}

As mentioned earlier, 59 studies developed DSL/DSML, metamodel, tool, or extension. 22 of these studies developed a metamodel, 15 studies developed DSL/DSML, 18 studies developed a tool (including 2 frameworks and 1 platform), and 4 tools presented extensions as shown in Figure \ref{fig:Frequency_and_type_of_the_developed_tool_language}. These developed tools/languages were addressed in a detailed way in the correlation analysis done between RQ1.2 and RQ2 which is presented in section \ref{RQ1}.

\textbf{RQ2.2.1: Is/are the developed tool(s) available and/or accessible?}

From the abovementioned 59 studies which consider the development of a tool/language, only 10 studies \cite{Angelo2018,Ataide2018,Alrimawi2018,Goncalves2016a,Zhang2014,Jiang2018,Huang2018,Goncalves2017,Weissnegger2016a,B2018} provided public access (mostly with a web link) to the developed tool/language. 

\textbf{RQ2.2.2: What is/are the framework(s) or programming language(s) for the development of this/these tools?}

\begin{figure} [H]
\centering
	\includegraphics[scale=1.1]{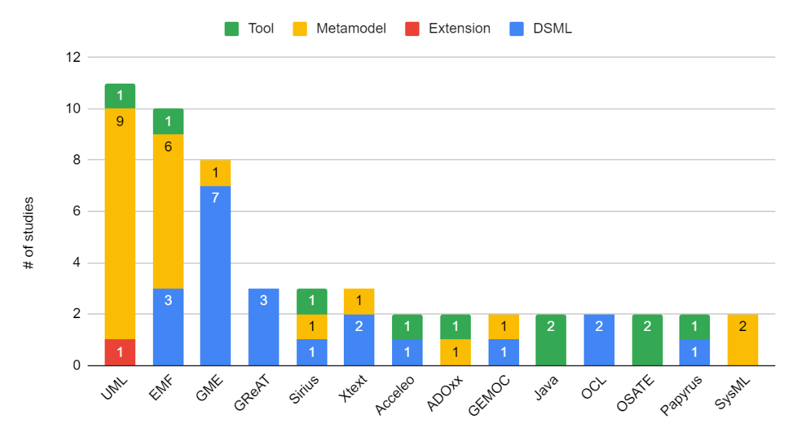}
	\caption{Used languages/frameworks for developing DSMLs, metamodels and/or tools.}
	\label{fig:Used_languages_frameworks}
\end{figure}

As indicated in Figure \ref{fig:Used_languages_frameworks}, UML is the most used language, followed by EMF and GME. Figure \ref{fig:Used_languages_frameworks} shows the correlation between RQ2.2 and RQ2.2.2. It is clear that UML is mostly used for building metamodels, where EMF is used for building both metamodels and DSLs, and GME is mostly used for building DSLs. Other presented tools are; GreAT that is used alongside with GME \cite{Bao2016,Zhang2014}, Sirius used for building the graphical concrete syntax \cite{Guan2018,Huang2018,Lavigne2018} and finally, Xtext used for developing DSL grammar in \cite{Nagele2017}, and for building the textual concrete syntax as in \cite{Huang2018}. These tools/languages are listed in Table \ref{tbl:SystemDesign} and discussed in detail in section \ref{RQ1}.

\vspace{2mm}

\begin{figure} [H]
\centering
	\includegraphics[scale=1.1]{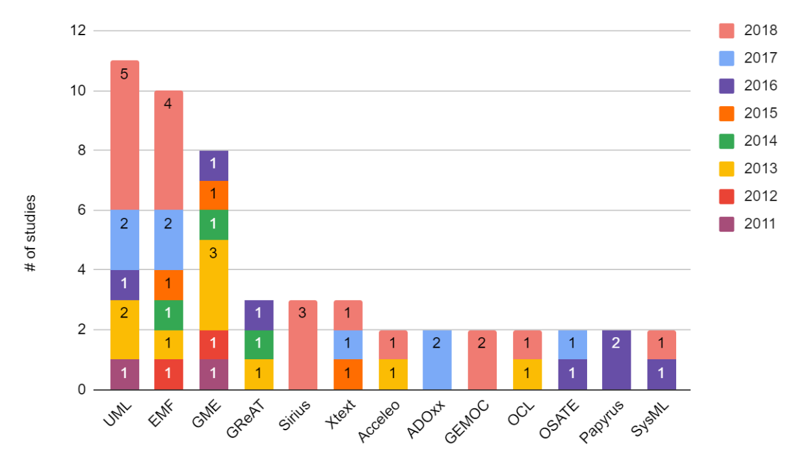}
	\caption{Distribution of languages/frameworks used for developing DSMLs, metamodels and/or tools over publication years.}
	\label{fig:Distribution_of_languages_frameworks_over_publication_years}
\end{figure}

\begin{figure} [H]
\centering
	\includegraphics[scale=1.1]{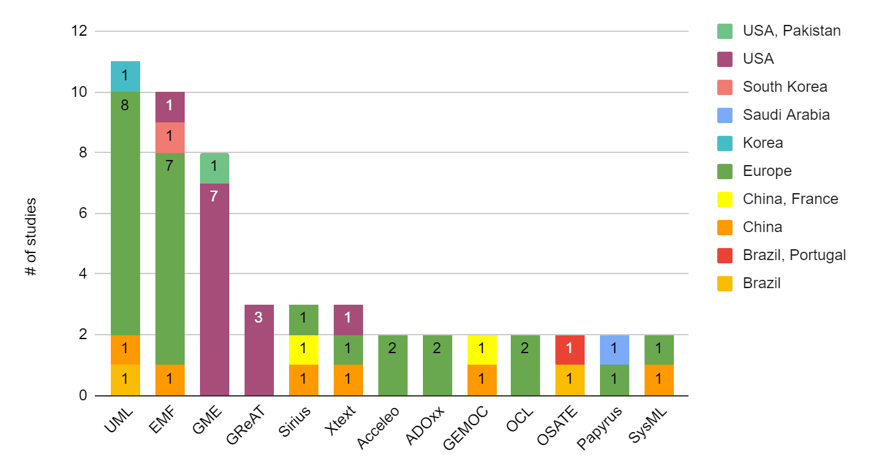}
	\caption{Distribution of languages/frameworks used for developing DSMLs, metamodels and/or tools over author affiliation countries.}
	\label{fig:Distribution_of_languages_frameworks_over_authors_affiliation_country}
\end{figure}

GME seems to be the third most used language for building DSLs and metamodels. However, it is worth mentioning that the results of distributing RQ2.2.2 over the publication years (see Figure \ref{fig:Distribution_of_languages_frameworks_over_publication_years}) show that GME is not used for the last 2 years (2017 and 2018) of the examined period by any of the primary studies. Further, the results of RQ2.2.2 were distributed over authors’ country of affiliation as depicted in Figure \ref{fig:Distribution_of_languages_frameworks_over_authors_affiliation_country}. The study found out that GME and its tool GReAT were only used by authors/researchers affiliated to the USA, where on the other hand, UML and EMF were mostly used by authors/researchers affiliated to Europe.

\subsubsection{Addressed CPS components}

 In this section, the results and findings for \textbf{“RQ3: What is/are the CPS component(s) addressed in the study?”} are presented. 

\vspace{2mm}

\begin{figure} [H]
\centering
	\includegraphics[scale=1.2]{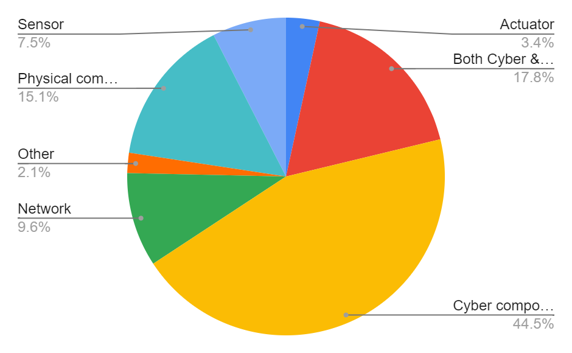}
	\caption{Used languages/frameworks for developing DSMLs, Metamodels, or tools.}
	\label{fig:Reported_CPS_components}
\end{figure}

According to \cite{gunes2014survey}, a CPS mainly consists of 5 components, which are, Physical components, Cyber components, Sensors, Actuators, and Network. Amongst the 140 primary studies, only 6 papers were left undetermined (the addressed CPS component by the 6 papers could not be determined) and 9 studies addressed more than 1 CPS component. Figure \ref{fig:Reported_CPS_components} shows the categories of CPS components. The full list of studies and their supporting CPS components is given in Table \ref{Tab:CPS_components}. 

\begin{itemize}
    \item Cyber Component: 65 studies (44.5\%) addressed this component, i.e. the software aspect of the system. Examples are Controllers (e.g. \cite{Aziz2016,Ataide2018}), Development Artifacts related with transformation \cite{Nagele2017}, simulation \cite{Son2012a}, System verification \cite{Zhou2018}.System behavior covering timing behavior \cite{Guo2018}, System safety properties \cite{Peters2015}, and System requirements \cite{Zhang2011,Hu2013}.
    
    \item Physical Component: Reported by 22 studies (15.1\%). These studies addressed the physical and hardware components of the system, e.g. Physical Dynamics (environment behavior)  \cite{Cheng2014,Cicirelli2018}, Power plant \cite{Liu2017a,Thramboulidis,Koutsoumpas2015}, Hardware \cite{Chen2016,Zhang2014c}.
    
    \item Both Cyber \& Physical components: Reported by 26 studies (17.8\%). This category contains the studies discussing modeling both cyber and physical aspects of the system. \cite{Broenink2016} reported about modeling a controller (cyber component), and a plant (physical component). Another example is \cite{Zhang2013a} where the authors modeled a lunar rover robot's body (physical component) and its navigation system (cyber component).

	\item Network: Reported by 14 studies (9.6\%). Studies in this group addressed issues like; sensor network  \cite{Zhang2014}, network security \cite{Maksuti2017}, physical attacks \cite{Cheh2017}, Security requirements and attacks \cite{Motii2017}.
	
	\item Sensors:  11 studies (7.5\%) reported this component. Studies in this group addressed the different operations of sensors, like sensor designing  \cite{Goncalves2016a}, sensor management \cite{Mamun2013}, sensor data analysis \cite{Banerjee2014}, and sensor failures \cite{Nannapaneni2016}. 

	\item Actuators: Reported by only 5 studies (3.4\%). This component is less addressed one compared to the other CPS components. For instance, \cite{Goncalves2016a} covered actuator modeling and design, while \cite{Sanislav2017} discussed actuator failure.
    
    \item Other: Studies, which do not fit any of the above categories, are grouped under this category. They consider Business processes \cite{Bocciarelli2017}, workflows (process) \cite{Seiger2015}, data \cite{Plasse2017}.
\end{itemize}

\begin{table}[]
\caption{Addressed CPS components and the corresponding studies}
\label{Tab:CPS_components}
\begin{tabular}{|l|l|p{0.4\textwidth}|}
\hline
\textbf{Category}                  & \textbf{\# of studies} & \textbf{Relevant studies}                                                                                                                                                                                                                                                                                                                                                                                                                                                                                                                                                                                                                                                                                                                       \\ \hline
Cyber component                   & 65                     & \cite{Aziz2016,Ataide2018,Son2012a,Guo2018,Peters2015,Garamvolgyi2018,Walch2017a,Nikiforova2017a,Nagele2017,Latombe2015a,Gerking2015,Zhou2018,Zhang2013b,Jeon2012,Buckl2010,Bao2016,Simko2013a,Chen2017,Housh2018,Lynch2016,Matinnejad2018,Gritzner2018,Brito2017,Kothari2018,Neema2014,Zhang2014d,Jiang2018,Pereira2016,Mezhuyev2013,Ollinger2013,Lavigne2018,Goncalves2017,Tariq2014,Li2015,Whitsitt2014,Dell2014,Barbieri2016,Jarus2016,Whalen2014,Orojloo2017,Petnga2016,VanAcker2015,Brandenbourger2016,SampathKumar2015,Cohen2015,Tan2013,Barve2018,Navet2016,Bakirtzis2018,Sinha2016,Alshareef2018,Rashid2018,Low2017,Zhang2011,Reijnen2017,Pohlmann2014,Passarini2014,Silva2014,Hu2013,Pradhan2015,Fan2018,Chen2018,Murugesan,B2018,Guan2018} \\ \hline
Both Cyber \& Physical components & 26                     & \cite{Martins2015,Zhang2013a,Silva2015,Sapienza2014,Bougouffa2018,Zhang2013c,DiNatale2016,Huang2018,Hartmann2015,Liu2011,An2011,He2018,Tuo2017,Chen2011,Xin2015,Kuesap2008,Pagliari2018,Neema2018,Kacem2017,Broenink2016,Morozov2018,Hahn2015,Lezoche2018,Cheng2015,Graja,Ali2018}                                                                                                                                                                                                                                                                                                                                                                                                                                                                    \\ \hline
Physical component                & 22                     & \cite{Angelo2018,Thramboulidis2018,Liu2017a,Guo2017,Pajic2012,Chen2016,Zhang2014c,Parveen2018,Chen2015,Weissnegger2016a,Bunting2016,Lora2017,VanDeMortel-Fronczak2014,Ciavotta2018,Feldmann2013,Cheng2014,Irisarri2016,Thramboulidis,Koutsoumpas2015,Rocchetto2017,Blackburn2014,Cicirelli2018}                                                                                                                                                                                                                                                                                                                                                                                                                                                  \\ \hline
Network                           & 14                     & \cite{Alrimawi2018,Tariq2012,Maksuti2017,Cicirelli2016,Koutsoukos2012,Drago2013,Housh2018,Zhang2014,Cicirelli2017a,Cheh2017,Mordecai2018,Motii2017,Tundis2017,Yan2009}                                                                                                                                                                                                                                                                                                                                                                                                                                                                                                                                                                                \\ \hline
Sensor                            & 11                     & \cite{Tariq2012,Goncalves2016a,Mamun2013,Cicirelli2016,Zhang2013b,Zhang2014,Cicirelli2017a,Banerjee2014,Sanislav2017,Dell2014,Nannapaneni2016}                                                                                                                                                                                                                                                                                                                                                                                                                                                                                                                                                                                                        \\ \hline
Actuator                          & 5                      & \cite{Goncalves2016a,Cicirelli2016,Cicirelli2017a,Sanislav2017,Dell2014}                                                                                                                                                                                                                                                                                                                                                                                                                                                                                                                                                                                                                                                                              \\ \hline
Other                             & 3                      & \cite{Bocciarelli2017,Seiger2015,Plasse2017}                                                                                                                                                                                                                                                                                                                                                                                                                                                                                                                                                                                                                                                                                                           \\ \hline
\end{tabular}
\end{table}

Further, in this SLR, a correlation analysis of the MDE activities and CPS components is scrutinized so as to provide an understanding of the addressed MDE activities in each CPS component, see the appendixes (Table \ref{App:Q1vsQ4}). Despite the fact that the correlation analysis cannot indicate the CPS domain wholly, for instance, one can see that in the Cyber component, most research works concentrated on  
Transformation (22 studies), V\&V (18 studies), simulation	(17 studies), code generating and system design (16 studies each), while Requirement analysis and System analysis were addressed only by 4 and 3 studies respectively.

Similarly, in the physical component, the research work converged on Simulation (10 studies), System design and V\&V (8 studies each), and Modeling (7 studies). While again System analysis and Requirement analysis are less addressed 3 and 1 study for each respectively. In terms of the sensor component, most research works concentrated on System design (6 studies), however, System analysis for sensors was not addressed by any study. Regarding the actuator component, which is the least addressed CPS component, it is interesting to note that Code generation, Simulation, Requirement analysis, and System analysis were not addressed by any study.

One can deduce from this that in terms of the CPS components, Actuator is the least addressed component. While in terms of the MDE activities, Requirement analysis and System analysis are the least addressed activities for every CPS component.

\subsubsection{Targeted CPS application domains}

In this section, the results and findings for \textbf{“RQ4: Does the study present any application domain?”} and its sub-questions are presented. 	

\vspace{2mm}

\begin{figure} [H]
\centering
	\includegraphics[scale=1.15]{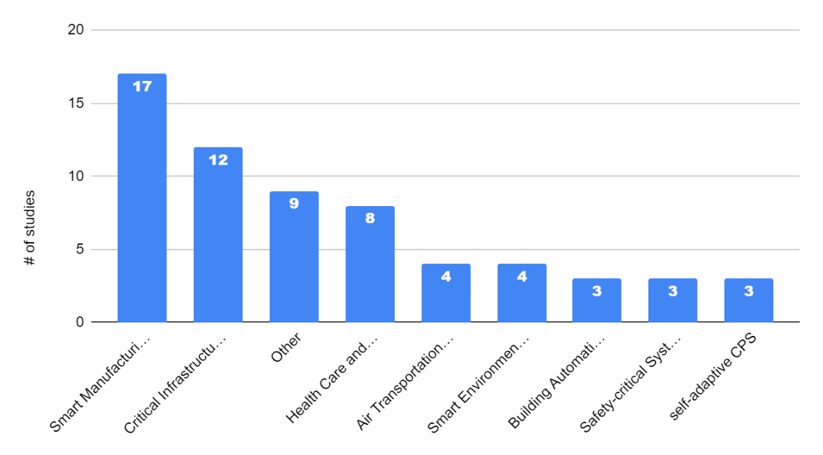}
	\caption{Reported CPS application domains targeted by the studies.}
	\label{fig:CPS_application_domains}
\end{figure}

Figure \ref{fig:CPS_application_domains} depicts the reported CPS application domains. Here, the study is reporting about the CPS domains targeted by the primary studies. There are various CPS domains, such as Critical Infrastructure, Smart Manufacturing, Air Transportation, Emergency Response, Intelligent Transportation, Health Care and Medicine \cite{gunes2014survey}. 63 studies out of total 140 studies (about 45\%), addressed a specific CPS domain, while the rest of them addressed CPS in general. This implies that the modeling activities presented in those studies can be applied to any domain of CPS. CPS application domains are correlated with the evaluation methods presented by the examined studies. Results of this correlation are presented in Table \ref{Tab:CPS_Application_Domain}.

\begin{itemize}
    \item Smart Manufacturing: Addressed by 17 out of total 63 studies (26.98\%). Studies under this category aim at optimizing productivity in factories (smart factories). Applications included in these studies take into account Industry 4.0/Cyber-physical production systems (CPPS) \cite{Maksuti2017,Bougouffa2018,Bocciarelli2017,Ciavotta2018,Brandenbourger2016,Irisarri2016,Thramboulidis,Lezoche2018}, Industrial Applications \cite{Walch2017a,Sinha2016,Rocchetto2017}, Automation Systems \cite{Buckl2010,Zhang2013c}, evolvable production systems \cite{Chen2016,Chen2015}, and Assembly systems (ASs) \cite{Thramboulidis2018}.
    
    \item Critical Infrastructure: 12 studies (19.05\%) reported under this category. It refers to the public infrastructures and valuable properties. Applications grouped under this category cover smart grids  \cite{Hartmann2015,Motii2017,Tundis2017,Chen2011,Xin2015,Koutsoumpas2015,Barve2018,Blackburn2014}, Irrigation Networks \cite{Tariq2012}, railway networks \cite{Drago2013,Kuesap2008}, water distribution systems \cite{Housh2018}.
    
    \item Health Care and Medicine (HC\&M): 8 studies (12.70\%) presented under this category. Included sub-categories are Medical Cyber-Physical Systems (MCPS) \cite{Silva2015,Banerjee2014,Silva2014,Murugesan}, medical best practice guideline \cite{Guo2018,Guo2017}, smart medical devices \cite{Parveen2018,Whalen2014}.
    
    \item Smart Environments: Addressed by 4 studies (6.35\%). The smart environment is a physical environment in which sensing, actuating, network, and computation capabilities are enriched. The followings are the studies grouped under this category:  \cite{Cicirelli2016,Cicirelli2017a,Seiger2015,Cicirelli2018}.
    
    \item Air Transportation: 4 studies (6.35\%) reported under this category. Applications are; Unmanned Aerial Vehicles  \cite{Goncalves2017,Goncalves2016}, Air Traffic Control (ATC) \cite{Mordecai2018}, Aerospace CPS \cite{Zhang2013a}.
    
    \item Safety-critical Systems: Reported by 3 studies (4.76\%). Safety-critical systems are systems whose failure or malfunction can have a severe loss, in terms of human or economic consequences. Studies of this cluster include \cite{Weissnegger2016a,Cohen2015,Zhao2017}.

    \item Reported by 3 studies (4.76\%). Studies in this category aim at providing optimum automation and control to buildings' heating, air conditioning, lighting, etc. by deploying sensors, actuators, and control systems. Studies classified under this group are \cite{Alrimawi2018,Nagele2017,Cheng2015}.

    \item Self-adaptive Systems: 3 studies (4.76\%) presented under this category. Self-adaptive systems are systems that modifies their own behavior during the runtime using feedback due to the constant changes in the system. The followings are the studies grouped under this cluster:  \cite{Angelo2018,Mamun2013,Zhang2013b}.

    \item Other: Studies which did not fit any of the aforementioned categories are grouped under this category. They are as follows: Distributed cyber-physical systems \cite{Ataide2018}, smart contracts \cite{Garamvolgyi2018}, networked control systems \cite{Koutsoukos2012}, Racing sailboats \cite{Lavigne2018}, Intelligent Transportation \cite{Bunting2016}, smart systems \cite{Lora2017},    Material Handling Applications \cite{An2011}, cloud-based CPS \cite{Dell2014}, complex systems \cite{Chen2018}.
    
\end{itemize}

\begin{table}[H]
\caption{CPS application domain correlated with the evaluations presented by the studies.}
\label{Tab:CPS_Application_Domain}
\begin{tabular}{|l|l|p{7cm}|}
\hline
\textbf{Domain}                           & \textbf{Evaluation type} & \textbf{Description}                                                                                                                                                                                                                                                                                                                                   \\ \hline
\multirow{4}{*}{Smart Manufacturing} & Case study & IKEA Gregor office chair \cite{Thramboulidis2018}, assembly production system \cite{Chen2016}, assembly system \cite{Chen2015}, Petroamazonas EP Oil Company \cite{Irisarri2016}, liqueur plant \cite{Thramboulidis}, industrial water process system \cite{Sinha2016}, enterprise production line \cite{Lezoche2018} \\ \cline{2-3} 
                                          & Empirical study          & OMiRob \cite{Walch2017a}
                                          \\ \cline{2-3} 
                                          & Example                  & 
                                          robot packaging system \cite{Merschak2018}, Pick and Place Unit \cite{Bougouffa2018}, Vehicular Ad-hoc NETwork \cite{Zhang2013c}, pneumatic stopper  unit \cite{Brandenbourger2016}, water treatment plant SWaT \cite{Rocchetto2017}.                                                                                            \\ \cline{2-3} 
                                          & Use case                 & 
end-to-end communication use case for an Industry 4.0 application \cite{Maksuti2017}, White-goods production \cite{Ciavotta2018}                                                                                                                                                                                     \\ \hline
\multirow{4}{*}{Critical Infrastructure}  & Case study               & flood level prediction \cite{Tariq2012}, SCADA system \cite{Motii2017}, secondary-voltage control system \cite{Xin2015}

\\ \cline{2-3} 
                                          & Empirical study
& Smart Grid \cite{Hartmann2015}, Water Distribution System \cite{Housh2018}
\\ \cline{2-3} 
                                          & Example                  &
Railway network \cite{Drago2013}, monitoring of smart grids \cite{Tundis2017}, smart meter \cite{Chen2011}, process plant design \cite{Blackburn2014}                                                                                                                                                                          \\ \cline{2-3} 
                                          & Use case                 & Virtual Power Plant \cite{Koutsoumpas2015}
\\ \hline
\multirow{3}{*} {Health Care and Medicine} & Case study   
&
Simplified stroke \cite{Guo2018}, simplified cardiac arrest \cite{Guo2017}, Holter Monitor \cite{Parveen2018}, Clinical scenario \cite{Silva2014}, Generic Patient Controlled Analgesia Infusion Pump (GPCA) system \cite{Murugesan}.                                                                                           \\ \cline{2-3} 
                                          & Empirical study          & clinical scenarios \cite{Silva2015}
\\ \cline{2-3} 
                                          & Use case                 & patient-controlled analgesia infusion pump \cite{Whalen2014}
\\ \hline
\multirow{2}{*} {Smart Environment}        & Case study               & smart environment scenario \cite{Cicirelli2016}, smart office \cite{Cicirelli2017a,Cicirelli2018}  
\\ \cline{2-3} 
                                          & Example                  & newspaper fetching task \cite{Seiger2015}
\\ \hline
\multirow{3}{*}{Air Transportation}       & Case study               & lunar rover system \cite{Zhang2013a}, Unmanned Aerial Vehicle \cite{Goncalves2017}

\\ \cline{2-3} 
                                          & Example                  & VTOL Unmanned Aerial Vehicle \cite{Goncalves2016}
\\ \cline{2-3} 
                                          & Use case                 & Malaysia Airlines Boeing 777 carrying flight MH-370 \cite{Mordecai2018}
\\ \hline
\multirow{2}{*}{Safety-critical Systems}  & Case study               & battery management system \cite{Weissnegger2016a}, railway signaling system \cite{Zhao2017}                                                                                                                                                                                                                         \\ \cline{2-3} 
                                          & Empirical study          & rocket system and its payload \cite{Cohen2015}
\\ \hline
\multirow{3}{*}{Building Automation}      & Case study               & energy-aware building \cite{Cheng2015}
\\ \cline{2-3} 
                                          & Example                  & Smart Building \cite{Alrimawi2018}, Room Thermostat \cite{Nagele2017}
\\ \cline{2-3} 
                        & Case Study               & Smart Power Grid \cite{Angelo2018}, self-driving miniature vehicle \cite{Mamun2013}, Water Monitoring \cite{Zhang2013b}
\\ \hline
\end{tabular}
\end{table}

\subsubsection{Conducted evaluations for the proposed solutions}

In this section, the results and findings for \textbf{“RQ5: Is there any evaluation presented in the study?”} and its sub-questions are presented. 

\vspace{2mm}

\begin{figure} [H]
\centering
	\includegraphics[scale=1.1]{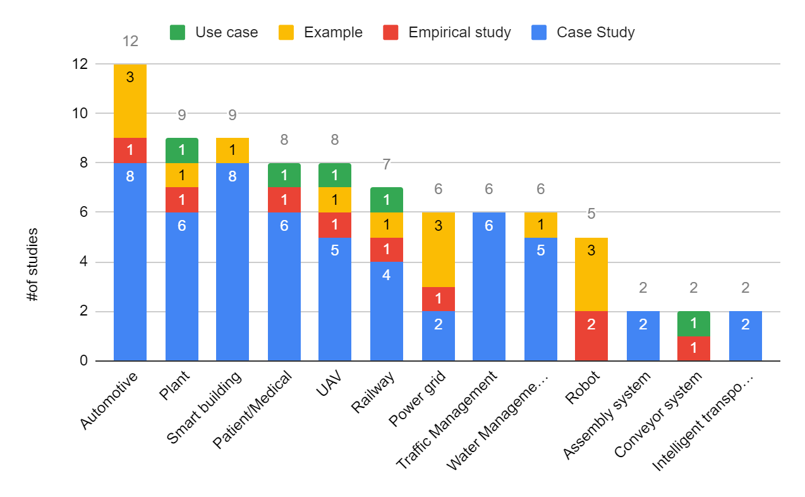}
	\caption{Reported CPS application domains targeted by the studies.}
	\label{fig:Evaluation_type}
\end{figure}

Out of the 140 studies considered in this study, 129 studies (92.1\%) evaluate their proposed solution. Among these studies, 70 of them (54.3\%) perform this evaluation by means of a case study, 31 of them (24\%) present an example, 17 studies (13.18\%) conduct an empirical study, and 11 studies (8.53\%) present a use case. This SLR groups the presented these evaluations performed by the primary studies into specific clusters to find out some patterns about them. 82 studies out of the 129 studies fit into the clusters shown in (Figure \ref{fig:Evaluation_type}), while the other 47 studies which do not fit in any of the clusters are grouped under "Other" cluster – not shown in the chart. The raw data related to this analysis is available online \footnote{\url{https://dx.doi.org/10.21227/zbkz-6461
}}.

\subsubsection{Addressed CPS challenges}

In this section, the challenges which the primary studies addressed are reported according to the \textbf{“RQ6: Does the study address any challenge(s)?”} in addition to its sub-questions.

\begin{figure} []
\centering
	\includegraphics[scale=1.3]{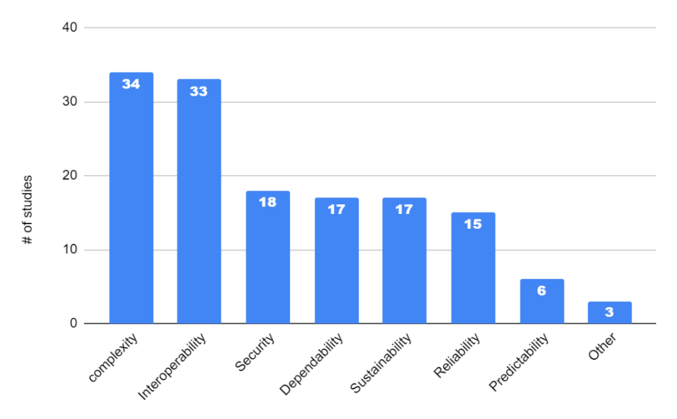}
	\caption{Categorized CPS challenges reported by the studies.}
	\label{fig:CPS_challenges}
\end{figure}

 107 studies out of 140 studies (76.43\%) reported the CPS challenge(s) they faced. Reported CPS challenges are shown in (Figure \ref{fig:CPS_challenges}). It is worth mentioning that several studies addressed more than one CPS challenge. In order to relate to the challenges presented by the studies to one another, the categorization of CPS challenges presented in \cite{gunes2014survey} was also followed in this study, moreover, we also added the complexity of CPS development and management as a new category in in addition to the existing categorization defined in \cite{gunes2014survey}. Reported CPS challenges and their corresponding studies are shown in (Table \ref{Tab:CPS_Challenges}).

\begin{itemize}
    \item \textbf{Complexity:} 34 studies (22.82\%) were classified under this category. It is reasonable that complexity was the most reported challenge, due to the nature of the CPS development process that requires complex engineering work. Some of the addressed complexity challenges include: complexity of design, timing behavior specification, execution complexity, co-simulation construction, architecture complexity, interaction complexity, semantics complexity, interdependency complexity, requirements complexity.
    \item \textbf{Interoperability:} also means Heterogeneity. 33 studies (22.15\%) classified under this category. To develop a CPS, the collaboration of different disciplines is a must. Thus, CPS combines different components (i.e. hardware, software, sensors, network, etc.), hence, managing and coordinating all these disciplines and operations are challenging. Scalability and composability are two important types of interoperability challenge. Scalability is challenging since the system ought to keep functioning adequately when new features are added. To provide the composability, CPS development should consider combining several components within a system and managing their interrelationships.
    \item \textbf{Security:} Reported by 18 studies (12.08\%). Studies in this category are concerned about the 3 security aspects of the CPS. Firstly, integrity needs to be supplied to protect the correctness of information from being manipulated or modified. An example for the CPS integrity problem would be compromising a sensor/actuator and injecting false data. Second aspect is confidentiality, that refers to allowing only authorized individuals to get access to the data. Third aspect is availability which means keeping the CPS components on service, e.g. preventing cyber- attacks (like denial of service) that may limit or block the availability of the system.
    \item \textbf{Dependability:} The ability of the system to keep functioning as required. 17 studies (11.41\%) were covered under this category. It encompasses aspects like safety, and maintainability. The system must be maintainable simply when a failure occurs.
    \item \textbf{Sustainability:}  17 studies (11.41\%) were covered under this category. It refers to challenges like adaptability, efficiency in using resources, reconfigurability, uncertainty, performance measurement, and optimization.
    \item \textbf{Reliability:} 15 studies (10.07\%) were covered under this category. Reliability means that the CPS should function correctly not only in closed and fixed environments but also in open and uncertain environments. Challenges to address are; fault tolerance, robustness, timing uncertainty etc.
    \item \textbf{Predictability:} 6 studies (4.03\%) were in this group. Predictability refers to the degree to which the system's behavior/functionality and outcomes are predictable and they satisfy the system requirements. For instance; predicting system’s stochastic behavior and accuracy, that is, the degree to which the system's measured outcomes need to be accurate. 
    \item \textbf{Other:} This category contains other challenges which are concurrency, latency and remote monitoring.
\end{itemize}

\begin{table}[H]
\caption{CPS challenges and their corresponding studies.}
\label{Tab:CPS_Challenges}
\begin{tabular}{|l|l|p{0.6\textwidth}|}
\hline
\textbf{CPS challenge} & \textbf{\# of studies} & \textbf{Relevant studies}                                                   \\ \hline
Complexity             & 34                     & \cite{Aziz2016,Peters2015,Nikiforova2017a,Nagele2017,Goncalves2016a,Gerking2015,Zhang2013b,Koutsoukos2012,Jeon2012,Buckl2010,Simko2013a,Sapienza2014,Lynch2016,Thramboulidis2018,Guo2017,Gritzner2018,DiNatale2016,Neema2014,Lavigne2018,Goncalves2017,Liu2011,VanDeMortel-Fronczak2014,Li2015,He2018,Seiger2015,Barbieri2016,Petnga2016,Kuesap2008,Koutsoumpas2015,Zhang2011,Reijnen2017,Pohlmann2014,Heinzemann2017slrcps,Murugesan} \\ \hline
interoperability       & 33                     & \cite{Aziz2016,Walch2017a,Nikiforova2017a,Merschak2018,Zhang2013a,Koutsoukos2012,Jeon2012,Simko2013a,Zhang2013c,Matinnejad2018,Gritzner2018,Kothari2018,Neema2014,Huang2018,Hartmann2015,Tuo2017,Tundis2017,Ciavotta2018,Dell2014,Barbieri2016,Goncalves2016,VanAcker2015,Brandenbourger2016,Thramboulidis,Barve2018,Sinha2016,Kacem2017,Alshareef2018,Morozov2018,Pohlmann2014,Pradhan2015,Zhao2017,Guan2018}                         \\ \hline
security               & 18                     & \cite{Alrimawi2018,Guo2018,Martins2015,Maksuti2017,Drago2013,Housh2018,Liu2017a,Cheh2017,Motii2017,Chen2011,Orojloo2017,Yan2009,Tan2013,Bakirtzis2018,Neema2018,Rocchetto2017,Rashid2018,Hahn2015}                                                                                                                                                                                                                                    \\ \hline
Dependability          & 17                     & \cite{Mamun2013,Zhang2013a,Buckl2010,Chen2017,Jiang2018,Bunting2016,Liu2011,Sanislav2017,An2011,Motii2017,VanDeMortel-Fronczak2014,Nannapaneni2016,Navet2016,Low2017,Hu2013,Fan2018,Murugesan}                                                                                                                                                                                                                                         \\ \hline
sustainability         & 17                     & \cite{Tariq2012,Garamvolgyi2018,Maksuti2017,Zhou2018,Zhang2013b,Silva2015,Bao2016,Thramboulidis2018,Ollinger2013,An2011,Bocciarelli2017,Tundis2017,Jarus2016,Feldmann2013,Xin2015,Plasse2017,Lezoche2018}                                                                                                                                                                                                                              \\ \hline
Reliability            & 15                     & \cite{Gerking2015,Zhou2018,Zhang2013a,Drago2013,Buckl2010,Silva2015,Bao2016,Chen2016,Weissnegger2016a,Bunting2016,Hartmann2015,Banerjee2014,Bocciarelli2017,Tan2013,Koutsoumpas2015}                                                                                                                                                                                                                                                   \\ \hline
Predictability         & 6                      & \cite{Angelo2018,Silva2015,Kothari2018,Banerjee2014,Cheng2014,Alshareef2018}                                                                                                                                                                                                                                                                                                                                                    \\ \hline
Other                  & 3                      & \cite{Latombe2015a,Mordecai2018,Irisarri2016}                                                                                                                                                                                                                                                                                                                                                                                        \\ \hline
\end{tabular}
\end{table}

Further, in this SLR, a correlation analysis of the CPS domains and its challenges is scrutinized so as to provide an understanding of the challenges addressed in each CPS application domain, see appendixes (Table \ref{App:domains_Vs_challenges}). Despite the fact that the correlation analysis cannot indicate the CPS domain wholly, for instance, one can see that in the smart manufacturing application domain, most research works converged on interoperability and sustainability challenges. Similarly, in the critical infrastructure application domain, most research works concentrated on security, sustainability, and interoperability challenges. However, it is interesting to note that the latency and the predictability challenges of both domains were not addressed by any of the examined papers.

\textbf{RQ6.2: Did the study reports challenges addressed while developing the approach/tool?} 

\vspace{2mm}

RQ6.2: Did the study reports challenges addressed while developing the approach/tool?
Only 15\%, that is, 21 studies out of the 140 studies reported about the limitations they faced. Studies reported limitations faced are; \cite{Alrimawi2018,Martins2015,Merschak2018,Koutsoukos2012,Silva2015,Zhang2014,Matinnejad2018,Jiang2018,Bunting2016,Hartmann2015,Whitsitt2014,Orojloo2017,Irisarri2016,Yan2009,Koutsoumpas2015,Alshareef2018,Rocchetto2017,Morozov2018,Pohlmann2014,Murugesan,B2018}.

\newpage
\section{Discussion} \label{discussion}

In this section, discussion of the findings achieved as the result of the applied research workflow of this SLR study is given along with its implications. Threats to the validity of the study is also discussed in this section.
At first, the quantitative analysis revealed that the number of published research papers in this field continues to increase year after year. USA affiliated researchers are the most interested researchers in this field (39 studies), followed by China (23 studies). Moreover, most preferred publication venues are conferences (64.75\%, 90 studies) by far. 

RQ 1.1 revealed that the metamodeling was the most used approach by the researchers. Model-based and DSL approaches follow the metamodeling. Also, modeling approaches were correlated with the authors' affiliation country in an attempt to determine which of the modeling approaches are mostly used in different countries. The study found out that, metamodeling and model-based approaches are mostly adopted by researchers affiliated to Europe, while DSL based approach was adopted mostly by USA affiliated researchers.

Although, in terms of the number of studies, metamodeling is the most adopted modeling approach, yet component-based approach is the most reported modeling approach in terms of the number of activities it is used for, which covered 9 activities namely: Adaptability, Analysis, Correctness, Development, Efficiency, Flexibility, Security, Simulation, and V\&V. 

As far as the purpose of modeling is concerned (RQ1.1.1), the most-reported purpose of modeling was development, that is, developing either DSL, metamodel, tool or automating the development process of a system. Other reported modeling purposes were Analysis (like safety analysis, performance analysis, requirement analysis, etc.), V\&V (DSML validation, metamodel verification, behavior verification, etc.), and security (threat modeling, attack modeling, analyzing cyber-attacks, etc.). 

Regarding model-driven engineering activity/phase addressed with RQ 1.2, the most considered MDE activity was system design. Researchers developed DSLs (15 studies), metamodels (22 studies) and tools (18 studies). Since the total number of DSL studies is quite low (10.71\%, only 15 studies out of 140) in a complex domain like CPS, this underpins the necessity for conducting more research to design DSLs to address different aspects of CPS development lifecycle. DSLs can provide a higher level of abstraction for complex systems such as CPS which may lead to increase the performance and to decrease the time and the cost of CPS development. Simulation was the second most reported MDE activity (40 studies, 16.81\%). Apart from 1 study that developed a tool \cite{Tundis2017} and 2 studies that developed metamodels \cite{Ciavotta2018,Merschak2018} for simulation purposes, the rest of the studies (37 studies) used existing simulation tools/languages. Therefore, one can observe that there is a research gap in developing domain-specific simulation tools/languages for CPS.

Furthermore, RQ 1.2 revealed that M2M transformation gains more attention in terms of the existing / developed tools, and languages in comparison with the other transformation types M2T and T2M. In addition, it is observed that languages like GenERTiCA and Xtext were used for the implementation of more than one transformation type. Also, it is worth mentioning that tools like UPP2SF and STU can be used as a complementary tool for M2M transformation. Other complementary languages for modeling CPS are Modelica and AADL, where Modelica is used for modeling the physical world and AADL for modeling the cyber part, and the transformation between these two languages do not require any third-party language or tool \cite{Zhang2013a,Zhang2014d}. V\&V was reported by 35 studies (14.71\%,). However, apart from 2 studies \cite{B2018,Weissnegger2016a} which developed a tool and one study that developed an ontology \cite{Lynch2016}, the rest of the studies used only existing tools. 

Results of R.Q 2.1 showed that the top 10 most used languages/tools in the field of applying MDE paradigm on CPS are; Simulink, AADL, UML, UPPAAL, SysML, MATLAB, Petri nets, Ptolemy II, Modelica and Papyrus. RQ 2.2 revealed that 59 studies out of 140 developed DSL, metamodel or tool. However, only 10 of the 59 studies reported the availability of these developed tools/languages (i.e. can be downloaded in their paper) according to the results for RQ 2.2.1. Therefore, this is also another alarming fact to consider by the CPS community that is interested in applying MDE for CPS; particularly, if they intend to have an impact on the industry. The results of RQ2.2.2 revealed that UML, EMF, and GME were the most used tools/environments which these 59 studies used while developing their DSLs, metamodels, and tools. However, GME was not present for the last 2 years (2017 and 2018) in any study. Findings also revealed that GME was mostly used by the researchers affiliated to the USA, while UML and EMF were mostly used by the researchers affiliated to Europe.

Regarding the addressed CPS components, most of the papers focused on the cyber and physical components of CPS (R.Q 3). There is limited work on the other components (sensor, network, actuator), Especially, the actuator is the component that received the least attention by the researchers on this topic. Results for R.Q 4 showed that 63 studies out of 140 (45\%) addressed a specific CPS domain. Smart manufacturing is the most addressed CPS domain by the researchers (26.98\%, 17 studies out of 63). The other domains followed are Critical Infrastructure, Health Care and Medicine, Air Transportation, Smart Environment, Building Automation, Safety-critical Systems, and self-adaptive CPS respectively.

For the evaluation method, R.Q 5 results revealed that the majority of the studies (54.26\%, 70 studies) presented case study(s) as the major evaluation method for their proposed solution. On the other hand, only 17 studies (13.18\%) presented an empirical evaluation for their proposed solution. That is, conducting empirical evaluations for this topic is an area which still requires much attention.

Results for R.Q 6 showed that a variety of CPS challenges were addressed. However, the most addressed challenges were complexity and interoperability. The much focus for these two challenges can be related to the heterogeneous nature of CPS. CPS combines different components and requires the interaction of different researchers from different backgrounds. Thus, it informs why researchers interested in this field should pay more attention to reducing the complexity and interoperability aspects of CPS. Other challenges addressed were; Security, Dependability, Sustainability, Reliability, Flexibility and Predictability.

Finally, the results of the study also showed several research gaps that the researchers in the community may take into account. First of all, the development of domain-specific languages, and domain-specific simulation and verification tools for CPS needs to be provided. Applying MDE on the different types of actuator components used inside CPS also needs further investigation since their role on building many CPS is vital and they are less addressed in the current MDE studies compared to the other CPS components. Moreover, conducting empirical evaluations in this field is missing which is critical on the assessment of the proposed modeling approaches especially on their usability for both the construction and the execution of CPS. 

\subsection{Threats to validity}

Threats to validity for this SLR study are classified according to categories proposed in \cite{wohlin2012experimentation}, and hence they include four types, namely construct, internal, external and conclusion validity threats.

\vspace{3mm}
Construct validity
\vspace{3mm}

It represents how the SLR study truly reflects the intent of the researchers, and what is asked by the research questions. To define the research questions, it is important to stress that the process proposed by \cite{de2007scientific} and \cite{siddaway2014systematic} and using guidelines defined by \cite{kitchenham2007guidelines} were followed in this study.

Furthermore, another aspect of construct validity is to assure that all relevant studies on the selected topic are found adequately. The possibility of missing primary studies is a common threat to the validity of any SLR. Both the terms MDE and CPS are well-established concepts, and thus, the terms are sufficiently good enough to be used as keywords. Therefore, to mitigate this risk, a good-enough search string through several iterations was formed, and adequate coverage of literature was achieved. General publication databases, which index most well-reputed publication venues, were extensively searched in this study as well. The list of publication venues shown in the appendixes (Table \ref{App:venues})  indicates that the coverage of the search is enough. Also, to improve the results, the forward snowballing sampling method was used, and it has proved to be effective.

\vspace{3mm}
Internal validity
\vspace{3mm}

This relates to the degree to which the design and the conduct of the SLR study are likely to prevent systematic errors. Internal validity is a prerequisite for external validity \cite{kitchenham2007guidelines}. Therefore, both qualitative and quantitative analysis were used to minimize threats. The use of a rigorous protocol and data extraction form mitigates this kind of threats to validity. Moreover, threats originating from personal bias or lack of understanding of the study were reduced by conducting data extraction phase iteratively. For this purpose, one researcher extracted data from the primary studies and answered quality and self-assessment questions. The other two researchers (expert in CPS and MDE) reviewed the extracted data from studies with low self-assessment rates below 50

\vspace{3mm}
External validity
\vspace{3mm}

According to \cite{wohlin2012experimentation}, external threats concern the generalizability of the SLR results, that is, the degree to which the primary studies is representative of the reviewed topic. In this study, the set of primary studies may not be representative of the entire set of existing studies on the topic, MDE for CPS. However, this threat was mitigated as follows; Firstly, the search strategy consisted of manual and automatic search, then followed by the forward snowballing. The forward snowballing enabled finding studies which were not captured by the search strings in the digital libraries. Secondly, the inclusion and exclusion criteria of the protocol created in this study support refining the set of primary studies which leads to include only studies which meet the topic. Only studies in English were included. Papers written in other languages concerning the same topic may exist. However, this threat is considered as having minimal effect.

\vspace{3mm}
Conclusion validity
\vspace{3mm}

All relevant primary studies cannot be identified \cite{kitchenham2007guidelines}. To alleviate this threat, the research protocol of this study was designed and validated carefully to minimize the risk of excluding relevant studies. Search strings were formed in a way that only a very small number of relevant studies could be missed, and a manageable quantity of irrelevant studies could be included. Besides the automatic search, a manual search and a forward snowballing were performed. The protocol was rigorously defined to be reusable by other researchers for reproducing the same study, i.e. the protocol is available online \footnote{\url{https://dx.doi.org/10.21227/zbkz-6461
}}.
\newpage
\section{Conclusion} \label{conclusion}

CPS have proven to offer tremendous opportunities in almost all areas of industry and society. Due to its inherent heterogeneity and complexity, developing and managing such systems is known to be a challenge for the developers. Thus, numerous researches were conducted and are still being conducted in this domain. 

The aim of this study is to identify the current features of the use of MDE for CPS. For this purpose, an SLR of the papers in the field, published between 2010 and 2018, was performed. The initial search retrieved 646 papers of which 140 were included in this study by following the defined selection strategy through a multi-stage process. A key feature of this SLR is that it is not restricted to a particular CPS domain. This broad scope in the search gives deeper insights into the state-of-the-art of using MDE for CPS. Furthermore, the study presented bibliometrics analysis to attain an understanding of the active researchers, publication trends per year, and publication venues in the area. Findings contribute new knowledge that can be used to improve CPS development using MDE.

The study points out that MDE for CPS is an active research area with an increasing number of publications over the years. Results showed that conferences account for the most frequently used publication venue. In terms of CPS domains, smart manufacturing is the most addressed CPS domain. Furthermore, the study showed the areas that have been covered, and approaches, techniques, languages, and tools that have been proposed. Regarding the CPS components, the effort was mostly put on the cyber and physical components, where the other components (networks, sensors, and actuators) did not get much attention. Study results revealed that solutions based on UML and Eclipse-based tools were mostly preferred.

Finally, the study also provided guidelines for assisting researchers to plan future work by pointing out research areas which need more attention. For instance, designing and modeling actuators used in CPS, code generation for an actuator, verification of actuator code, conducting an empirical evaluation, and developing domain-specific simulation tools requires further investigation.

\nocite{*}


\newpage
\addcontentsline{toc}{section}{References}
\bibliographystyle{Template/spmpsci}      
\bibliography{References}   

\begin{thebibliography}{100}
\providecommand{\url}[1]{{#1}}
\providecommand{\urlprefix}{URL }
\expandafter\ifx\csname urlstyle\endcsname\relax
  \providecommand{\doi}[1]{DOI~\discretionary{}{}{}#1}\else
  \providecommand{\doi}{DOI~\discretionary{}{}{}\begingroup
  \urlstyle{rm}\Url}\fi

\bibitem{agner2013brazilian}
Agner, L.T.W., Soares, I.W., Stadzisz, P.C., Sim{\~a}O, J.M.: A brazilian
  survey on uml and model-driven practices for embedded software development.
\newblock Journal of Systems and Software \textbf{86}(4), 997--1005 (2013)

\bibitem{Ali2018}
Ali, N., Hong, J.E.: Failure detection and prevention for cyber-physical
  systems using ontology-based knowledge base.
\newblock Computers \textbf{7}(4), 68 (2018)

\bibitem{de2007scientific}
de~Almeida~Biolchini, J.C., Mian, P.G., Natali, A.C.C., Conte, T.U., Travassos,
  G.H.: Scientific research ontology to support systematic review in software
  engineering.
\newblock Advanced Engineering Informatics \textbf{21}(2), 133--151 (2007)

\bibitem{Alrimawi2018}
Alrimawi, F., Pasquale, L., Mehta, D., Nuseibeh, B.: I've seen this before:
  sharing cyber-physical incident knowledge.
\newblock In: 2018 IEEE/ACM 1st International Workshop on Security Awareness
  from Design to Deployment (SEAD), pp. 33--40. IEEE (2018)

\bibitem{Alshareef2018}
Alshareef, A., Sarjoughian, H.S.: Model-driven time-accurate devs-based
  approaches for cps design.
\newblock In: Proceedings of the Model-driven Approaches for Simulation
  Engineering Symposium, p.~8. Society for Computer Simulation International
  (2018)

\bibitem{An2011}
An, K., Trewyn, A., Gokhale, A., Sastry, S.: Model-driven performance analysis
  of reconfigurable conveyor systems used in material handling applications.
\newblock In: 2011 IEEE/ACM Second International Conference on Cyber-Physical
  Systems, pp. 141--150. IEEE (2011)

\bibitem{Angelo2018}
Angelo, M., Napolitano, A., Caporuscio, M.: Cyphef: a model-driven engineering
  framework for self-adaptive cyber-physical systems.
\newblock In: Proceedings of the 40th International Conference on Software
  Engineering: Companion Proceeedings, pp. 101--104. ACM (2018)

\bibitem{Ataide2018}
Ata{\'\i}de, A., Barros, J.P., Brito, I.S., Gomes, L.: Towards automatic code
  generation for distributed cyber-physical systems: A first prototype for
  arduino boards.
\newblock In: 2017 22nd IEEE International Conference on Emerging Technologies
  and Factory Automation (ETFA), pp. 1--4. IEEE (2017)

\bibitem{Aziz2016}
Aziz, M.W., Rashid, M.: Domain specific modeling language for cyber physical
  systems.
\newblock In: 2016 International Conference on Information Systems Engineering
  (ICISE), pp. 29--33. IEEE (2016)

\bibitem{Bakirtzis2018}
Bakirtzis, G., Carter, B.T., Elks, C.R., Fleming, C.H.: A model-based approach
  to security analysis for cyber-physical systems.
\newblock In: 2018 Annual IEEE International Systems conference (SysCon), pp.
  1--8. IEEE (2018)

\bibitem{Banerjee2014}
Banerjee, A., Gupta, S.K.: Model based code generation for medical cyber
  physical systems.
\newblock In: Proceedings of the 1st Workshop on Mobile Medical Applications,
  pp. 22--27. ACM (2014)

\bibitem{Bao2016}
Bao, S., Porter, J., Gokhale, A.: Reasoning for cps education using surrogate
  simulation models.
\newblock In: 2016 IEEE 40th Annual Computer Software and Applications
  Conference (COMPSAC), vol.~1, pp. 764--773. IEEE (2016)

\bibitem{Barbieri2016}
Barbieri, G., Fantuzzi, C.: Design of cyber-physical systems: Definition and
  metamodel for reusable resources.
\newblock In: 2016 IEEE 21st International Conference on Emerging Technologies
  and Factory Automation (ETFA), pp. 1--9. IEEE (2016)

\bibitem{barivsicsystematic}
Bari{\v{s}}ic, A., Savic, D., Al-Ali, R., Ruchkin, I., Blouin, D., Cicchetti,
  A., Eslampanah, R., Nikiforova, O., Abshir, M., Challenger, M., et~al.:
  Systematic literature review on multi-paradigm modelling for cyber-physical
  systems

\bibitem{Barve2018}
Barve, Y., Neema, H., Rees, S., Sztipanovits, J.: Towards a design studio for
  collaborative modeling and co-simulations of mixed electrical energy systems.
\newblock In: 2018 IEEE International Science of Smart City Operations and
  Platforms Engineering in Partnership with Global City Teams Challenge
  (SCOPE-GCTC), pp. 24--29. IEEE (2018)

\bibitem{Blackburn2014}
Blackburn, M., Denno, P.: Virtual design and verification of cyber-physical
  systems: Industrial process plant design.
\newblock Procedia Computer Science \textbf{28}, 883--890 (2014)

\bibitem{Bocciarelli2017}
Bocciarelli, P., D'Ambrogio, A., Giglio, A., Paglia, E.: A bpmn extension for
  modeling cyber-physical-production-systems in the context of industry 4.0.
\newblock In: 2017 IEEE 14th International Conference on Networking, Sensing
  and Control (ICNSC), pp. 599--604. IEEE (2017)

\bibitem{Bougouffa2018}
Bougouffa, S., Me{\ss}zmer, K., Cha, S., Trunzer, E., Vogel-Heuser, B.:
  Industry 4.0 interface for dynamic reconfiguration of an open lab size
  automated production system to allow remote community experiments.
\newblock In: 2017 IEEE International Conference on Industrial Engineering and
  Engineering Management (IEEM), pp. 2058--2062. IEEE (2017)

\bibitem{brambilla2017model}
Brambilla, M., Cabot, J., Wimmer, M.: Model-driven software engineering in
  practice.
\newblock Synthesis Lectures on Software Engineering \textbf{3}(1), 1--207
  (2017)

\bibitem{Brandenbourger2016}
Brandenbourger, B., Vathoopan, M., Zoitl, A.: Behavior modeling of automation
  components using cross-domain interdependencies.
\newblock In: 2016 IEEE 21st International Conference on Emerging Technologies
  and Factory Automation (ETFA), pp. 1--4. IEEE (2016)

\bibitem{Brito2017}
Brito, I.S., Barros, J.P., Gomes, L.: From requirements to code (re2code)—a
  model-based approach for controller implementation.
\newblock In: 2016 IEEE 14th International Conference on Industrial Informatics
  (INDIN), pp. 1224--1230. IEEE (2016)

\bibitem{Broenink2016}
Broenink, J.F., Vos, P.J.D., Lu, Z., Bezemer, M.M.: A co-design approach for
  embedded control software of cyber-physical systems.
\newblock In: 2016 11th System of Systems Engineering Conference (SoSE), pp.
  1--5. IEEE (2016)

\bibitem{Buckl2010}
Buckl, C., Sojer, D., Knoll, A.: Ftos: Model-driven development of
  fault-tolerant automation systems.
\newblock In: 2010 IEEE 15th Conference on Emerging Technologies \& Factory
  Automation (ETFA 2010), pp. 1--8. IEEE (2010)

\bibitem{Bunting2016}
Bunting, M., Zeleke, Y., McKeever, K., Sprinkle, J.: A safe autonomous vehicle
  trajectory domain specific modeling language for non-expert development.
\newblock In: Proceedings of the International Workshop on Domain-Specific
  Modeling, pp. 42--48. ACM (2016)

\bibitem{casalaro2015model}
Casalaro, G.L., Cattivera, G.: Model-driven engineering for mobile robot
  systems: A systematic mapping study (2015)

\bibitem{abshir-dataset}
Challenger, M.A.M.G.K.M.: Dataset for: Systematic literature review on
  model-driven engineering for cyber-physical systems (2021).
\newblock \doi{10.21227/zbkz-6461}.
\newblock \urlprefix\url{https://dx.doi.org/10.21227/zbkz-6461}

\bibitem{Cheh2017}
Cheh, C., Keefe, K., Feddersen, B., Chen, B., Temple, W.G., Sanders, W.H.:
  Developing models for physical attacks in cyber-physical systems.
\newblock In: Proceedings of the 2017 Workshop on Cyber-Physical Systems
  Security and PrivaCy, pp. 49--55. ACM (2017)

\bibitem{Chen2017}
Chen, D., Lu, Z.: A model-based approach to dynamic self-assessment for
  automated performance and safety awareness of cyber-physical systems.
\newblock In: International Symposium on Model-Based Safety and Assessment, pp.
  227--240. Springer (2017)

\bibitem{Chen2015}
Chen, D., Maffei, A., Ferreirar, J., Akillioglu, H., Khabazzi, M.R., Zhang, X.:
  A virtual environment for the management and development of cyber-physical
  manufacturing systems.
\newblock IFAC-PapersOnLine \textbf{48}(7), 29--36 (2015)

\bibitem{Chen2016}
Chen, D., Panfilenko, D.V., Khabbazi, M.R., Sonntag, D.: A model-based approach
  to qualified process automation for anomaly detection and treatment.
\newblock In: 2016 IEEE 21st International Conference on Emerging Technologies
  and Factory Automation (ETFA), pp. 1--8. IEEE (2016)

\bibitem{Chen2018}
Chen, R., Liu, Y., Ye, X.: Ontology based behavior verification for complex
  systems.
\newblock In: ASME 2018 International Design Engineering Technical Conferences
  and Computers and Information in Engineering Conference. American Society of
  Mechanical Engineers Digital Collection (2018)

\bibitem{Chen2011}
Chen, T.M., Sanchez-Aarnoutse, J.C., Buford, J.: Petri net modeling of
  cyber-physical attacks on smart grid.
\newblock IEEE Transactions on Smart Grid \textbf{2}(4), 741--749 (2011)

\bibitem{Cheng2014}
Cheng, B., Du, D.: Towards a stochastic occurrence-based modeling approach for
  stochastic cpss.
\newblock In: 2014 Theoretical Aspects of Software Engineering Conference, pp.
  162--169. IEEE (2014)

\bibitem{Cheng2015}
Cheng, B., Wang, X., Liu, J., Du, D.: Modana: An integrated framework for
  modeling and analysis of energy-aware cpss.
\newblock In: 2015 IEEE 39th Annual Computer Software and Applications
  Conference, vol.~2, pp. 127--136. IEEE (2015)

\bibitem{Ciavotta2018}
Ciavotta, M., Bettoni, A., Izzo, G.: Interoperable meta model for
  simulation-in-the-loop.
\newblock In: 2018 IEEE Industrial Cyber-Physical Systems (ICPS), pp. 702--707.
  IEEE (2018)

\bibitem{Cicirelli2018}
Cicirelli, F., Fortino, G., Guerrieri, A., Mercuri, A., Spezzano, G., Vinci,
  A.: A metamodel framework for edge-based smart environments.
\newblock In: 2018 IEEE International Conference on Cloud Engineering (IC2E),
  pp. 286--291. IEEE (2018)

\bibitem{Cicirelli2016}
Cicirelli, F., Fortino, G., Guerrieri, A., Spezzano, G., Vinci, A.: A
  meta-model framework for the design and analysis of smart cyber-physical
  environments.
\newblock In: 2016 IEEE 20th International Conference on Computer Supported
  Cooperative Work in Design (CSCWD), pp. 687--692. IEEE (2016)

\bibitem{Cicirelli2017a}
Cicirelli, F., Fortino, G., Guerrieri, A., Spezzano, G., Vinci, A.:
  Metamodeling of smart environments: from design to implementation.
\newblock Advanced Engineering Informatics \textbf{33}, 274--284 (2017)

\bibitem{Cohen2015}
Cohen, R., Long, A.T.B., Jobredeaux, R., Feron, E.: A credible autocoding
  application within a rocket and its payload.
\newblock In: 2015 IEEE/AIAA 34th Digital Avionics Systems Conference (DASC),
  pp. 8C4--1. IEEE (2015)

\bibitem{Dell2014}
Dell, J., Greiner, T., Rosenstiel, W.: Model-based platform design and
  evaluation of cloud-based cyber-physical systems (ccps).
\newblock In: 2014 12th IEEE International Conference on Industrial Informatics
  (INDIN), pp. 376--381. IEEE (2014)

\bibitem{DiNatale2016}
Di~Natale, M., Morelli, M., Cremona, F.: Matching execution architecture models
  with functional models to analyze the time performance of cps systems.
\newblock In: 2015 International Conference on Complex Systems Engineering
  (ICCSE), pp. 1--6. IEEE (2015)

\bibitem{Kuesap2008}
Ding, J., Atif, Y., Andler, S.F., Lindstr{\"o}m, B., Jeusfeld, M.: Cps-based
  threat modeling for critical infrastructure protection.
\newblock ACM SIGMETRICS Performance Evaluation Review \textbf{45}(2), 129--132
  (2017)

\bibitem{Drago2013}
Drago, A., Marrone, S., Mazzocca, N., Tedesco, A., Vittorini, V.: Model-driven
  estimation of distributed vulnerability in complex railway networks.
\newblock In: 2013 IEEE 10th International Conference on Ubiquitous
  Intelligence and Computing and 2013 IEEE 10th International Conference on
  Autonomic and Trusted Computing, pp. 380--387. IEEE (2013)

\bibitem{Fan2018}
Fan, C.F., Chan, C.C., Yu, H.Y., Yih, S.: A simulation platform for
  human-machine interaction safety analysis of cyber-physical systems.
\newblock International journal of industrial ergonomics \textbf{68}, 89--100
  (2018)

\bibitem{Feldmann2013}
Feldmann, S., R{\"o}sch, S., Sch{\"u}tz, D., Vogel-Heuser, B.: Model-driven
  engineering and semantic technologies for the design of cyber-physical
  systems.
\newblock IFAC Proceedings Volumes \textbf{46}(7), 210--215 (2013)

\bibitem{france2007model}
France, R., Rumpe, B.: Model-driven development of complex software: A research
  roadmap.
\newblock In: 2007 Future of Software Engineering, pp. 37--54. IEEE Computer
  Society (2007)

\bibitem{Garamvolgyi2018}
Garamv{\"o}lgyi, P., Kocsis, I., Gehl, B., Klenik, A.: Towards model-driven
  engineering of smart contracts for cyber-physical systems.
\newblock In: 2018 48th Annual IEEE/IFIP International Conference on Dependable
  Systems and Networks Workshops (DSN-W), pp. 134--139. IEEE (2018)

\bibitem{garousi2013systematic}
Garousi, V., Mesbah, A., Betin-Can, A., Mirshokraie, S.: A systematic mapping
  study of web application testing.
\newblock Information and Software Technology \textbf{55}(8), 1374--1396 (2013)

\bibitem{Gerking2015}
Gerking, C., Schafer, W., Dziwok, S., Heinzemann, C.: Domain-specific model
  checking for cyber-physical systems.
\newblock In: MoDeVVa\@MoDELS, pp. 18--27 (2015)

\bibitem{Goncalves2016a}
Gon{\c{c}}alves, F.S., Becker, L.B.: Model driven engineering approach to
  design sensing and actuation subsystems.
\newblock In: 2016 IEEE 21st International Conference on Emerging Technologies
  and Factory Automation (ETFA), pp. 1--8. IEEE (2016)

\bibitem{Goncalves2017}
Gon{\c{c}}alves, F.S., Pereira, D., Tovar, E., Becker, L.B.: Formal
  verification of aadl models using uppaal.
\newblock In: 2017 VII Brazilian Symposium on Computing Systems Engineering
  (SBESC), pp. 117--124. IEEE (2017)

\bibitem{Goncalves2016}
Gon{\c{c}}alves, F.S., Raffo, G.V., Becker, L.B.: Managing cps complexity:
  Design method for unmanned aerial vehicles.
\newblock IFAC-PapersOnLine \textbf{49}(32), 141--146 (2016)

\bibitem{Graja}
Graja, I., Kallel, S., Guermouche, N., Kacem, A.H.: Bpmn4cps: A bpmn extension
  for modeling cyber-physical systems.
\newblock In: 2016 IEEE 25th International Conference on Enabling Technologies:
  Infrastructure for Collaborative Enterprises (WETICE), pp. 152--157. IEEE
  (2016)

\bibitem{B2018}
Graja, I., Kallel, S., Guermouche, N., Kacem, A.H.: Verification of the
  consistency of time-aware cyber-physical processes.
\newblock In: International Conference on Service-Oriented Computing, pp.
  67--79. Springer (2017)

\bibitem{Gritzner2018}
Gritzner, D., Greenyer, J.: Generating correct, compact, and efficient plc code
  from scenario-based assume-guarantee specifications.
\newblock Procedia Manufacturing 24 (2018) \textbf{24}, 153--158 (2018)

\bibitem{Guan2018}
Guan, C., Ao, Y., Du, D., Mallet, F.: xshs: An executable domain-specific
  modeling language for modeling stochastic and hybrid behaviors of
  cyber-physical systems.
\newblock In: 2018 25th Asia-Pacific Software Engineering Conference (APSEC),
  pp. 683--687. IEEE (2018)

\bibitem{gunes2014survey}
Gunes, V., Peter, S., Givargis, T., Vahid, F.: A survey on concepts,
  applications, and challenges in cyber-physical systems.
\newblock KSII Transactions on Internet \& Information Systems \textbf{8}(12)
  (2014)

\bibitem{Guo2017}
Guo, C., Fu, Z., Ren, S., Jiang, Y., Rahmaniheris, M., Sha, L.: Pattern-based
  statechart modeling approach for medical best practice guidelines-a case
  study.
\newblock In: 2017 IEEE 30th International Symposium on Computer-Based Medical
  Systems (CBMS), pp. 117--122. IEEE (2017)

\bibitem{Guo2018}
Guo, C., Fu, Z., Zhang, Z., Ren, S., Sha, L.: Model and integrate medical
  resource available times and relationships in verifiably correct executable
  medical best practice guideline models.
\newblock In: Proceedings of the 9th ACM/IEEE International Conference on
  Cyber-Physical Systems, pp. 253--262. IEEE Press (2018)

\bibitem{Hahn2015}
Hahn, A., Thomas, R.K., Lozano, I., Cardenas, A.: A multi-layered and
  kill-chain based security analysis framework for cyber-physical systems.
\newblock International Journal of Critical Infrastructure Protection
  \textbf{11}, 39--50 (2015)

\bibitem{Hartmann2015}
Hartmann, T., Moawad, A., Fouquet, F., Nain, G., Klein, J., Le~Traon, Y.:
  Stream my models: Reactive peer-to-peer distributed models@ run. time.
\newblock In: 2015 ACM/IEEE 18th International Conference on Model Driven
  Engineering Languages and Systems (MODELS), pp. 80--89. IEEE (2015)

\bibitem{He2018}
He, X.: Modeling and analyzing cyber physical systems using high level petri
  nets.
\newblock In: 2018 IEEE International Conference on Software Quality,
  Reliability and Security Companion (QRS-C), pp. 469--476. IEEE (2018)

\bibitem{Heinzemann2017slrcps}
Heinzemann, C., Becker, S., Volk, A.: Transactional execution of hierarchical
  reconfigurations in cyber-physical systems.
\newblock Software \& Systems Modeling \textbf{18}(1), 157--189 (2019)

\bibitem{Housh2018}
Housh, M., Ohar, Z.: Model-based approach for cyber-physical attack detection
  in water distribution systems.
\newblock Water research \textbf{139}, 132--143 (2018)

\bibitem{Hu2013}
Hu, X., Liu, S., Chen, G., Jiang, C.: Dependability modelling and evaluation of
  cyber-physical systems: A model-driven perspective.
\newblock In: 1st International Workshop on Cloud Computing and Information
  Security. Atlantis Press (2013)

\bibitem{Huang2018}
Huang, P., Jiang, K., Guan, C., Du, D.: Towards modeling cyber-physical systems
  with sysml/marte/pccsl.
\newblock In: 2018 IEEE 42nd Annual Computer Software and Applications
  Conference (COMPSAC), vol.~1, pp. 264--269. IEEE (2018)

\bibitem{Irisarri2016}
Irisarri, E., Garc{\'\i}a, M.V., P{\'e}rez, F., Est{\'e}vez, E., Marcos, M.: A
  model-based approach for process monitoring in oil production industry.
\newblock In: 2016 IEEE 21st International Conference on Emerging Technologies
  and Factory Automation (ETFA), pp. 1--4. IEEE (2016)

\bibitem{Jarus2016}
Jarus, N., Sarvestani, S.S., Hurson, A.R.: Models, metamodels, and model
  transformation for cyber-physical systems.
\newblock In: 2016 Seventh International Green and Sustainable Computing
  Conference (IGSC), pp. 1--8. IEEE (2016)

\bibitem{Jeon2012}
Jeon, J., Chun, I., Kim, W.: Metamodel-based cps modeling tool.
\newblock In: Embedded and Multimedia Computing Technology and Service, pp.
  285--291. Springer (2012)

\bibitem{Jiang2018}
Jiang, Y., Song, H., Yang, Y., Liu, H., Gu, M., Guan, Y., Sun, J., Sha, L.:
  Dependable model-driven development of cps: From stateflow simulation to
  verified implementation.
\newblock ACM Transactions on Cyber-Physical Systems \textbf{3}(1), 12 (2019)

\bibitem{Kacem2017}
Kacem, M.A.H., Simeu-Abazi, Z., Gascard, E., Lemasson, G., Maisonnasse, J.:
  Application of a modeling approach on a cyber-physical system" robair".
\newblock IFAC-PapersOnLine \textbf{50}(1), 14,230--14,235 (2017)

\bibitem{kitchenham2007guidelines}
Kitchenham, B., Charters, S.: Guidelines for performing systematic literature
  reviews in software engineering  (2007)

\bibitem{Kothari2018}
Kothari, S., Awadhutkar, P., Tamrawi, A., Mathews, J.: Modeling lessons from
  verifying large software systems for safety and security.
\newblock In: Proceedings of the 2017 Winter Simulation Conference, p. 109.
  IEEE Press (2017)

\bibitem{Koutsoukos2012}
Koutsoukos, X., Kottenstette, N., Hall, J., Eyisi, E., Leblanc, H., Porter, J.,
  Sztipanovits, J.: A passivity approach for model-based compositional design
  of networked control systems.
\newblock ACM Transactions on Embedded Computing Systems (TECS) \textbf{11}(4),
  75 (2012)

\bibitem{Koutsoumpas2015}
Koutsoumpas, V.: A model-based approach for the specification of a virtual
  power plant operating in open context.
\newblock In: Proceedings of the First International Workshop on Software
  Engineering for Smart Cyber-Physical Systems, pp. 26--32. IEEE Press (2015)

\bibitem{SampathKumar2015}
Kumar, V.R.S., Shanmugavel, M., Ganapathy, V., Shirinzadeh, B.: Unified
  meta-modeling framework using bond graph grammars for conceptual modeling.
\newblock Robotics and Autonomous Systems \textbf{72}, 114--130 (2015)

\bibitem{Latombe2015a}
Latombe, F., Cr{\'e}gut, X., Combemale, B., Deantoni, J., Pantel, M.: Weaving
  concurrency in executable domain-specific modeling languages.
\newblock In: Proceedings of the 2015 ACM SIGPLAN International Conference on
  Software Language Engineering, pp. 125--136. ACM (2015)

\bibitem{Lavigne2018}
Lavigne, E., Guillou, G., Babau, J.P.: Avs, a model-based racing sailboat
  simulator: application to wind integration.
\newblock IFAC-PapersOnLine \textbf{51}(10), 88--94 (2018)

\bibitem{lee2008cyber}
Lee, E.A.: Cyber physical systems: Design challenges.
\newblock In: 2008 11th IEEE International Symposium on Object and
  Component-Oriented Real-Time Distributed Computing (ISORC), pp. 363--369.
  IEEE (2008)

\bibitem{lee2015past}
Lee, E.A.: The past, present and future of cyber-physical systems: A focus on
  models.
\newblock Sensors \textbf{15}(3), 4837--4869 (2015)

\bibitem{Lezoche2018}
Lezoche, M., Panetto, H.: Cyber-physical systems, a new formal paradigm to
  model redundancy and resiliency.
\newblock Enterprise Information Systems pp. 1--22 (2018)

\bibitem{Li2015}
Li, S., Li, D., Li, F., Zhou, N.: Cpsicgf: A code generation framework for cps
  integration modeling.
\newblock Microprocessors and Microsystems \textbf{39}(8), 1234--1244 (2015)

\bibitem{liebel2018model}
Liebel, G., Marko, N., Tichy, M., Leitner, A., Hansson, J.: Model-based
  engineering in the embedded systems domain: an industrial survey on the
  state-of-practice.
\newblock Software \& Systems Modeling \textbf{17}(1), 91--113 (2018)

\bibitem{Liu2011}
Liu, J., Zhang, L.: Aspect-oriented mda development method for non-functional
  properties of cyber physical systems.
\newblock In: 2011 Second International Conference on Networking and
  Distributed Computing, pp. 149--153. IEEE (2011)

\bibitem{Liu2017a}
Liu, X., Zhang, J., Zhu, P.: Modeling cyber-physical attacks based on
  probabilistic colored petri nets and mixed-strategy game theory.
\newblock International Journal of Critical Infrastructure Protection
  \textbf{16}, 13--25 (2017)

\bibitem{Lora2017}
Lora, M., Fraccaroli, E., Fummi, F.: Virtual prototyping of smart systems
  through automatic abstraction and mixed-signal scheduling.
\newblock In: 2017 22nd Asia and South Pacific Design Automation Conference
  (ASP-DAC), pp. 232--237. IEEE (2017)

\bibitem{Low2017}
Low, T.M., Franchetti, F.: High assurance code generation for cyber-physical
  systems.
\newblock In: 2017 IEEE 18th International Symposium on High Assurance Systems
  Engineering (HASE), pp. 104--111. IEEE (2017)

\bibitem{Lynch2016}
Lynch, K., Ramsey, R., Ball, G., Schmit, M., Collins, K.: Ontology-driven
  metamodel validation in cyber-physical systems.
\newblock In: Information Technology: New Generations, pp. 1255--1258. Springer
  (2016)

\bibitem{Maksuti2017}
Maksuti, S., Bicaku, A., Tauber, M., Palkovits-Rauter, S., Haas, S., Delsing,
  J.: Towards flexible and secure end-to-end communication in industry 4.0.
\newblock In: 2017 IEEE 15th International Conference on Industrial Informatics
  (INDIN), pp. 883--888. IEEE (2017)

\bibitem{Mamun2013}
Mamun, M.A.A., Berger, C., Hansson, J.: Mde-based sensor management and
  verification for a self-driving miniature vehicle.
\newblock In: Proceedings of the 2013 ACM workshop on Domain-specific modeling,
  pp. 1--6. ACM (2013)

\bibitem{Martins2015}
Martins, G., Bhatia, S., Koutsoukos, X., Stouffer, K., Tang, C., Candell, R.:
  Towards a systematic threat modeling approach for cyber-physical systems.
\newblock In: 2015 Resilience Week (RWS), pp. 1--6. IEEE (2015)

\bibitem{Matinnejad2018}
Matinnejad, R., Nejati, S., Briand, L., Bruckmann, T.: Test generation and test
  prioritization for simulink models with dynamic behavior.
\newblock IEEE Transactions on Software Engineering  (2018)

\bibitem{Merschak2018}
Merschak, S., Hehenberger, P., Witters, M., Gadeyne, K.: A hierarchical
  meta-model for the design of cyber-physical production systems.
\newblock In: 2018 19th International Conference on Research and Education in
  Mechatronics (REM), pp. 36--41. IEEE (2018)

\bibitem{Mezhuyev2013}
Mezhuyev, V., Samet, R.: Geometrical meta-metamodel for cyber-physical
  modelling.
\newblock In: 2013 International Conference on Cyberworlds, pp. 89--93. IEEE
  (2013)

\bibitem{Mordecai2018}
Mordecai, Y.: Conceptual modeling of cyber-physical gaps in air traffic
  control.
\newblock Procedia Computer Science \textbf{140}, 21--28 (2018)

\bibitem{Morozov2018}
Morozov, D., Lezoche, M., Panetto, H.: Multi-paradigm modelling of
  cyber-physical systems.
\newblock IFAC-PapersOnLine \textbf{51}(11), 1385--1390 (2018)

\bibitem{Motii2017}
Motii, A., Hamid, B., Lanusse, A., Bruel, J.M.: Guiding the selection of
  security patterns for real-time systems.
\newblock In: 2016 21st International Conference on Engineering of Complex
  Computer Systems (ICECCS), pp. 155--164. IEEE (2016)

\bibitem{Murugesan}
Murugesan, A., Heimdahl, M.P., Whalen, M.W., Rayadurgam, S., Komp, J., Duan,
  L., Kim, B.G., Sokolsky, O., Lee, I.: From requirements to code: Model based
  development of a medical cyber physical system.
\newblock In: Software Engineering in Health Care, pp. 96--112. Springer (2014)

\bibitem{Nagele2017}
N{\"a}gele, T., Hooman, J.: Rapid construction of co-simulations of
  cyber-physical systems in hla using a dsl.
\newblock In: 2017 43rd Euromicro Conference on Software Engineering and
  Advanced Applications (SEAA), pp. 247--251. IEEE (2017)

\bibitem{Nannapaneni2016}
Nannapaneni, S., Mahadevan, S., Pradhan, S., Dubey, A.: Towards
  reliability-based decision making in cyber-physical systems.
\newblock In: 2016 IEEE International Conference on Smart Computing
  (SMARTCOMP), pp. 1--6. IEEE (2016)

\bibitem{Navet2016}
Navet, N., Fejoz, L.: Cpal: High-level abstractions for safe embedded systems.
\newblock In: Proceedings of the International Workshop on Domain-Specific
  Modeling, pp. 35--41. ACM (2016)

\bibitem{Neema2018}
Neema, H., Potteiger, B., Koutsoukos, X., Karsai, G., Volgyesi, P.,
  Sztipanovits, J.: Integrated simulation testbed for security and resilience
  of cps.
\newblock In: Proceedings of the 33rd Annual ACM Symposium on Applied
  Computing, pp. 368--374. ACM (2018)

\bibitem{Neema2014}
Neema, S., Simko, G., Levendovszky, T., Porter, J., Agrawal, A., Sztipanovits,
  J.: Formalization of software models for cyber-physical systems.
\newblock In: Proceedings of the 2nd FME Workshop on Formal Methods in Software
  Engineering, pp. 45--51. ACM (2014)

\bibitem{Nikiforova2017a}
Nikiforova, O., El~Marzouki, N., Gusarovs, K., Vangheluwe, H., Bures, T.,
  Al-Ali, R., Iacono, M., Esquivel, P.O., Leon, F.: The two-hemisphere
  modelling approach to the composition of cyber-physical systems.
\newblock In: 12th International Conference on Software Technologies, ICSOFT
  2017, pp. 286--293. SciTePress (2017)

\bibitem{Ollinger2013}
Ollinger, L., Wehrmeister, M.A., Pereira, C.E., Z{\"u}hlke, D.: An integrated
  concept for the model-driven engineering of distributed automation
  architectures on embedded systems.
\newblock IFAC Proceedings Volumes \textbf{46}(7), 222--227 (2013)

\bibitem{Orojloo2017}
Orojloo, H., Azgomi, M.A.: A game-theoretic approach to model and quantify the
  security of cyber-physical systems.
\newblock Computers in Industry \textbf{88}, 44--57 (2017)

\bibitem{Pagliari2018}
Pagliari, L., Mirandola, R., Trubiani, C.: Multi-modeling approach to
  performance engineering of cyber-physical systems design.
\newblock In: 2017 22nd International Conference on Engineering of Complex
  Computer Systems (ICECCS), pp. 142--145. IEEE (2017)

\bibitem{Pajic2012}
Pajic, M., Jiang, Z., Lee, I., Sokolsky, O., Mangharam, R.: From verification
  to implementation: A model translation tool and a pacemaker case study.
\newblock In: 2012 IEEE 18th Real Time and Embedded Technology and Applications
  Symposium, pp. 173--184. IEEE (2012)

\bibitem{Palachi2013}
Palachi, E., Cohen, C., Takashi, S.: Simulation of cyber physical models using
  sysml and numerical solvers.
\newblock In: 2013 IEEE International Systems Conference (SysCon), pp.
  671--675. IEEE (2013)

\bibitem{Parveen2018}
Parveen, R., Pradhan, S., Goveas, N.: Design and feasibility study of health
  related devices using cots components.
\newblock In: 2018 IEEE Industrial Cyber-Physical Systems (ICPS), pp. 529--533.
  IEEE (2018)

\bibitem{Passarini2014}
Passarini, R.F., Becker, L.B., Farines, J.M.: The assisted transformation of
  models: Supporting cyber-physical systems design by extracting architectural
  aspects and operating modes from simulink functional models.
\newblock In: 2013 III Brazilian Symposium on Computing Systems Engineering,
  pp. 47--52. IEEE (2013)

\bibitem{Pereira2016}
Pereira, F., Gomes, L.: The iopt-flow framework pairing petri nets and
  data-flows for embedded controller development.
\newblock In: IECON 2016-42nd Annual Conference of the IEEE Industrial
  Electronics Society, pp. 4832--4837. IEEE (2016)

\bibitem{Peters2015}
Peters, J., Wille, R., Przigoda, N., K{\"u}hne, U., Drechsler, R.: A generic
  representation of ccsl time constraints for uml/marte models.
\newblock In: Proceedings of the 52nd Annual Design Automation Conference, p.
  122. ACM (2015)

\bibitem{petersen2008systematic}
Petersen, K., Feldt, R., Mujtaba, S., Mattsson, M.: Systematic mapping studies
  in software engineering.
\newblock In: Ease, vol.~8, pp. 68--77 (2008)

\bibitem{Petnga2016}
Petnga, L., Austin, M.: An ontological framework for knowledge modeling and
  decision support in cyber-physical systems.
\newblock Advanced Engineering Informatics \textbf{30}(1), 77--94 (2016)

\bibitem{pinsonneault1993survey}
Pinsonneault, A., Kraemer, K.: Survey research methodology in management
  information systems: an assessment.
\newblock Journal of management information systems \textbf{10}(2), 75--105
  (1993)

\bibitem{Plasse2017}
Plasse, J., Noble, J., Myers, K.: An adaptive modeling framework for bivariate
  data streams with applications to change detection in cyber-physical systems.
\newblock In: 2017 IEEE International Conference on Data Mining Workshops
  (ICDMW), pp. 1074--1081. IEEE (2017)

\bibitem{Pohlmann2014}
Pohlmann, U., Trsek, H., D{\"u}rkop, L., Dziwok, S., Oesters{\"o}tebier, F.:
  Application of an intelligent network architecture on a cooperative
  cyber-physical system: An experience report.
\newblock In: Proceedings of the 2014 IEEE Emerging Technology and Factory
  Automation (ETFA), pp. 1--6. IEEE (2014)

\bibitem{Pradhan2015}
Pradhan, S.M., Dubey, A., Gokhale, A., Lehofer, M.: Chariot: A domain specific
  language for extensible cyber-physical systems.
\newblock In: Proceedings of the workshop on domain-specific modeling, pp.
  9--16. ACM (2015)

\bibitem{Qian2013}
Qian, Z., Yu, H.: A taopn approach to modeling and scheduling cyber-physical
  systems.
\newblock In: 2013 International Conference on Information Science and
  Applications (ICISA), pp. 1--7. IEEE (2013)

\bibitem{queiroz2014development}
Queiroz, P.G.G., Braga, R.T.V.: Development of critical embedded systems using
  model-driven and product lines techniques: A systematic review.
\newblock In: 2014 Eighth Brazilian Symposium on Software Components,
  Architectures and Reuse, pp. 74--83. IEEE (2014)

\bibitem{rashid2015toward}
Rashid, M., Anwar, M.W., Khan, A.M.: Toward the tools selection in model based
  system engineering for embedded systems—a systematic literature review.
\newblock Journal of Systems and Software \textbf{106}, 150--163 (2015)

\bibitem{Rashid2018}
Rashid, N., Wan, J., Quiros, G., Canedo, A., Al~Faruque, M.A.: Modeling and
  simulation of cyberattacks for resilient cyber-physical systems.
\newblock In: 2017 13th IEEE Conference on Automation Science and Engineering
  (CASE), pp. 988--993. IEEE (2017)

\bibitem{Reijnen2017}
Reijnen, F., Goorden, M., van~de Mortel-Fronczak, J., Rooda, J.: Supervisory
  control synthesis for a waterway lock.
\newblock In: 2017 IEEE Conference on Control Technology and Applications
  (CCTA), pp. 1562--1563. IEEE (2017)

\bibitem{Rocchetto2017}
Rocchetto, M., Tippenhauer, N.O.: Towards formal security analysis of
  industrial control systems.
\newblock In: Proceedings of the 2017 ACM on Asia Conference on Computer and
  Communications Security, pp. 114--126. ACM (2017)

\bibitem{sanislav2017dependability}
Sanislav, T., Mois, G.: A dependability analysis model in the context of
  cyber-physical systems.
\newblock In: 2017 18th International Carpathian Control Conference (ICCC), pp.
  146--150. IEEE (2017)

\bibitem{Sanislav2017}
Sanislav, T., Mois, G.: A dependability analysis model in the context of
  cyber-physical systems.
\newblock In: 2017 18th International Carpathian Control Conference (ICCC), pp.
  146--150. IEEE (2017)

\bibitem{Sapienza2014}
Sapienza, G., Crnkovic, I., Potena, P.: Architectural decisions for hw/sw
  partitioning based on multiple extra-functional properties.
\newblock In: 2014 IEEE/IFIP Conference on Software Architecture, pp. 175--184.
  IEEE (2014)

\bibitem{Seiger2015}
Seiger, R., Keller, C., Niebling, F., Schlegel, T.: Modelling complex and
  flexible processes for smart cyber-physical environments.
\newblock Journal of Computational Science \textbf{10}, 137--148 (2015)

\bibitem{siddaway2014systematic}
Siddaway, A.: What is a systematic literature review and how do i do one.
\newblock University of Stirling (I), 1 (2014)

\bibitem{Silva2015}
Silva, L., Almeida, H., Perkusich, A., Perkusich, M.: A model-based approach to
  support validation of medical cyber-physical systems.
\newblock Sensors \textbf{15}(11), 27,625--27,670 (2015)

\bibitem{Silva2014}
Silva, L.C., Perkusich, M., Bublitz, F.M., Almeida, H.O., Perkusich, A.: A
  model-based architecture for testing medical cyber-physical systems.
\newblock In: Proceedings of the 29th Annual ACM Symposium on Applied
  Computing, pp. 25--30. ACM (2014)

\bibitem{Simko2013a}
Simko, G., Lindecker, D., Levendovszky, T., Jackson, E., Neema, S.,
  Sztipanovits, J.: A framework for unambiguous and extensible specification of
  dsmls for cyber-physical systems.
\newblock In: 2013 20th IEEE International Conference and Workshops on
  Engineering of Computer Based Systems (ECBS), pp. 30--39. IEEE (2013)

\bibitem{Sinha2016}
Sinha, R., Pang, C., Mart{\'\i}nez, G.S., Kuronen, J., Vyatkin, V.:
  Requirements-aided automatic test case generation for industrial
  cyber-physical systems.
\newblock In: 2015 20th International Conference on Engineering of Complex
  Computer Systems (ICECCS), pp. 198--201. IEEE (2015)

\bibitem{skoglund2009reference}
Skoglund, M., Runeson, P.: Reference-based search strategies in systematic
  reviews.
\newblock In: EASE (2009)

\bibitem{Son2012a}
Son, H.S., Kim, W.Y., Kim, R.Y., Min, H.G.: Metamodel design for model
  transformation from simulink to ecml in cyber physical systems.
\newblock In: Computer Applications for Graphics, Grid Computing, and
  Industrial Environment, pp. 56--60. Springer (2012)

\bibitem{Tan2013}
Tan, F., Wang, Y., Wang, Q., Bu, L., Zheng, R., Suri, N.: Guaranteeing
  proper-temporal-embedding safety rules in wireless cps: A hybrid formal
  modeling approach.
\newblock In: 2013 43rd Annual IEEE/IFIP International Conference on Dependable
  Systems and Networks (DSN), pp. 1--12. IEEE (2013)

\bibitem{Tariq2014}
Tariq, M.U., Florence, J., Wolf, M.: Design specification of cyber-physical
  systems: Towards a domain-specific modeling language based on simulink,
  eclipse modeling framework, and giotto.
\newblock In: ACESMB@ MoDELS, pp. 6--15 (2014)

\bibitem{Tariq2012}
Tariq, M.U., Nasir, H.A., Muhammad, A., Wolf, M.: Model-driven performance
  analysis of large scale irrigation networks.
\newblock In: 2012 IEEE/ACM Third International Conference on Cyber-Physical
  Systems, pp. 151--160. IEEE (2012)

\bibitem{Thramboulidis2018}
Thramboulidis, K., Kontou, I., Vachtsevanou, D.C.: Towards an iot-based
  framework for evolvable assembly systems.
\newblock IFAC-PapersOnLine \textbf{51}(11), 182--187 (2018)

\bibitem{Thramboulidis}
Thramboulidis, K., Vachtsevanou, D.C., Solanos, A.: Cyber-physical
  microservices: An iot-based framework for manufacturing systems.
\newblock In: 2018 IEEE Industrial Cyber-Physical Systems (ICPS), pp. 232--239.
  IEEE (2018)

\bibitem{Tundis2017}
Tundis, A., Egert, R., M{\"u}hlh{\"a}user, M.: Applying a properties modeling
  approach for monitoring smart grids.
\newblock In: 2017 IEEE 14th International Conference on Networking, Sensing
  and Control (ICNSC), pp. 714--719. IEEE (2017)

\bibitem{Tuo2017}
Tuo, M., Zhou, X., Yang, G., Fu, N.: An approach for safety analysis of
  cyber-physical system based on model transformation.
\newblock In: 2016 IEEE International Conference on Internet of Things
  (iThings) and IEEE Green Computing and Communications (GreenCom) and IEEE
  Cyber, Physical and Social Computing (CPSCom) and IEEE Smart Data
  (SmartData), pp. 636--639. IEEE (2016)

\bibitem{VanAcker2015}
Van~Acker, B., Denil, J., Vangheluwe, H., De~Meulenaere, P.: Managing
  heterogeneity in model-based systems engineering of cyber-physical systems.
\newblock In: 2015 10th International Conference on P2P, Parallel, Grid, Cloud
  and Internet Computing (3PGCIC), pp. 617--622. IEEE (2015)

\bibitem{VanDeMortel-Fronczak2014}
Van Mortel-Fronczak, J.M., van~der Heijden, M.H., Huisman, R.G., Reniers, M.A.:
  Supervisor synthesis in model-based automotive systems engineering.
\newblock In: ICCPS'14: ACM/IEEE 5th International Conference on Cyber-Physical
  Systems (with CPS Week 2014), pp. 187--198. IEEE Computer Society (2014)

\bibitem{Walch2017a}
Walch, M.: Knowledge-driven enrichment of cyber-physical systems for industrial
  applications using the kbr modelling approach.
\newblock In: 2017 IEEE International Conference on Agents (ICA), pp. 84--89.
  IEEE (2017)

\bibitem{Weissnegger2016a}
Weissnegger, R., Schuss, M., Kreiner, C., Pistauer, M., R{\"o}mer, K., Steger,
  C.: Simulation-based verification of automotive safety-critical systems based
  on east-adl.
\newblock Procedia computer science \textbf{83}, 245--252 (2016)

\bibitem{Whalen2014}
Whalen, M.W., Murugesan, A., Rayadurgam, S., Heimdahl, M.P.: Structuring
  simulink models for verification and reuse.
\newblock In: Proceedings of the 6th International Workshop on Modeling in
  Software Engineering, pp. 19--24. ACM (2014)

\bibitem{Whitsitt2014}
Whitsitt, S., Sprinkle, J., Lysecky, R.: Generating model transformations for
  mending dynamic constraint violations in cyber physical systems.
\newblock In: Proceedings of the 14th Workshop on Domain-Specific Modeling, pp.
  35--40. ACM (2014)

\bibitem{wohlin2014guidelines}
Wohlin, C.: Guidelines for snowballing in systematic literature studies and a
  replication in software engineering.
\newblock In: Proceedings of the 18th international conference on evaluation
  and assessment in software engineering, p.~38. Citeseer (2014)

\bibitem{wohlin2012experimentation}
Wohlin, C., Runeson, P., H{\"o}st, M., Ohlsson, M.C., Regnell, B., Wessl{\'e}n,
  A.: Experimentation in software engineering.
\newblock Springer Science \& Business Media (2012)

\bibitem{Xin2015}
Xin, S., Guo, Q., Sun, H., Zhang, B., Wang, J., Chen, C.: Cyber-physical
  modeling and cyber-contingency assessment of hierarchical control systems.
\newblock IEEE Transactions on Smart Grid \textbf{6}(5), 2375--2385 (2015)

\bibitem{Yan2009}
Yan, W., Xue, Y., Li, X., Weng, J., Busch, T., Sztipanovits, J.: Integrated
  simulation and emulation platform for cyber-physical system security
  experimentation.
\newblock In: Proceedings of the 1st international conference on High
  Confidence Networked Systems, pp. 81--88. ACM (2012)

\bibitem{Zhang2013b}
Zhang, K., Sprinkle, J.: Model-based software synthesis for self-reconfigurable
  sensor network in water monitoring.
\newblock In: 2013 20th IEEE International Conference and Workshops on
  Engineering of Computer Based Systems (ECBS), pp. 40--48. IEEE (2013)

\bibitem{Zhang2014}
Zhang, K., Sprinkle, J.: A closed-loop model-based design approach based on
  automatic verification and transformation.
\newblock In: Proceedings of the 14th Workshop on Domain-Specific Modeling, pp.
  1--6. ACM (2014)

\bibitem{Zhang2011}
Zhang, L.: Aspect-oriented mda approach for non-functional properties of
  distributed cyber physical systems.
\newblock In: 2011 10th International Symposium on Distributed Computing and
  Applications to Business, Engineering and Science, pp. 284--288. IEEE (2011)

\bibitem{Zhang2013c}
Zhang, L.: An integration approach to specify and model automotive cyber
  physical systems.
\newblock In: 2013 International Conference on Connected Vehicles and Expo
  (ICCVE), pp. 568--573. IEEE (2013)

\bibitem{Zhang2013a}
Zhang, L.: View oriented approach to specify and model aerospace cyber-physical
  systems.
\newblock In: 2013 IEEE 11th International Conference on Dependable, Autonomic
  and Secure Computing, pp. 296--303. IEEE (2013)

\bibitem{Zhang2014c}
Zhang, L.: A framework to specify big data driven complex cyber physical
  control systems.
\newblock In: 2014 IEEE International Conference on Information and Automation
  (ICIA), pp. 548--553. IEEE (2014)

\bibitem{Zhang2014d}
Zhang, L., Feng, S.: Integration design and model transformation for cyber
  physical systems.
\newblock In: 2014 IEEE 5th International Conference on Software Engineering
  and Service Science, pp. 754--757. IEEE (2014)

\bibitem{Zhao2017}
Zhao, H., Apvrille, L., Mallet, F.: Multi-view design for cyber-physical
  systems (2017)

\bibitem{Zhou2018}
Zhou, Y., Gong, X., Li, J., Li, B.: Verifying cps for self-adaptability.
\newblock In: 2018 IEEE/ACIS 17th International Conference on Computer and
  Information Science (ICIS), pp. 166--172. IEEE (2018)

\end{thebibliography}

\addcontentsline{toc}{section}{Appendixes}
\pagebreak
\section*{Appendixes} \label{appendixes}



\begin{longtable}{|p{0.3\textwidth}|p{0.1\textwidth}|p{0.3\textwidth}|p{0.1\textwidth}|}
\caption{List of Authors}
\label{App:authors}\\
\hline
\textbf{Author(s)} & \textbf{\# of papers} & \textbf{Author(s)} & \textbf{\# of papers} \\ \hline
\endfirsthead
\multicolumn{4}{c}%
{{\bfseries Table \thetable\ continued from previous page}} \\
\hline
\textbf{Author(s)} & \textbf{\# of papers} & \textbf{Author(s)} & \textbf{\# of papers} \\ \hline
\endhead
Janos Sztipanovits & 6 & Jon Mathews & 1 \\ \hline
Lichen Zhang & 6 & Jordan Noble & 1 \\ \hline
Dehui Du & 4 & Jorgen Hansson & 1 \\ \hline
Jonathan Sprinkle & 4 & JOSEPH HALL & 1 \\ \hline
Andrea Vinci & 3 & Joshua Plasse & 1 \\ \hline
Aniruddha Gokhale & 3 & Jozef Hooman & 1 \\ \hline
Antonio Guerrieri & 3 & Juan Carlos Sanchez-Aarnoutse & 1 \\ \hline
DeJiu Chen & 3 & Judith Peters & 1 \\ \hline
Franco Cicirelli & 3 & Jufu Liu & 1 \\ \hline
Giancarlo Fortino & 3 & Juha Kuronen & 1 \\ \hline
Giandomenico Spezzano & 3 & Julien Deantoni & 1 \\ \hline
Leandro Buss Becker & 3 & K. Meßzmer & 1 \\ \hline
Lui Sha & 3 & Kaiqiang Jiang & 1 \\ \hline
Luís Gomes & 3 & Kary Myers & 1 \\ \hline
Xenofon Koutsoukos & 3 & Kay Römera & 1 \\ \hline
Abhishek Dubey & 2 & Keith Stouffer & 1 \\ \hline
Ahmed Hadj Kacem & 2 & Ken Keefe & 1 \\ \hline
Angelo Perkusich & 2 & Kennon McKeever & 1 \\ \hline
Anitha Murugesan & 2 & Kevin Lynch & 1 \\ \hline
Bei Cheng & 2 & Klaas Gadeyne & 1 \\ \hline
Christian Heinzemann & 2 & Konstantins Gusarovs & 1 \\ \hline
Chunhui Guo & 2 & Kuesap J & 1 \\ \hline
Chunlin Guan & 2 & Kyle Collins & 1 \\ \hline
Danai C. Vachtsevanou & 2 & Kyoungho An & 1 \\ \hline
Fernando Silvano Gonçalves & 2 & Lars Dürkop & 1 \\ \hline
Gabor Simko & 2 & Leandro B.Becker & 1 \\ \hline
Hans Vangheluwe & 2 & Lei Bu & 1 \\ \hline
Hervé Panetto & 2 & Leonard Petngaa & 1 \\ \hline
Himanshu Neema & 2 & Li B & 1 \\ \hline
Hyggo O. Almeida & 2 & Lian Duan & 1 \\ \hline
Imen Graja & 2 & Liliana Pasquale & 1 \\ \hline
Insup Lee & 2 & Lionel C. Briand & 1 \\ \hline
Isabel Sofia Brito & 2 & Lisa Ollinger & 1 \\ \hline
João Paulo Barros & 2 & Loïc Fejoz & 1 \\ \hline
JOSEPH PORTER & 2 & Lorenzo Pagliari & 1 \\ \hline
Kleanthis Thramboulidis & 2 & Ludovic Apvrille & 1 \\ \hline
Kun Zhang & 2 & M.A. Goorden & 1 \\ \hline
Lenardo C. Silva & 2 & Maarten M. Bezemer & 1 \\ \hline
Marilyn Wolf & 2 & Maarten Witters & 1 \\ \hline
Mario Lezoche & 2 & Madhavan Shanmugavel & 1 \\ \hline
Michael W. Whalen & 2 & Mahmood R. Khabbazi & 1 \\ \hline
Mirko Perkusich & 2 & Mahmood R.Khabazzi & 1 \\ \hline
Muhammad Umer Tariq & 2 & Marc Pantel & 1 \\ \hline
Nawal Guermouche & 2 & Marcelo V. García & 1 \\ \hline
Oleg Sokolsky & 2 & Marco A.Wehrmeister & 1 \\ \hline
Sandeep Neema & 2 & Marco Di Natale & 1 \\ \hline
Sanjai Rayadurgam & 2 & Marco Rocchetto & 1 \\ \hline
Shangping Ren & 2 & Marga Marcos & 1 \\ \hline
Slim Kallel & 2 & Mark Austinb & 1 \\ \hline
Stefan Dziwok & 2 & Mark Blackburn & 1 \\ \hline
Tihamer Levendovszky & 2 & Markus Pistauerb & 1 \\ \hline
Yu Jiang & 2 & Markus Schussa & 1 \\ \hline
Zhicheng Fu & 2 & Markus Tauber & 1 \\ \hline
Abdurrahman Alshareef & 1 & Martin H. R. van der Heijden & 1 \\ \hline
Abubakr Muhammad & 1 & Martin Lehofer & 1 \\ \hline
Adam Hahn & 1 & Maryam Rahmaniheris & 1 \\ \hline
Adam Trewyn & 1 & Mashor Housh & 1 \\ \hline
Agnes Lanusse & 1 & Mats P. E. Heimdahl & 1 \\ \hline
Ahmed Tamrawi & 1 & Mats P.E. Heimdahl & 1 \\ \hline
Akshay Agrawal & 1 & Matt Bunting & 1 \\ \hline
Alessandro Mercuri & 1 & Matt Schmit & 1 \\ \hline
Alexandros Solanos & 1 & Matteo Morelli & 1 \\ \hline
Ali R. Hurson & 1 & Mauro Caporuscio & 1 \\ \hline
Alois Knoll & 1 & Mauro Iacono & 1 \\ \hline
Alois Zoitl & 1 & Max Mühlhäuser & 1 \\ \hline
Alvaro Cardenas & 1 & Md Abdullah Al Mamun & 1 \\ \hline
Anas Motii & 1 & Michael Walch & 1 \\ \hline
Andrea Bettoni & 1 & Michel A. Reniers & 1 \\ \hline
Andrea D'Ambrogio & 1 & Michele Ciavotta & 1 \\ \hline
Andrea Giglio & 1 & Michele Lora & 1 \\ \hline
Andrea Tundis & 1 & Milan Vathoopan & 1 \\ \hline
Andreas Volk & 1 & Ming Gu & 1 \\ \hline
Anh-Toan Bui Long & 1 & Mingfu Tuo & 1 \\ \hline
Ani Bicaku & 1 & Mirko D'Angelo & 1 \\ \hline
Annalisa Napolitano & 1 & Miroslav Pajic & 1 \\ \hline
Annarita Drago & 1 & Mohamed Amine Haj Kacem & 1 \\ \hline
Annarita Tedesco & 1 & Mohammad Abdollahi Azgomi & 1 \\ \hline
Antonio Maffei & 1 & Mohammad Abdullah Al Faruque & 1 \\ \hline
Arquimedes Canedo & 1 & Muhammad Rashid & 1 \\ \hline
Artur Ataíde & 1 & Muhammad Waqar Aziz & 1 \\ \hline
Assaad Moawad & 1 & Na-Bangchang K & 1 \\ \hline
Attila Klenik & 1 & Nafiul Rashid & 1 \\ \hline
Ayan Banerjee & 1 & Nan Zhou & 1 \\ \hline
B. Vogel-Heuser & 1 & Nathan Jarus & 1 \\ \hline
B.Vogel-Heuser & 1 & Nazakat Ali & 1 \\ \hline
Baek-Gyu Kim & 1 & Neena Goveas & 1 \\ \hline
Bashar Nuseibeh & 1 & Neeraj Suri & 1 \\ \hline
Benjamin Brandenbourger & 1 & NICHOLAS KOTTENSTETTE & 1 \\ \hline
Benjámin Gehl & 1 & Nicola Mazzocca & 1 \\ \hline
Benoit Combemale & 1 & Nicolas Navet & 1 \\ \hline
Bert Van Acker & 1 & Nils Ole Tippenhauer & 1 \\ \hline
Bijan Shirinzadeh & 1 & Nils Przigoda & 1 \\ \hline
Binbin Chen & 1 & Ning Fu & 1 \\ \hline
Bixin Li & 1 & Nisrine El Marzouki & 1 \\ \hline
Boming Zhang & 1 & Numata I & 1 \\ \hline
Bradley Potteiger & 1 & Oksana Nikiforova & 1 \\ \hline
Brahim Hamid & 1 & Paolo Bocciarelli & 1 \\ \hline
Brett Feddersen & 1 & Pasqualina Potena & 1 \\ \hline
Bryan T. Carter & 1 & Paul De Meulenaere & 1 \\ \hline
Carl R. Elks & 1 & Payas Awadhutkar & 1 \\ \hline
Carlos E.Pereira & 1 & Peidong Zhu & 1 \\ \hline
Carmen Cheh & 1 & Peter Denno & 1 \\ \hline
Catia Trubiani & 1 & Péter Garamvölgyi & 1 \\ \hline
Cesare Fantuzzi & 1 & Peter Hehenberger & 1 \\ \hline
Chaim Cohen & 1 & Peter Volgyesi & 1 \\ \hline
Cheeyee Tang & 1 & Peter-Jan D. Vos & 1 \\ \hline
Chen Chen & 1 & Ping Huang & 1 \\ \hline
Cheng Pang & 1 & Priscill Orue Esquivel & 1 \\ \hline
Chin-Feng Fan & 1 & Qinglai Guo & 1 \\ \hline
Ching-Chieh Chan & 1 & Qixin Wang & 1 \\ \hline
Christian Berger & 1 & Raffaela Mirandola & 1 \\ \hline
Christian Buckl & 1 & Rahul Mangharam & 1 \\ \hline
Christian Kreinera & 1 & Ralph Weissneggerab & 1 \\ \hline
Christian Steger & 1 & Randall Ramsey & 1 \\ \hline
Christine Keller & 1 & Raphael Cohen & 1 \\ \hline
Christopher Gerking & 1 & Refik Samet & 1 \\ \hline
Cody H. Fleming & 1 & Reza Matinnejad & 1 \\ \hline
Cuiqing Jiang & 1 & Richard Candell & 1 \\ \hline
D.Schütz & 1 & Rima Al-Ali & 1 \\ \hline
Daniel Gritznera & 1 & Rizwan Parveen & 1 \\ \hline
Daniel Sonntag & 1 & Robert Wille & 1 \\ \hline
David Lindecker & 1 & Robert YoungChul Kim & 1 \\ \hline
David Pereira & 1 & Rolf Drechsler & 1 \\ \hline
Deepak Mehta & 1 & Rolf Egert & 1 \\ \hline
Detlef Zühlke & 1 & Romain Jobredeaux & 1 \\ \hline
Di Li & 1 & Roman Lysecky & 1 \\ \hline
Dmitri Valeri Panfilenko & 1 & Rong Zheng & 1 \\ \hline
Dmitry Morozov & 1 & Ronny Seiger & 1 \\ \hline
Dominik Sojer & 1 & Roopak Sinha & 1 \\ \hline
E. Trunzer & 1 & Rosane Fátima Passarini & 1 \\ \hline
Eduardo Tovar & 1 & Roshan K. Thomas & 1 \\ \hline
Edurne Irisarri & 1 & Rudolf G. M. Huisman & 1 \\ \hline
Eldad Palachi & 1 & Ruirui CHEN & 1 \\ \hline
Elisabet Estévez & 1 & S. Bougouffa & 1 \\ \hline
EMEKA EYISI & 1 & S. Cha & 1 \\ \hline
Emiliano Paglia & 1 & S.Feldmann & 1 \\ \hline
Emilien Lavigne & 1 & S.Rösch & 1 \\ \hline
Enrico Fraccaroli & 1 & Sahra Sedigh Sarvestani & 1 \\ \hline
Eric Feron & 1 & Saideep Nannapaneni & 1 \\ \hline
Eric Gascard & 1 & Sajal Bhatia & 1 \\ \hline
Ethan Jackson & 1 & Sakairi Takashi & 1 \\ \hline
F.F.H. Reijnen & 1 & Sampath Kumar & 1 \\ \hline
Fabio Cremona & 1 & Sandeep K. S. Gupta & 1 \\ \hline
Faeq Alrimawi & 1 & Sankaran Mahadevan & 1 \\ \hline
Fang Li & 1 & Sarah Haas & 1 \\ \hline
Federico Pérez & 1 & Satarug S & 1 \\ \hline
Felix Oestersötebier & 1 & Sean Whitsitt & 1 \\ \hline
Feng Tan & 1 & Shibahara S & 1 \\ \hline
Fernando Pereira & 1 & Shiva Nejati & 1 \\ \hline
Fernando S.Gonçalves & 1 & Shivakumar Sastry & 1 \\ \hline
Florent Latombe & 1 & Shixi Liu & 1 \\ \hline
Florian Niebling & 1 & Shubham Pradhan & 1 \\ \hline
Florin Leon & 1 & Shuguang Feng & 1 \\ \hline
Franco Fummi & 1 & Shujun Xin & 1 \\ \hline
Francois Fouquet & 1 & Shunxing Bao & 1 \\ \hline
Franz Franchetti & 1 & Silia Maksuti & 1 \\ \hline
Frederic Mallet & 1 & Silke Palkovits-Rauter & 1 \\ \hline
Frédéric Mallet & 1 & Simon Merschak & 1 \\ \hline
Frederico M. Bublitz & 1 & Song Li & 1 \\ \hline
Gabor Karsai & 1 & Stefano Marrone & 1 \\ \hline
Gabriele Izzo & 1 & Steffen Becker & 1 \\ \hline
Gaetana Sapienza & 1 & Stephen Rees & 1 \\ \hline
Gang Yang & 1 & Subhav M. Pradhan & 1 \\ \hline
George Ball & 1 & Subhav Pradhan & 1 \\ \hline
George Mois & 1 & Suresh Kothari & 1 \\ \hline
Georgios Bakirtzis & 1 & Swu Yih & 1 \\ \hline
Gerardo Santillán Martínez & 1 & Takeda K & 1 \\ \hline
Germain Lemasson & 1 & Teodora Sanislav & 1 \\ \hline
Giacomo Barbieri & 1 & Thomas Bruckmann & 1 \\ \hline
Goncalo Martins & 1 & Thomas Greiner & 1 \\ \hline
Goulven Guillou & 1 & Thomas Hartmann & 1 \\ \hline
Gregory Nain & 1 & Thomas M. Chen & 1 \\ \hline
Guilherme V.Raffo & 1 & Thomas Nägele & 1 \\ \hline
Guilin Chen & 1 & Thomas Schlegel & 1 \\ \hline
Gustavo Quiros & 1 & Timothy Busch & 1 \\ \hline
Hakan Akillioglu & 1 & Tomas Bures & 1 \\ \hline
Hamed Orojloo & 1 & Tze Meng Low & 1 \\ \hline
Han Liu & 1 & Ulrich Kühne & 1 \\ \hline
Hang-Gi Min & 1 & Uwe Pohlmann & 1 \\ \hline
Hasan Arshad Nasir & 1 & Valeria Vittorini & 1 \\ \hline
HEATH LEBLANC & 1 & Valeriy Vyatkin & 1 \\ \hline
Henning Trsek & 1 & Vasileios Koutsoumpas & 1 \\ \hline
Hessam S. Sarjoughian & 1 & Veera Ragavan & 1 \\ \hline
Hongbin Sun & 1 & Velappa Ganapathy & 1 \\ \hline
Houbing Song & 1 & Vitaliy Mezhuyev & 1 \\ \hline
Hsiang-Yu Yu & 1 & Wei Yan & 1 \\ \hline
Hui Zhao & 1 & Wilhelm Schafer & 1 \\ \hline
Huiqun Yu & 1 & William G. Temple & 1 \\ \hline
Hyun Seung Son & 1 & William H. Sanders & 1 \\ \hline
Imre Kocsis & 1 & Wolfgang Rosenstiel & 1 \\ \hline
InGeol Chun & 1 & WonTae Kim & 1 \\ \hline
Ioanna Kontou & 1 & Woo Yeol Kim & 1 \\ \hline
Ivan Lozano & 1 & Xavier Crégut & 1 \\ \hline
Ivica Crnkovic & 1 & Xiao Wang & 1 \\ \hline
J.E. Rooda & 1 & Xiaojing Hu & 1 \\ \hline
J.M. van de Mortel-Fronczak & 1 & Xiaoping YE & 1 \\ \hline
Jacques Florence & 1 & Xiaowei Li & 1 \\ \hline
Jacques Klein & 1 & Xiaoxue Liu & 1 \\ \hline
Jaeho Jeon & 1 & Xingshe Zhou & 1 \\ \hline
Jan F. Broenink & 1 & Xinhai Zhang & 1 \\ \hline
Jang-Eui Hong & 1 & Xudong He & 1 \\ \hline
Jean-Marie Farines & 1 & Xufang Gong & 1 \\ \hline
Jean-Michel Bruel & 1 & Yaniv Mordecai & 1 \\ \hline
Jean-Philippe Baba & 1 & Yegeta Zeleke & 1 \\ \hline
Jerker Delsing & 1 & Yi Ao & 1 \\ \hline
Jérôome Maisonnasse & 1 & Ying Zhou & 1 \\ \hline
Jiaguang Sun & 1 & Yixiao Yang & 1 \\ \hline
Jiakai Li & 1 & Yogesh Barve & 1 \\ \hline
Jiang Wan & 1 & Yong Guan & 1 \\ \hline
Jianhui Wang & 1 & Yuan Xue & 1 \\ \hline
Jiannian Weng & 1 & Yufei Wang & 1 \\ \hline
Jiexin Zhang & 1 & Yusheng LIU & 1 \\ \hline
Jingyong Liu & 1 & Yves Le Traon & 1 \\ \hline
Joachim Denil & 1 & Zhenyu Zhang & 1 \\ \hline
Joanna M. van de Mortel-Fronczak & 1 & Zhihao Jiang & 1 \\ \hline
João Ferreirar & 1 & Zhilin Qian & 1 \\ \hline
Joe Porter & 1 & Zhonghai Lu & 1 \\ \hline
Joel Greenyer & 1 & Zhou Lu & 1 \\ \hline
Johannes Dell & 1 & Zineb Simeu-Abazi & 1 \\ \hline
John Buford & 1 & Ziv Ohar & 1 \\ \hline
John Komp & 1 &  &  \\ \hline
\end{longtable}

\begin{longtable}{|l|l|l|l|}
\caption{List of Countries, based on the authors' affiliation.}
\label{app:Countries}\\
\hline
\textbf{Author country} & \textbf{\# of papers} & \textbf{Author country} & \textbf{\# of papers} \\ \hline
\endfirsthead
\multicolumn{4}{c}%
{{\bfseries Table \thetable\ continued from previous page}} \\
\hline
\textbf{Author country} & \textbf{\# of papers} & \textbf{Author country} & \textbf{\# of papers} \\ \hline
\endhead
USA & 39 & Czech Republic & 1 \\ \hline
China & 23 & Finland & 1 \\ \hline
Germany & 16 & Hong Kong & 1 \\ \hline
Italy & 13 & Hungary & 1 \\ \hline
France & 12 & Iran & 1 \\ \hline
Brazil & 7 & Ireland & 1 \\ \hline
Sweden & 7 & Israel & 1 \\ \hline
Netherlands & 5 & Jordan & 1 \\ \hline
Austria & 4 & Korea & 1 \\ \hline
Luxembourg & 4 & Latvia & 1 \\ \hline
Portugal & 4 & Malaysia & 1 \\ \hline
Spain & 3 & Morocco & 1 \\ \hline
Belgium & 2 & New Zealand & 1 \\ \hline
Greece & 2 & Pakistan & 1 \\ \hline
India & 2 & Saudi Arabia & 1 \\ \hline
Isreal & 2 & Singapore & 1 \\ \hline
Japan & 2 & Switzerland & 1 \\ \hline
Romania & 2 & Taiwan & 1 \\ \hline
South Korea & 2 & Thailand & 1 \\ \hline
Tunisia & 2 & Turkey & 1 \\ \hline
UK & 2 & Ukraine & 1 \\ \hline
Australia & 1 &  &  \\ \hline
Canada & 1 &  &  \\ \hline
\end{longtable}

\pagebreak


\begin{longtable}{|l|p{0.7\textwidth}|p{0.1\textwidth}|}
\caption{List of publication venues}
\label{App:venues}\\
\hline
\textbf{Venue Type} & \textbf{publication venue} & \textbf{\# of studies} \\ \hline
\endfirsthead
\multicolumn{3}{c}%
{{\bfseries Table \thetable\ continued from previous page}} \\
\hline
Venue Type & \textbf{publication venue} & \textbf{\# of studies} \\ \hline
\endhead
\textbf{Conference} & International Conference on Emerging Technologies and Factory Automation (ETFA) & 6 \\ \hline
\textbf{} & International Conference on Cyber-Physical Systems & 4 \\ \hline
\textbf{} & Industrial Cyber-Physical Systems (ICPS) & 3 \\ \hline
\textbf{} & International Conference on Engineering of Complex Computer Systems (ICECCS) & 3 \\ \hline
\textbf{} & International Conference on Industrial Informatics (INDIN) & 3 \\ \hline
\textbf{} & Annual Computer Software and Applications Conference (COMPSAC) & 2 \\ \hline
\textbf{} & Brazilian Symposium on Computing Systems Engineering (SBESC) & 2 \\ \hline
\textbf{} & International Conference on Networking, Sensing and Control (ICNSC) & 2 \\ \hline
\textbf{} & International Systems Conference (SysCon) & 2 \\ \hline
\textbf{} & Symposium on Applied Computing & 2 \\ \hline
\textbf{} & ACM SIGPLAN International Conference on Software Language Engineering & 1 \\ \hline
\textbf{} & ACM Transactions on Embedded Computing Systems (TECS) & 1 \\ \hline
\textbf{} & Annual Conference of the IEEE Industrial Electronics Society & 1 \\ \hline
\textbf{} & Annual Design Automation Conference & 1 \\ \hline
\textbf{} & Asia and South Pacific Design Automation Conference (ASP-DAC) & 1 \\ \hline
\textbf{} & Asia Conference on Computer and Communications Security & 1 \\ \hline
\textbf{} & Asia-Pacific Software Engineering Conference (APSEC) & 1 \\ \hline
\textbf{} & Computer Software and Applications Conference & 1 \\ \hline
\textbf{} & Computers and Information in Engineering Conference & 1 \\ \hline
\textbf{} & Conference on Automation Science and Engineering (CASE) & 1 \\ \hline
\textbf{} & Conference on Control Technology and Applications (CCTA) & 1 \\ \hline
\textbf{} & Conference on Emerging Technologies \& Factory Automation (ETFA 2010) & 1 \\ \hline
\textbf{} & Digital Avionics Systems Conference (DASC) & 1 \\ \hline
\textbf{} & Emerging Technology and Factory Automation (ETFA) & 1 \\ \hline
\textbf{} & Euromicro Conference on Software Engineering and Advanced Applications (SEAA) & 1 \\ \hline
\textbf{} & IEEE International Conference and Workshops on Engineering of Computer Based Systems (ECBS) & 1 \\ \hline
\textbf{} & IEEE International Conference on Industrial Engineering and Engineering Management (IEEM) & 1 \\ \hline
\textbf{} & IEEE International Conference on Information and Automation (ICIA) & 1 \\ \hline
\textbf{} & IEEE International Symposium on Computer-Based Medical Systems (CBMS) & 1 \\ \hline
\textbf{} & IEEE Transactions on Software Engineering & 1 \\ \hline
\textbf{} & IEEE/IFIP Conference on Software Architecture & 1 \\ \hline
\textbf{} & International Carpathian Control Conference (ICCC) & 1 \\ \hline
\textbf{} & International Conference and Workshops on Engineering of Computer Based Systems (ECBS) & 1 \\ \hline
\textbf{} & International Conference on Agents (ICA) & 1 \\ \hline
\textbf{} & International Conference on Cloud Engineering (IC2E) & 1 \\ \hline
\textbf{} & International Conference on Complex Systems Engineering (ICCSE) & 1 \\ \hline
\textbf{} & International Conference on Computer and Information Science (ICIS) & 1 \\ \hline
\textbf{} & International Conference on Computer Supported Cooperative Work in Design (CSCWD) & 1 \\ \hline
\textbf{} & International Conference on Connected Vehicles and Expo (ICCVE) & 1 \\ \hline
\textbf{} & International Conference on Cyberworlds & 1 \\ \hline
\textbf{} & International Conference on Data Mining Workshops (ICDMW) & 1 \\ \hline
\textbf{} & International Conference on Dependable Systems and Networks (DSN) & 1 \\ \hline
\textbf{} & International Conference on Dependable Systems and Networks Workshops (DSN-W) & 1 \\ \hline
\textbf{} & International Conference on Dependable, Autonomic and Secure Computing & 1 \\ \hline
\textbf{} & International Conference on Enabling Technologies: Infrastructure for Collaborative Enterprises (WETICE) & 1 \\ \hline
\textbf{} & international conference on High Confidence Networked Systems & 1 \\ \hline
\textbf{} & International Conference on ICT in Education, Research, and Industrial Applications & 1 \\ \hline
\textbf{} & International Conference on Information Science and Applications (ICISA) & 1 \\ \hline
\textbf{} & International Conference on Information Systems Engineering (ICISE) & 1 \\ \hline
\textbf{} & International Conference on Internet of Things (iThings), Green Computing and Communications (GreenCom), Physical and Social Computing (CPSCom), IEEE Smart Data (SmartData) & 1 \\ \hline
\textbf{} & International Conference on Model Driven Engineering Languages and Systems (MODELS) & 1 \\ \hline
\textbf{} & International Conference on Networking and Distributed Computing & 1 \\ \hline
\textbf{} & International Conference on P2P, Parallel, Grid, Cloud and Internet Computing (3PGCIC) & 1 \\ \hline
\textbf{} & International Conference on Research and Education in Mechatronics (REM) & 1 \\ \hline
\textbf{} & International Conference on Smart Computing (SMARTCOMP) & 1 \\ \hline
\textbf{} & International Conference on Software Engineering & 1 \\ \hline
\textbf{} & International Conference on Software Engineering and Service Science & 1 \\ \hline
\textbf{} & International Conference on Software Quality, Reliability and Security Companion (QRS-C) & 1 \\ \hline
\textbf{} & International Conference on Software Technologies & 1 \\ \hline
\textbf{} & International Conference on Ubiquitous Intelligence, International Conference on Autonomic and Trusted Computing & 1 \\ \hline
\textbf{} & International Green and Sustainable Computing Conference (IGSC) & 1 \\ \hline
\textbf{} & International Science of Smart City Operations and Platforms Engineering in Partnership with Global City Teams Challenge (SCOPE-GCTC) & 1 \\ \hline
\textbf{} & International Symposium on Distributed Computing and Applications to Business, Engineering and Science & 1 \\ \hline
\textbf{} & International Symposium on High Assurance Systems Engineering (HASE) & 1 \\ \hline
\textbf{} & Model-driven Approaches for Simulation Engineering Symposium & 1 \\ \hline
\textbf{} & Real Time and Embedded Technology and Applications Symposium & 1 \\ \hline
\textbf{} & Resilience Week (RWS) & 1 \\ \hline
\textbf{} & System of Systems Engineering Conference (SoSE) & 1 \\ \hline
\textbf{} & Theoretical Aspects of Software Engineering (TASE) & 1 \\ \hline
\textbf{} & Winter Simulation Conference & 1 \\ \hline
\textbf{Journal} & Advanced Engineering Informatics & 2 \\ \hline
\textbf{} & IEEE Transactions on Smart Grid & 2 \\ \hline
\textbf{} & IFAC Symposium on Information Control Problems in Manufacturing INCOM & 2 \\ \hline
\textbf{} & International Journal of Critical Infrastructure Protection & 2 \\ \hline
\textbf{} & ACM Transactions on Cyber-Physical Systems & 1 \\ \hline
\textbf{} & Biochemical and Biophysical Research Communications & 1 \\ \hline
\textbf{} & Computers in Industry & 1 \\ \hline
\textbf{} & Cyber Physical Systems and Deep Learning & 1 \\ \hline
\textbf{} & Cyber-Physical \& Human-Systems CPHS & 1 \\ \hline
\textbf{} & Embedded and Multimedia Computing Technology and Service & 1 \\ \hline
\textbf{} & Enterprise Information Systems & 1 \\ \hline
\textbf{} & Fakultät für Elektrotechnik und Informatik & 1 \\ \hline
\textbf{} & IFAC Conference on Embedded Systems, Computational Intelligence and Telematics in Control CESCIT & 1 \\ \hline
\textbf{} & IFAC International Workshop on Dependable Control of Discrete Systems & 1 \\ \hline
\textbf{} & IFAC World Congress & 1 \\ \hline
\textbf{} & Information Technology: New Generations & 1 \\ \hline
\textbf{} & International Conference on Ambient Systems, Networks and Technologies (ANT), International Conference on Sustainable Energy Information Technology (SEIT) & 1 \\ \hline
\textbf{} & International Journal of Industrial Ergonomics & 1 \\ \hline
\textbf{} & International Symposium on Model-Based Safety and Assessment & 1 \\ \hline
\textbf{} & International Workshop on Model-based Architecting and Construction of Embedded Systems & 1 \\ \hline
\textbf{} & Journal of Computational Science & 1 \\ \hline
\textbf{} & Microprocessors and Microsystems & 1 \\ \hline
\textbf{} & Procedia Computer Science & 1 \\ \hline
\textbf{} & Robotics and Autonomous Systems & 1 \\ \hline
\textbf{} & Software \& Systems Modeling & 1 \\ \hline
\textbf{} & Water Research & 1 \\ \hline
\textbf{Workshop} & workshop on Domain-specific modeling & 6 \\ \hline
\textbf{} & IFAC Workshop on Intelligent Manufacturing Systems & 2 \\ \hline
\textbf{} & Advances in Intelligent Systems Research & 1 \\ \hline
\textbf{} & International Workshop on Modeling in Software Engineering & 1 \\ \hline
\textbf{} & International Workshop on Security Awareness from Design to Deployment (SEAD) & 1 \\ \hline
\textbf{} & International Workshop on Software Engineering for Smart Cyber-Physical Systems & 1 \\ \hline
\textbf{} & Workshop on Cyber-Physical Systems Security and PrivaCy & 1 \\ \hline
\textbf{} & Workshop on Formal Methods in Software Engineering & 1 \\ \hline
\textbf{} & Workshop on Mobile Medical Applications & 1 \\ \hline
\textbf{} & Workshop on Model-Driven Engineering, Verification and Validation & 1 \\ \hline
\end{longtable}

\pagebreak


\begin{longtable}{|l|l|l|}
\caption{List of modeling activities together with the approaches used and the studies reported them}
\label{App:modelingactivities}\\
\hline
\multicolumn{1}{|c|}{\textbf{Activity}} & \multicolumn{1}{c|}{\textbf{Approach}} & \multicolumn{1}{c|}{\textbf{Papers'  Key}}                                                                                          \\ \hline
\endhead
\multirow{13}{*}{Development}           & DSL                                    & \cite{Angelo2018,Aziz2016,Zhang2013b,Koutsoukos2012,Tariq2014,Bunting2016,Pradhan2015,Guan2018}                    \\ \cline{2-3} 
 & Metamodeling                           & \cite{Latombe2015a,Cicirelli2016,Cicirelli2017a,Mezhuyev2013,Chen2015,Barbieri2016,SampathKumar2015,Cicirelli2018} \\ \cline{2-3} 
                                        & Model-based approach                   & \cite{Goncalves2016a,Buckl2010,Chen2016,VanDeMortel-Fronczak2014,Reijnen2017,Passarini2014,Murugesan}              \\ \cline{2-3} 
                                        & Component based approach               & \cite{Nikiforova2017a,Li2015,Pohlmann2014}                                                                         \\ \cline{2-3} 
                                        & UML                                    & \cite{Thramboulidis2018}                                                                                           \\ \cline{2-3} 
                                        & equation based                         & \cite{Mezhuyev2013,Jarus2016}                                                                                      \\ \cline{2-3} 
                                        & petri nets                             & \cite{Pereira2016}                                                                                                 \\ \cline{2-3} 
                                        & Integrated approaches                  & \cite{Zhang2014d,Broenink2016}                                                                                     \\ \cline{2-3} 
                                        & architecture based approach            & \cite{Ollinger2013,Goncalves2016}                                                                                  \\ \cline{2-3} 
                                        & Aspect-Oriented approach               & \cite{Liu2011}                                                                                                     \\ \cline{2-3} 
                                        & Process based modeling                 & \cite{Graja}                                                                                                       \\ \cline{2-3} 
                                        & text based modeling                    & \cite{Navet2016}                                                                                                   \\ \cline{2-3} 
                                        & Other                                  & \cite{Thramboulidis,Pereira2016}                                                                                   \\ \hline
\multirow{11}{*}{Analysis}              & DSL                                    & \cite{Tariq2012,An2011,Dell2014}                                                                                   \\ \cline{2-3} 
                                        & Metamodeling                           & \cite{Sapienza2014,Cicirelli2017a,Cheh2017,Morozov2018,Zhao2017,Lezoche2018}                                       \\ \cline{2-3} 
                                        & Model-based approach                   & \cite{DiNatale2016,Chen2016,Mordecai2018,Tuo2017,Nannapaneni2016,Fan2018}                                          \\ \cline{2-3} 
                                        & Component based approach               & \cite{Sapienza2014,Kacem2017}                                                                                      \\ \cline{2-3} 
                                        & UML                                    & \cite{Drago2013,Dell2014,Bakirtzis2018,Cheng2015}                                                                  \\ \cline{2-3} 
                                        & equation based                         & \cite{Tariq2012,Qian2013,Xin2015}                                                                                  \\ \cline{2-3} 
                                        & petri nets                             & \cite{Qian2013,Liu2017a,Brito2017,He2018,Hu2013}                                                                   \\ \cline{2-3} 
                                        & Simulation                             & \cite{Lora2017,Neema2018}                                                                                          \\ \cline{2-3} 
                                        & Integrated approaches                  & \cite{Zhang2013a,Zhang2013c,Goncalves2017,Pagliari2018}                                                            \\ \cline{2-3} 
                                        & Aspect-Oriented approach               & \cite{Qian2013}                                                                                                    \\ \cline{2-3} 
                                        & Other                                  & \cite{Whitsitt2014,Alshareef2018,He2018}                                                                           \\ \hline
\multirow{11}{*}{V\&V}                  & DSL                                    & \cite{Angelo2018,Mamun2013,Gerking2015,Zhang2014,Blackburn2014}                                                    \\ \cline{2-3} 
                                        & Model-based approach                   & \cite{Pajic2012,Kothari2018,Tuo2017,Whalen2014,Cohen2015}                                                          \\ \cline{2-3} 
                                        & Component based approach               & \cite{Parveen2018,Silva2014}                                                                                       \\ \cline{2-3} 
                                        & Simulation                             & \cite{Silva2015,Matinnejad2018,Jiang2018}                                                                          \\ \cline{2-3} 
                                        & UML                                    & \cite{Zhou2018}                                                                                                    \\ \cline{2-3} 
                                        & equation based                         & \cite{Silva2015,Plasse2017}                                                                                        \\ \cline{2-3} 
                                        & petri nets                             & \cite{Guo2017}                                                                                                     \\ \cline{2-3} 
                                        & Ontology based                         & \cite{Lynch2016,Chen2018}                                                                                          \\ \cline{2-3} 
                                        & architecture based approach            & \cite{Guo2018}                                                                                                     \\ \cline{2-3} 
                                        & Process based modeling                 & \cite{B2018}                                                                                                       \\ \cline{2-3} 
                                        & Other                                  & \cite{Zhang2014,Neema2014}                                                                                         \\ \hline
\multirow{11}{*}{Security}              & Metamodeling                           & \cite{Alrimawi2018,Martins2015,Maksuti2017}                                                                        \\ \cline{2-3} 
                                        & Model-based approach                   & \cite{Chen2017,Banerjee2014}                                                                                       \\ \cline{2-3} 
                                        & Component based approach               & \cite{Hahn2015,Heinzemann2017slrcps}                                                                               \\ \cline{2-3} 
                                        & Simulation                             & \cite{Housh2018,Jiang2018,Rashid2018}                                                                              \\ \cline{2-3} 
                                        & UML                                    & \cite{Garamvolgyi2018,Motii2017}                                                                                   \\ \cline{2-3} 
                                        & equation based                         & \cite{Orojloo2017}                                                                                                 \\ \cline{2-3} 
                                        & petri nets                             & \cite{Liu2017a,Chen2011}                                                                                           \\ \cline{2-3} 
                                        & Integrated approaches                  & \cite{Kuesap2008}                                                                                                  \\ \cline{2-3} 
                                        & pattern-based modeling                 & \cite{Motii2017,Tan2013}                                                                                           \\ \cline{2-3} 
                                        & text based modeling                    & \cite{Rocchetto2017}                                                                                               \\ \cline{2-3} 
                                        & Other                                  & \cite{Yan2009}                                                                                                     \\ \hline
\multirow{8}{*}{Simulation}             & DSL                                    & \cite{Nagele2017,Koutsoukos2012,Bao2016,Simko2013a}                                                                \\ \cline{2-3} 
                                        & Metamodeling                           & \cite{Merschak2018,Ciavotta2018}                                                                                   \\ \cline{2-3} 
                                        & Model-based approach                   & \cite{Pajic2012}                                                                                                   \\ \cline{2-3} 
                                        & Component based approach               & \cite{Zhang2014c,Lavigne2018}                                                                                      \\ \cline{2-3} 
                                        & Simulation                             & \cite{Merschak2018,Jiang2018,Lavigne2018,Weissnegger2016a,VanAcker2015,Barve2018,Palachi2013}                      \\ \cline{2-3} 
                                        & UML                                    & \cite{Huang2018}                                                                                                   \\ \cline{2-3} 
                                        & equation based                         & \cite{Simko2013a}                                                                                                  \\ \cline{2-3} 
                                        & Integrated approaches                  & \cite{Broenink2016}                                                                                                \\ \hline
\multirow{5}{*}{Monitoring}             & DSL                                    & \cite{Mamun2013}                                                                                                   \\ \cline{2-3} 
                                        & Metamodeling                           & \cite{Chen2015,Bocciarelli2017,Irisarri2016}                                                                       \\ \cline{2-3} 
                                        & Model-based approach                   & \cite{Chen2017}                                                                                                    \\ \cline{2-3} 
                                        & Process based modeling                 & \cite{Bocciarelli2017}                                                                                             \\ \cline{2-3} 
                                        & Other                                  & \cite{Hartmann2015,Tundis2017}                                                                                     \\ \hline
\multirow{4}{*}{Time} & UML & \cite{Peters2015}                                                                                                  \\ \cline{2-3} 
                                        & text based modeling                    & \cite{Navet2016}                                                                                                   \\ \cline{2-3} 
                                        & Aspect-Oriented approach               & \cite{Zhang2011}                                                                                                   \\ \cline{2-3} 
                                        & Process based modeling                 & \cite{B2018}                                                                                                       \\ \hline
\multirow{3}{*}{Adaptability}           & Metamodeling                           & \cite{Maksuti2017}                                                                                                 \\ \cline{2-3} 
                                        & component based approach               & \cite{Feldmann2013}                                                                                                \\ \cline{2-3} 
                                        & equation based                         & \cite{Plasse2017}                                                                                                  \\ \hline
\multirow{2}{*}{Correctness}            & component based approach               & \cite{Gritzner2018,Brandenbourger2016}                                                                             \\ \cline{2-3} 
                                        & equation based                         & \cite{Low2017}                                                                                                     \\ \hline
\multirow{3}{*}{Integration}            & Integrated approaches                  & \cite{Zhang2013a}                                                                                                  \\ \cline{2-3} 
                                        & model-based approach                   & \cite{VanDeMortel-Fronczak2014}                                                                                    \\ \cline{2-3} 
                                        & Other                                  & \cite{Whitsitt2014}                                                                                                \\ \hline
\multirow{2}{*}{Efficiency}             & Ontology based                         & \cite{Lynch2016}                                                                                                   \\ \cline{2-3} 
                                        & component based approach               & \cite{Gritzner2018}                                                                                                \\ \hline
\multirow{2}{*}{Flexibility}            & model-based approach                   & \cite{Bougouffa2018}                                                                                               \\ \cline{2-3} 
                                        & component based approach               & \cite{Seiger2015}                                                                                                  \\ \hline
\multirow{13}{*}{Failure identification} & UML                                    & \cite{Ali2018}                                                                                                     \\ \cline{2-3} 
                                        & Ontology based                         & \cite{Ali2018}                                                                                                     \\ \hline
Uncertainty                             & Other                                  & \cite{Cheng2014,Koutsoumpas2015}                                                                                   \\ \hline
Complexity                              & Metamodeling                           & \cite{Seiger2015}                                                                                                  \\ \hline
dependability                           & UML                                    & \cite{Sanislav2017}                                                                                                \\ \hline
performance                             & model-based approach                   & \cite{Chen2017}                                                                                                    \\ \hline
Reliability                             & Simulation                             & \cite{Housh2018}                                                                                                   \\ \hline
resilience                              & Other                                  & \cite{Tundis2017}                                                                                                  \\ \hline
self-assessment                         & model-based approach                   & \cite{Chen2017}                                                                                                    \\ \hline
Test case Generation                    & model-based approach                   & \cite{Sinha2016}                                                                                                   \\ \hline
contingency assessment                  & equation based                         & \cite{Xin2015}                                                                                                     \\ \hline
Fault tolerance                         & equation based                         & \cite{Plasse2017}                                                                                                  \\ \hline
Decision making support                 & Ontology based                         & \cite{Petnga2016}                                                                                                  \\ \hline
                                    
\end{longtable}

\pagebreak

\begin{landscape}
\begin{table}[]
\centering
\caption{Addressed MDE activities in each CPS component}
\label{App:Q1vsQ4}
\begin{tabular}{|p{0.14\textwidth}|p{0.14\textwidth}|p{0.14\textwidth}|p{0.14\textwidth}|p{0.14\textwidth}|p{0.14\textwidth}|p{0.14\textwidth}|p{0.14\textwidth}|p{0.14\textwidth}|}
\hline
 & \textbf{code generating} & \textbf{modeling} & \textbf{requirement Analysis} & \textbf{simulation} & \textbf{system analysis} & \textbf{system design} & \textbf{Transformation} & \textbf{V\&V} \\ \hline
\textbf{Actuators} &  & \cite{Sanislav2017} &  &  &  & \cite{Cicirelli2016,Cicirelli2017a} & \cite{Goncalves2016a,Dell2014} & \cite{Sanislav2017} \\ \hline
\textbf{Both Cyber \& Physical component} & \cite{Bougouffa2018,Hartmann2015} & \cite{Zhang2013a,Silva2015,Zhang2013c,DiNatale2016,He2018,Chen2011,Xin2015,Kacem2017} & \cite{Liu2011,Pagliari2018} & \cite{DiNatale2016,An2011,Neema2018,Broenink2016,Hahn2015} & \cite{Sapienza2014,Zhang2013c,Liu2011,Morozov2018,Lezoche2018,Cheng2015,Ali2018} & \cite{Martins2015,Bougouffa2018,Huang2018,Hartmann2015,An2011,Kuesap2008,Graja} & \cite{Zhang2013a,Bougouffa2018,DiNatale2016,Tuo2017,Cheng2015} & \cite{Silva2015,He2018,Pagliari2018,Broenink2016} \\ \hline
\textbf{Cyber component} & \cite{Ataide2018,Peters2015,Garamvolgyi2018,Nagele2017,Buckl2010,Gritzner2018,Neema2014,Jiang2018,Pereira2016,Ollinger2013,Lavigne2018,Li2015,Cohen2015,Low2017,Reijnen2017,Murugesan} & \cite{Guo2018,Garamvolgyi2018,Zhou2018,Kothari2018,Li2015,Barbieri2016,Whalen2014,Orojloo2017,Petnga2016,Brandenbourger2016,Navet2016,Silva2014,Murugesan} & \cite{Tan2013,Sinha2016,Zhang2011,Murugesan} & \cite{Nagele2017,Bao2016,Simko2013a,Housh2018,Matinnejad2018,Neema2014,Jiang2018,Pereira2016,VanAcker2015,Barve2018,Navet2016,Alshareef2018,Rashid2018,Reijnen2017,Pohlmann2014,Silva2014,Fan2018} & \cite{Chen2017,Hu2013,Fan2018} & \cite{Aziz2016,Walch2017a,Nagele2017,Latombe2015a,Zhang2013b,Jeon2012,Buckl2010,Bao2016,Mezhuyev2013,Ollinger2013,Lavigne2018,Tariq2014,Barbieri2016,SampathKumar2015,Pradhan2015,Guan2018} & \cite{Son2012a,Guo2018,Nikiforova2017a,Gerking2015,Zhou2018,Buckl2010,Bao2016,Simko2013a,Brito2017,Zhang2014d,Jiang2018,Ollinger2013,Goncalves2017,Tariq2014,Whitsitt2014,Dell2014,Jarus2016,Bakirtzis2018,Alshareef2018,Passarini2014,Chen2018,B2018} & \cite{Guo2018,Peters2015,Walch2017a,Gerking2015,Zhou2018,Buckl2010,Lynch2016,Kothari2018,Neema2014,Jiang2018,Goncalves2017,Whalen2014,Cohen2015,Reijnen2017,Pohlmann2014,Silva2014,Murugesan,B2018} \\ \hline
\textbf{Network} & \cite{Koutsoukos2012,Zhang2014} & \cite{Cheh2017,Motii2017} & \cite{Motii2017} & \cite{Tariq2012,Koutsoukos2012,Housh2018,Mordecai2018,Tundis2017,Yan2009} &  & \cite{Alrimawi2018,Tariq2012,Maksuti2017,Cicirelli2016,Koutsoukos2012,Zhang2014,Cicirelli2017a,Yan2009} & \cite{Drago2013,Zhang2014,Motii2017} & \cite{Zhang2014} \\ \hline
\textbf{Physical component} & \cite{Pajic2012,Lora2017} & \cite{Liu2017a,Guo2017,Chen2016,Zhang2014c,Parveen2018,Cheng2014,Rocchetto2017} & \cite{Koutsoumpas2015} & \cite{Angelo2018,Pajic2012,Zhang2014c,Parveen2018,Weissnegger2016a,Lora2017,VanDeMortel-Fronczak2014,Ciavotta2018,Koutsoumpas2015,Blackburn2014} & \cite{Liu2017a,Chen2016,Rocchetto2017} & \cite{Angelo2018,Thramboulidis2018,Chen2015,Bunting2016,Feldmann2013,Irisarri2016,Thramboulidis,Cicirelli2018} & \cite{Thramboulidis2018,Liu2017a,Pajic2012,Chen2016,Weissnegger2016a} & \cite{Angelo2018,Guo2017,Pajic2012,Chen2015,Weissnegger2016a,Bunting2016,VanDeMortel-Fronczak2014,Blackburn2014} \\ \hline
\textbf{Sensors} & \cite{Zhang2014,Banerjee2014} & \cite{Sanislav2017} & \cite{Nannapaneni2016} & \cite{Tariq2012} &  & \cite{Tariq2012,Mamun2013,Cicirelli2016,Zhang2013b,Zhang2014,Cicirelli2017a} & \cite{Goncalves2016a,Mamun2013,Zhang2014,Dell2014} & \cite{Mamun2013,Zhang2014,Banerjee2014,Sanislav2017} \\ \hline
\end{tabular}
\end{table}
\end{landscape}

\pagebreak

\begin{landscape}
\begin{table}[]
\tiny
\caption{Correlation analysis between CPS application domains and its challenges}
\label{App:domains_Vs_challenges}
\begin{tabular}{|p{0.15\textwidth}|p{0.10\textwidth}|p{0.10\textwidth}|p{0.10\textwidth}|p{0.10\textwidth}|p{0.10\textwidth}|p{0.10\textwidth}|p{0.10\textwidth}|p{0.10\textwidth}|p{0.10\textwidth}|p{0.10\textwidth}|}
\hline
\textbf{} & \textbf{complexity} & \textbf{Dependability} & \textbf{Flexibility} & \textbf{Interoperability} & \textbf{Latency} & \textbf{Predictability} & \textbf{Reliability} & \textbf{Remote monitoring} & \textbf{Security} & \textbf{Sustainability} \\ \hline
\textbf{Air Transportation (AT)} & \cite{Goncalves2017} & \cite{Zhang2013a} &  & \cite{Zhang2013a,Goncalves2016} & \cite{Mordecai2018} &  & \cite{Zhang2013a} &  &  &  \\ \hline
\textbf{Building Automation (BA)} & \cite{Nagele2017} &  &  &  &  &  &  &  & \cite{Alrimawi2018} &  \\ \hline
\textbf{Critical Infrastructure (CI)} & \cite{Kuesap2008,Koutsoumpas2015} & \cite{Motii2017} &  & \cite{Hartmann2015,Tundis2017,Barve2018} &  &  & \cite{Drago2013,Hartmann2015,Koutsoumpas2015} &  & \cite{Drago2013,Housh2018,Motii2017,Chen2011} & \cite{Tariq2012,Tundis2017,Xin2015} \\ \hline
\textbf{Health Care and Medicine (HC\&M)} & \cite{Guo2017,Murugesan} & \cite{Murugesan} &  &  &  & \cite{Silva2015,Banerjee2014} & \cite{Silva2015,Banerjee2014} &  & \cite{Guo2018} & \cite{Silva2015} \\ \hline
\textbf{Safety-critical Systems} &  &  &  & \cite{Zhao2017} &  &  & \cite{Weissnegger2016a} &  &  &  \\ \hline
\textbf{self-adaptive CPS} & \cite{Zhang2013b} & \cite{Mamun2013} &  &  &  & \cite{Angelo2018} &  &  &  & \cite{Zhang2013b} \\ \hline
\textbf{Smart Environment (SE)} & \cite{Seiger2015} &  &  &  &  &  &  &  &  &  \\ \hline
\textbf{Smart Manufacturing (SM)} & \cite{Buckl2010,Thramboulidis2018} & \cite{Buckl2010} &  & \cite{Walch2017a,Merschak2018,Zhang2013c,Ciavotta2018,Brandenbourger2016,Thramboulidis,Sinha2016} &  &  & \cite{Buckl2010,Chen2016,Bocciarelli2017} & \cite{Irisarri2016} & \cite{Maksuti2017,Rocchetto2017} & \cite{Maksuti2017,Thramboulidis2018,Bocciarelli2017,Lezoche2018} \\ \hline
\end{tabular}
\end{table}
\end{landscape}

\end{document}